\documentclass[aps,pre,twocolumn,groupedaddress]{revtex4-2}

\usepackage{amsmath}
\usepackage{amssymb}
\usepackage{amsthm}
\usepackage{mathtools}
\usepackage{hyperref}

\newcommand\EqRef[1]{Eq.~\eqref{#1}}
\newcommand\ApRef[1]{Appendix~\ref{#1}}

\begin{document}


\title{Traveling Bands, Clouds, and Vortices of Chiral Active Matter}


\author{Nikita Kruk\(^{1}\)}
\author{Jos\'{e} A. Carrillo\(^{2}\)}
\author{Heinz Koeppl\(^{1,}\)}
\email{Author to whom correspondence should be addressed. \\ heinz.koeppl@bcs.tu-darmstadt.de}
\affiliation{\(^{1}\)Department of Electrical Engineering and Information Technology, Technische Universit\"{a}t Darmstadt, Rundeturmstrasse 12, 64283, Darmstadt, Germany}
\affiliation{\(^{2}\)Mathematical Institute, University of Oxford, Oxford OX2 6GG, UK}

\date{\today}

\begin{abstract}
	We consider stochastic dynamics of self-propelled particles with nonlocal normalized alignment interactions subject to phase lag. The role of the lag is to indirectly generate chirality into particle motion. To understand large scale behavior, we derive a continuum description of an active Brownian particle (ABP) flow with macroscopic scaling in the form of a partial differential equation (PDE) for a one-particle probability density function (DF). Due to indirect chirality, we find a new spatially homogeneous nonstationary analytic solution for this class of equations. Our development of kinetic and hydrodynamic theories towards such a solution reveals the existence of a wide variety of spatially nonhomogeneous patterns reminiscent of traveling bands, clouds, and vortical structures of linear active matter. Our model may thereby serve as the basis for understanding the nature of chiral active media and designing multiagent swarms with designated behavior.
\end{abstract}


\maketitle


\section{Introduction}

Synchronized motion of collectives of agents is a widespread phenomenon that can be encountered both in nature and in artificially manufactured systems. The most remarkable examples include bacterial swarming, flocking of birds, schooling of fish, human crowds, and robotic swarms \cite{vicsek:phys_rep}. It is remarkable that all these systems can exhibit similar synchronized behavior despite the inherent diversity of the constituent agents. In order to understand what defines such behavior, we study minimal models of collective motion.
Such models often describe systems that are far from equilibrium and are referred to as active matter. It has become a standard approach to analyze such systems with the Vicsek model (VM) \cite{vicsek:prl} in discrete time or its time continuous counterpart often referred to as an ABP model \cite{romanczuk:epjst}. Models of this type have been extensively analyzed and a number of spatially nonhomogeneous structures like large scale traveling bands or irregular high density clouds have been reported \cite{chate:epjb,mishra:pre,farrell:prl,nagai2015,solon:prl,keeffe:nature_comm}.


ABP models usually describe the motion of linear swimmers. This implies that particles prefer to move in a straightforward way rather than perform circular motion. Due to the lack of possibility for a particle to deliberately undertake circular motion in such models, there has recently been an increase of interest in a new class of models now known as chiral active matter \cite{degond:math_models_and_methods,denk:prl,chen:nature,liebchen:prl,levis:prr,lei:sci_adv,souslov:nature_physics,han:pnas,tociu:prx,nourhani:prl,narinder:prl}. The most prominent examples of such motion are bacterial swarming close to boundaries of a substrate \cite{lauga:biophysical,lemelle:bacteriology}, irregular vortex structures in dense suspensions of swimming bacteria \cite{sumino:nature}, swarming of magnetotactic bacteria in a rotating magnetic field \cite{erglis:biophysical_journal,cebers:jmmm}, swimming of sperm cells \cite{riedel:science,friedrich:pnas}, and shimmering behavior of giant honeybees against predatory wasps \cite{kastberger:plos_one}.

Despite rich diversity of patterns in linear swimmer models, their chiral counterparts have not yet been shown to possess the same variety of nonequilibrium dynamics. Inspired by results on the Kuramoto-Sakaguchi model \cite{kuramoto2002,abrams2004,omelchenko:nonlinearity,omelchenko:pre} for networks of phase oscillators, which we might regard as stationary particles, we generalized it to a self-propelled particle model and reported the existence of chimeric structures, i.e., the coexistence of synchronized and chaotic interacting particle groups even for a zero noise level \cite{kruk:aps2018}. However, we believe that as an ABP model, it might exhibit a much wider class of nonequilibrium behavior.

This paper investigates the continuum limit of a minimal ABP model with alignment interactions only. Its key components are nonlocality of interactions, alignment subject to a homogeneous phase lag, and stochasticity of particle's dynamics. The presence of the phase lag induces particle rotation. We consider its inclusion as an alternative to introducing chirality explicitly through a rotational frequency for each particle \cite{degond:math_models_and_methods,liebchen:prl}. In particular, for the latter models where frequencies are heterogeneous \cite{chen:nature,levis:prr}, rotational symmetry is already broken to start with, whereas our model exhibits spontaneous symmetry breaking.
We analyze our ABP model by deriving its kinetic and hydrodynamic descriptions and performing linear stability analysis of several spatially homogeneous solutions. As a result, we additionally find the existence of a large variety of spatially nonhomogeneous regimes, the most prominent of which are traveling bands of both high and low density, dense clouds, and vortices, as well as multiheaded localized self-propelled chimera states. To the best of our knowledge, most of these patterns have not yet been seen in chiral active particle systems. Note that rotating flocks in \cite{liebchen:prl} are internally homogeneous whereas our dense clouds are not. Moreover, the vortices reported here are stable, they do not disintegrate after several rotations as in \cite{denk:prl}, and particles may join and leave them. We also remark that the phenomenon in \cite{chen:nature} is qualitatively similar to our momentum wave solution but relies on a more complex model.

\begin{figure*}[t]
	\includegraphics[width=1.0\textwidth]{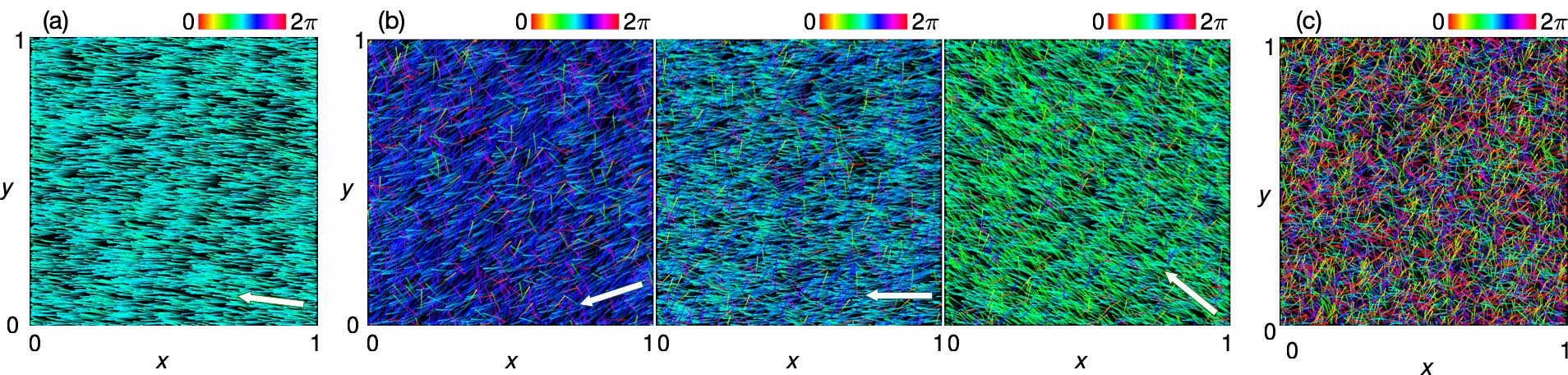}
	\caption
	{
		\label{fig:homogeneous_particle_dynamics} Examples of homogeneous particle dynamics according to \EqRef{eq:chimera_sde}. (a) Linear polar (Vicsek) phase with $\alpha=0$. (b) Nonlocalized chimera state at consecutive time points $t_1<t_2<t_3$ with $\alpha=1.3$ (see the Supplemental Material \cite{supplemental_material,bcs_youtube_channel,*figshare} for a corresponding movie). (c) Disordered motion with $\alpha=1.6$. Particles are colored according to their instantaneous orientations. White arrows indicate the mean direction of motion. Other parameters: $v_0=1.0$, $\sigma=1.0$, $\varrho=0.01$, $D_\varphi=0.01$, $N=5\cdot10^3$.
	}
\end{figure*}

\section{Particle Model}

Let $\mathbb{U} \coloneqq \mathbb{R}/(L\mathbb{Z})$ and $\mathbb{T} \coloneqq \mathbb{R}/(2\pi\mathbb{Z})$ be one-dimensional spaces with periodic boundaries extending from $[0,L]$ and $[0,2\pi]$, respectively.
We consider a system of $N$ particles moving in a two-dimensional space $\mathbb{U}^2$ of fixed size $L$ with periodic boundaries such that the coordinates of a particle $i=1,\dots,N$ are given by $r_i=(x_i, y_i)\in\mathbb{U}^2$. The speed of each particle is assumed to be constant $v_0\in\mathbb{R}_+$ and its velocity is determined by its directional phase $\varphi_i\in\mathbb{T}$. Particles interact with each other within a radius $\varrho$. Therefore, the set of all neighbors for a particle $i$ is defined as
\begin{equation*}
	B_{\varrho}^{i} := \{j \mid j \in \{1,\dots,N\}\backslash i,\; (x_{i} - x_{j}) ^ 2 + (y_{i} - y_{j}) ^ 2 \leq \varrho^2\}.
\end{equation*}
Particles evolve according to the following system of coupled stochastic differential equations (SDEs):
\begin{equation}
\label{eq:chimera_sde}
\begin{aligned}
	\mathrm{d}x_i &= v_0\cos\varphi_i\; \mathrm{d}t \\
	\mathrm{d}y_i &= v_0\sin\varphi_i\; \mathrm{d}t \\
	\mathrm{d}\varphi_i &= \frac{\sigma}{\vert B_{\varrho}^i \vert} \sum_{j \in B_{\varrho}^i} \sin(\varphi_j - \varphi_i - \alpha)\; \mathrm{d}t + \sqrt{2D_\varphi}\; \mathrm{d}W_i.
\end{aligned}
\end{equation}
According to the third equation, each particle adjusts its direction of motion to the average one over its nonlocal neighborhood $B_\varrho^i$, with $\vert B_{\varrho}^i \vert$ denoting the cardinality of the set of all neighbors. Particle interaction is controlled by a coupling strength parameter $\sigma\in\mathbb{R}_+$ and is additionally generalized by adding a phase lag parameter $\alpha\in\mathbb{T}$, which allows for rotation upon particle interaction. Note that this implicitly defines \EqRef{eq:chimera_sde} as a chiral active particle model as long as $\alpha\neq0$. Particles are subject to the external source of randomness with intensity $D_\varphi\in\mathbb{R}_+$, modeled by a family of independent Wiener processes. Our interest is to investigate the stochastic dynamics in the large $N$ limit by preserving nonlocality of particle interactions.

\begin{figure}[b]
	\includegraphics[width=0.5\textwidth]{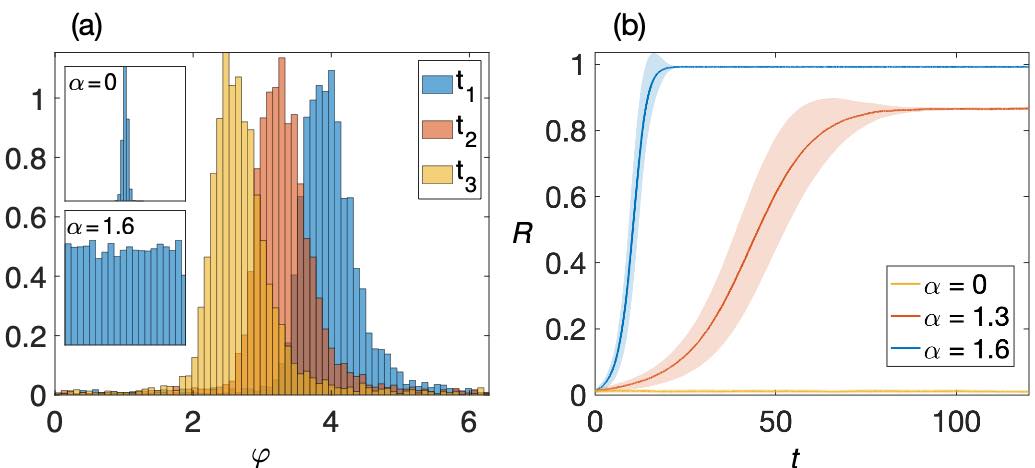}
	\caption
	{
		\label{fig:homogeneous_particle_dynamics_supplement} Description of homogeneous particle dynamics. (a) Particle phase distributions for the dynamics presented in Fig.~\ref{fig:homogeneous_particle_dynamics}. The main figure shows the distributions for respective snapshots in Fig.~\ref{fig:homogeneous_particle_dynamics}(b). The insets show the respective distributions for particle dynamics in Figs.~\ref{fig:homogeneous_particle_dynamics}(a),(c). (b) Evolution of the polar order parameter magnitude over time starting from random initial conditions for different values of phase lag $\alpha$. The curves show its mean value out of 100 experiments, with standard deviation shown as shaded regions. Model parameters are the same as in Fig.~\ref{fig:homogeneous_particle_dynamics}.
	}
\end{figure}

It has been found \cite{kruk:aps2018}, that \EqRef{eq:chimera_sde} gives rise to such phases as spatially homogeneous ordered motion (SHOM), which can be rotational or linear, spatially homogeneous disordered motion (SHDM), and (self-propelled) localized and nonlocalized chimera states. In order to make distinction between various phases, we consider a complex polar order parameter
\begin{equation*}
	R(t)e^{i\Theta(t)} = \frac{1}{N} \sum_{j=1}^{N} e^{i\varphi_j(t)}.
\end{equation*}
Its magnitude $R\in[0,1]$ indicates the extent to which particles align along the mean direction $\Theta\in\mathbb{T}$. If the polar order is absent, $R=0$; if particles become completely synchronized, $R=1$; otherwise, one observes partial synchronization with respect to particle orientations. In the stochastic setup with $D_\varphi>0$, polar order is characterized by some distribution over particle orientations, and one might say that the system exhibits partial synchronization since $R<1$. If the noise is removed from the system, polar order is characterized by a point mass distribution with respect to particle orientations. By considering sufficiently large phase lag values, it is possible to observe a phenomenon known as a chimera state. In such systems, particles decompose into two interacting populations. The first one is characterized by complete synchronization while the second one remains disordered. Next, we provide the details on the aforementioned phases.

For $\alpha=0$, one might consider \EqRef{eq:chimera_sde} as a time continuous variation of the Vicsek model \cite{degond:math_models_and_methods}. In this case, two solutions are possible, i.e., one observes the formation of polar order (cf. Figs.~\ref{fig:homogeneous_particle_dynamics}(a),\ref{fig:homogeneous_particle_dynamics_supplement}(b)), when noise is sufficiently small, or particles exhibit disordered motion (cf. Fig.~\ref{fig:homogeneous_particle_dynamics}(c)). We note that due to such normalized particle alignment as in \EqRef{eq:chimera_sde}, we have not observed the formation of traveling bands next to order-disorder transition known for the Vicsek model (see the discussion in Section~\ref{sec:stability_analysis_of_the_traveling_wave_solution}). In the deterministic case $D_\varphi=0$, particles orient in the same direction, resulting in complete synchronization with $R=1$ attaining its maximal value.

By introducing phase lag $\alpha$, the synchronized particle system starts to rotate with some constant group velocity $v$, which conversely depends on $\alpha$ (cf. Fig.~\ref{fig:homogeneous_particle_dynamics}(b) and a corresponding movie in \cite{supplemental_material,bcs_youtube_channel,*figshare}). By increasing $\alpha$ as well as by increasing $D_\varphi$, particles become less ordered. In \cite{kruk:aps2018}, we referred to such a spatially homogeneous rotating solution as the nonlocalized chimera state. For $|\alpha|\geq\pi/2$, particles do not synchronize (cf. Figs.~\ref{fig:homogeneous_particle_dynamics}(c),\ref{fig:homogeneous_particle_dynamics_supplement}(b)). Note that the chimera state is a purely deterministic construct, i.e., the separation of particles into synchronized and disordered populations occurs for sufficiently high values of phase lag in the absence of noise $D_\varphi=0$. Therefore, in the deterministic setup, one differentiates between complete synchronization of rotating particles (with $R=1$) and the nonlocalized chimera state (with $R<1$) where such synchronization coexists with disordered group of particles. In the stochastic setup, however, both solutions consist of partially synchronized particles whose phases follow some skewed unimodal distribution (cf. Fig.~\ref{fig:homogeneous_particle_dynamics_supplement}(a)) and we cannot differentiate between them anymore. In this paper, we investigate the stochastic particle dynamics only and we will generally refer to such solutions as SHOM. We would like to mention that for sufficiently large $\alpha$ by varying the radius of interaction $\varrho$, one might observe a spatially nonhomogeneous localized chimera state. In the rest of the main text, we will not discuss it and refer the interested reader to \cite{kruk:aps2018} for its detailed description and to \ApRef{sec:spatially_nonhomogeneous_particle_dynamics} for an example of such dynamics.


To reduce the number of independent parameters, we choose time and space units as $1/\sigma$ and $L$, respectively. Thus, the model has four control parameters, e.g., the particle velocity $\hat{v}_0 = v_0/(L\sigma)$, the radius of interaction $\varrho$, the phase lag $\alpha$, and the rotational diffusion rate relative to the coupling strength $\hat{D}_\varphi = D_\varphi/\sigma$. We will study a continuum limit of \EqRef{eq:chimera_sde}, where each particle is considered to be a point mass. In this case, we can find a limit with $N\rightarrow\infty$ with the system size fixed $L=\text{const}$. Therefore, we put $L=1$ without loss of generality. Similar limits for weakly interacting particle systems with a large radius of interaction are known in kinetic theory as Vlasov limits \cite{dobrushin,lancellotti:ttsp}. Note that a particle density, usually defined as $\rho_0 = N/L^2$, which plays an important role in the standard VM \cite{vicsek:prl}, does not arise here as an independent parameter due to the probabilistic interpretation of the density function in this setup (therefore, it is now fixed as $\rho_0 \equiv 1$). It should be treated as the average number of particles per unit length in the system of fixed size $L\times L=1$ \cite{kipnis1998scaling} divided into $\sqrt{N}\times\sqrt{N}$ units in two dimensions.

\section{Continuum Limit}

To understand mechanisms leading to spatially nonhomogeneous behavior in the large $N$ limit, we derive a continuum limit \cite{laney:comp_gas_dyn} of the Langevin dynamics \EqRef{eq:chimera_sde} within the framework of Fokker-Planck equations \cite{risken}, and look subsequently for solutions of a resulting PDE. The approach we follow here \cite{archer:jpa} (see \ApRef{sec:continuum_limit_derivation}), provides us with the hierarchy of evolution equations for $n$-particle DFs that incorporate interactions of any order. Admitting a molecular chaos assumption \cite{spohn:springer}, we close the hierarchy at the first order and obtain a differential equation for a desired one-particle DF $f=f(r,\varphi,t)$

\begin{equation*}
\begin{aligned}
	\partial_t &f = -v_0e(\varphi)\cdot\nabla_{r}f + D_\varphi \partial_{\varphi\varphi} f \\
	&- \partial_{\varphi} \left(\frac{f}{\vert C(r) \vert} \int_{C(r)} \sin(\varphi' \!-\! \varphi \!-\! \alpha)f(r',\varphi',t)\; \mathrm{d}r'\mathrm{d}\varphi' \right),
\end{aligned}
\end{equation*}
where $r=(x,y)\in\mathbb{U}^2$ is a position vector, $e(\varphi)=(\cos\varphi,\sin\varphi)\in\mathbb{S}^1\subset\mathbb{R}^2$ is a unit velocity vector in the direction of $\varphi\in\mathbb{T}$, $\nabla_{r}=(\partial_x,\partial_y)$ denotes a spatial gradient, and the nonlocal neighborhood domain is defined as $C(r)=\left\{ (r',\varphi')\in\mathbb{U}^2\times\mathbb{T} \mid \Vert r'-r \Vert \leq \varrho \right\}$. The normalization by the neighborhood mass corresponds to the respective normalization in the alignment term of the Langevin dynamics \EqRef{eq:chimera_sde} and it reads
\begin{equation*}
	\vert C(r) \vert = \int_{C(r)} f(r',\varphi',t)\; \mathrm{d}r'\mathrm{d}\varphi'.
\end{equation*}

The continuum limit equation has two spatially homogeneous fixed points, i.e., $f=f(\varphi,t)$. The first one is trivial and is a uniform probability DF $f(\varphi,t) = 1/(2\pi)$. It corresponds to disordered motion of a particle system. The second solution is a von Mis\'{e}s DF 
\begin{equation}
\label{eq:von_mises_df}
	f(\varphi,t) = \frac{\exp(R\cos(\varphi-\Theta)/\hat{D}_\varphi)}{2\pi I_0(R/\hat{D}_\varphi)},
\end{equation}
where $I_0$ is the modified Bessel function of the first kind, and the parameters $R$ and $\Theta$ are redefined as the magnitude and direction of the polar order parameter according to
\begin{equation*}
	R(t)e^{i\Theta(t)} = \int_{\mathbb{T}} e^{i\varphi}f(\varphi,t)\; \mathrm{d}\varphi.
\end{equation*}
The latter solution is valid only for $\alpha=0$ and is a solution to the time continuous VM \cite{degond:math_models_and_methods} or the Kuramoto model (KM) for coupled noisy phase oscillators \cite{bertini:journal_of_stat_phys,giacomin:nonlinearity,gupta:first_order}. It corresponds to polarized motion of particles, where the degree of polarization is given by $R$, and $\Theta$ is the direction of collective motion. Note that in the limit of zero noise $\hat{D}_\varphi\rightarrow0_+$, one obtains complete synchronization of a system.

\begin{figure}[t]
	\includegraphics[width=0.5\textwidth]{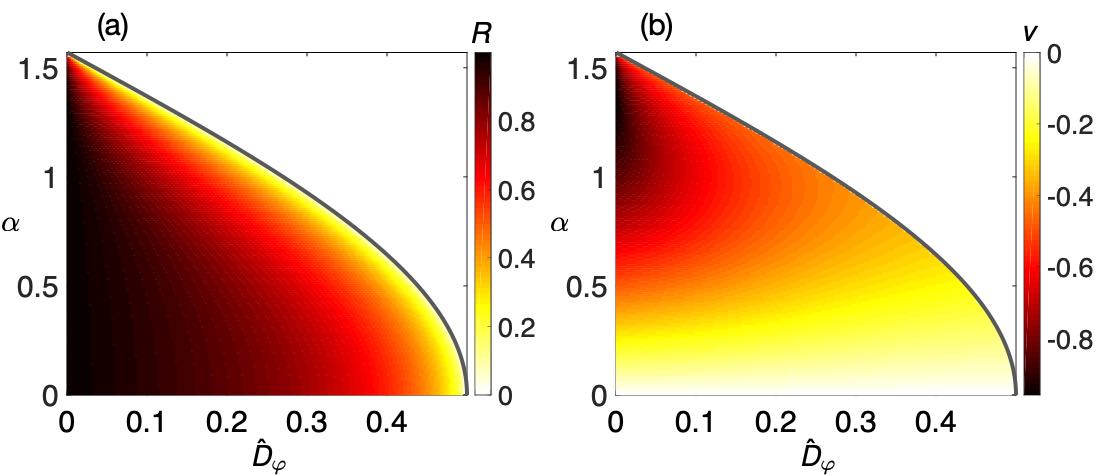}
	\caption
	{
		\label{fig:selfconsistent_equations_nonzero_alpha} Solution of the system of self-consistent equations (see \ApRef{sec:solutions_for_the_continuum_limit_pde}) comprising a DF as a traveling wave solution \EqRef{eq:solution_profile_uniform_nonzero_lag} and a complex order parameter defined by that solution.
		(a) The order parameter magnitude $R$ and (b) the group velocity $v$ versus the phase lag $\alpha$ and the noise strength $D_\varphi$. The dark gray line indicates the order-disorder transition line $D_\varphi=\frac{1}{2}\cos\alpha$. The critical group velocity along that line is $v=-\frac{1}{2}\sin\alpha$.
	}
\end{figure}

\begin{figure}[b]
	\includegraphics[width=0.5\textwidth]{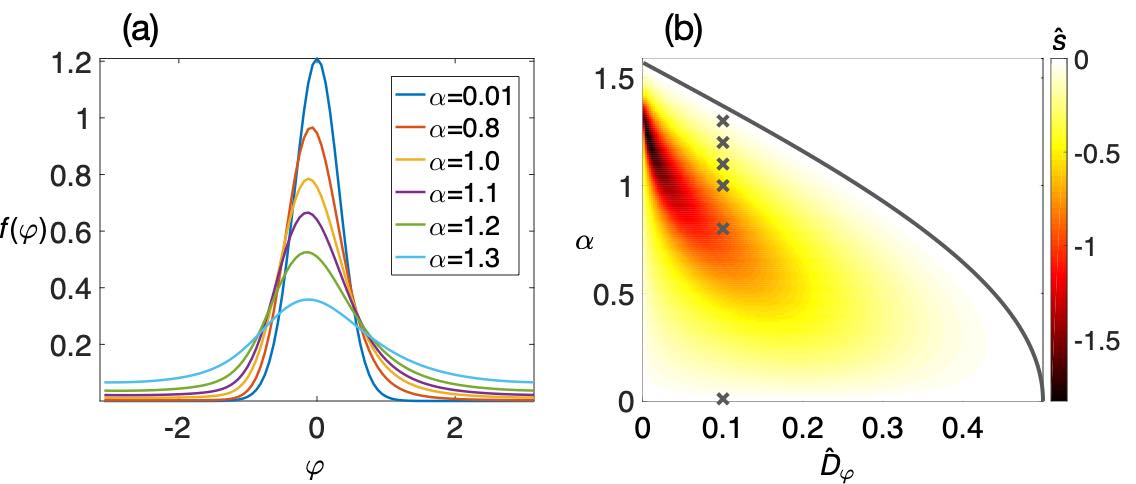}
	\caption
	{
		\label{fig:homogeneous_solution_nonzero_alpha}
		Examples of a spatially homogeneous traveling wave solution \EqRef{eq:solution_profile_uniform_nonzero_lag}. (a) DFs for the different values of the phase lag $\alpha$. (b) The circular skewness of \EqRef{eq:solution_profile_uniform_nonzero_lag} versus the phase lag and the noise, quantified by a circular skewness coefficient $\hat{s}=\mathbb{E}\left[\sin2(\varphi-\Theta)\right] / \left(1-\left\vert \mathbb{E}\left[e^{i\varphi}\right]\right\vert\right)^{3/2}$ \cite{mardia2009directional}. Crosses indicate the parameter values used to generate the DFs in (a). The dark gray line indicates the order-disorder transition line, to compare with Fig.~\ref{fig:selfconsistent_equations_nonzero_alpha}.
	}
\end{figure}

A more interesting regime is the one with nonzero phase lag, which introduces constant motion of a DF with some group velocity $v\in\mathbb{R}$, the sign of which conversely depends on $\alpha$. 
Introducing a traveling wave ansatz into the PDE and solving the resulting equation (see \ApRef{sec:solutions_for_the_continuum_limit_pde}), we find
\begin{equation}
\label{eq:solution_profile_uniform_nonzero_lag}
	f(\varphi,t) = c_0 E(\varphi,t)
	\left( 1 + c_1 \frac{\int_{0}^{\varphi-vt} E^{-1}(\varphi,0)\; \mathrm{d}\varphi}{\int_{\mathbb{T}} E^{-1}(\varphi,0)\; \mathrm{d}\varphi} \right),
\end{equation}
where $c_0\in\mathbb{R}$ is a normalization constant, $c_1=\exp(2\pi v / \hat{D}_\varphi) - 1$ accounts for a periodicity constraint $f(0,t)=f(2\pi,t)$, and
\begin{equation*}
	E(\varphi,t) = \exp\left[ -v\varphi/\hat{D}_\varphi + R \cos(\varphi - vt + \alpha) / \hat{D}_\varphi \right]
\end{equation*}
is an auxiliary function. \EqRef{eq:solution_profile_uniform_nonzero_lag} is a continuum limit representation of a nonlocalized chimera state reported in \cite{kruk:aps2018}. The solution depends on the order parameter magnitude $R$, which is in turn defined in terms of this DF. To be able to use this solution, we must solve the system of self-consistent equations for $f$ and the complex order parameter $R(t)e^{i\Theta(t)} = \int_{\mathbb{T}} e^{i\varphi}f(\varphi,t)\; \mathrm{d}\varphi$, the solution of which is presented in Fig.~\ref{fig:selfconsistent_equations_nonzero_alpha}. The resulting DF is a $2\pi$-periodic skewed function (cf. Fig.~\ref{fig:homogeneous_solution_nonzero_alpha}(a)). By expanding the self-consistent equations with respect to $R$ around $R=0$, we find a line indicating the onset of orientational order $\hat{D}_\varphi = \frac{1}{2}\cos\alpha$ as well as a critical group velocity $v = -\frac{1}{2}\sin\alpha$ from within the region of existence of \EqRef{eq:solution_profile_uniform_nonzero_lag}. One can check that in the Vicsek regime $\alpha=0$, \EqRef{eq:solution_profile_uniform_nonzero_lag} simplifies to the von Mis\'{e}s DF \EqRef{eq:von_mises_df}. Note that because \EqRef{eq:solution_profile_uniform_nonzero_lag} is not symmetric, first and second moments do not characterize it completely. The third moment allows us to quantify the extent, to which particle motion deviates from polar order (cf. Fig.~\ref{fig:homogeneous_solution_nonzero_alpha}(b)).

\section{Stability Analysis of the Traveling Wave Solution\label{sec:stability_analysis_of_the_traveling_wave_solution}}

To reveal the emergence of spatially nonhomogeneous patterns, we perform stability analysis of \EqRef{eq:solution_profile_uniform_nonzero_lag} as a solution to a spatially dependent PDE.
First, we discuss the hydrodynamic theory approach (see \ApRef{sec:stability_analysis_via_hydrodynamic_theory}), in which we elaborate the continuum limit description of a particle system in terms of a marginal DF $\rho(r,t)$ and a momentum field $w(r,t)$.

Under the large diffusion approximation (see \ApRef{subsec:stability_analysis_via_hydrodynamic_theory_stationary_solutions}), we find the following closed system of hydrodynamic equations
\begin{equation}
\label{eq:hydrodynamic_equations}
\begin{aligned}
	\partial_t \rho &= -\hat{v}_0 \nabla\cdot w, \\
	\partial_t w &= -\frac{\hat{v}_0}{2} \nabla\rho - D_\varphi w + \frac{\hat{v}_0^2}{16D_\varphi} \Delta w + \frac{\rho}{2} Q_{-\alpha} W \\
	&+ \frac{\hat{v}_0}{8D_\varphi} \Biggl\{ \frac{1}{2} Q_\alpha \left[ (W\cdot\nabla)w + (W_\perp\cdot\nabla)w_\perp \right] \\
	&+ Q_{-\alpha} [ \nabla(w\cdot W) - (W\cdot\nabla)w - (\nabla\cdot W)w \\
	&- W(\nabla\cdot w) - (w\cdot\nabla)W ] \Biggr\} - \frac{1}{8D_\varphi} w\Vert W\Vert^2,
\end{aligned}
\end{equation}
where $w_\perp=(-w_y,w_x)^T$ and $W_\perp=(-W_y,W_x)^T$ denote vectors orthogonal to $w$ and $W$, respectively. We have denoted a spatially averaged momentum field as
\begin{equation*}
	W = W(r,t) = \frac{\iint_{B(r;\varrho)} w(r',t)\; \mathrm{d}r'}{\iint_{B(r;\varrho)} \rho(r',t)\; \mathrm{d}r'}
\end{equation*}
arising due to the nonlocal interaction term in \EqRef{eq:chimera_sde}. The neighborhood domain is defined as $B(r;\varrho) = \left\{ r' \in \mathbb{U}^2 \mid \Vert r' - r \Vert \leq \varrho \right\}$. The matrix $Q_\alpha$ represents anticlockwise rotation by $\alpha$ radians. 
Note that the particle density $\rho_0$ does not appear in \EqRef{eq:hydrodynamic_equations} due to the type of the continuum limit we derived \cite{kipnis1998scaling}.
We have the following terms in the right hand side of the momentum equation.
The first term is a pressure gradient. The second and the last terms constitute the relaxation of the momentum field. The third term represents the damping of collective motion. The fourth term generates coupling between density and momentum fields. The rest of the terms appear as a result of the broken Galilean invariance. Up to the rotational operation and integration over a nonlocal neighborhood, they constitute all three combinations of one spatial gradient and two momenta, as described in \cite{toner:pre}.
\EqRef{eq:hydrodynamic_equations} allows for the stability analysis of the stationary solutions, i.e., either disordered or synchronized motion for $\alpha=0$. When $\alpha\neq0$, the particle flow is described by the nonstationary solution \EqRef{eq:solution_profile_uniform_nonzero_lag}, and we cannot apply the same stability analysis to it directly. Therefore, we rederive hydrodynamic equations in a moving reference frame in which such a solution becomes stationary. The form of those hydrodynamic equations is functionally similar to \EqRef{eq:hydrodynamic_equations} except for couplings between longitudinal and transversal directions as the result of applying a suitable ansatz (see \ApRef{sec:stability_analysis_via_hydrodynamic_theory}).

Apart from the impact of $\alpha$ and the ansatz, the apparent distinction of \EqRef{eq:hydrodynamic_equations} from the majority of equations of the Toner-Tu kind is the frequent appearance of the nonlocally averaged momentum field $W$. This is the result of the continuum limit approach that allowed us to preserve nonlocality of interactions. From \EqRef{eq:hydrodynamic_equations}, we see that both $\hat{v}_0$ and $\varrho$ influence the length scale. Therefore, if we rescale spatial variables and introduce a normalized radius $\tilde{\varrho} = \varrho/\hat{v}_0$, we conclude that there are three independent parameters in our model, i.e., the phase lag $\alpha$, the noise strength $\hat{D}_\varphi$, and the normalized radius $\tilde{\varrho}$.

The hydrodynamic equations in a moving reference frame have two stationary spatially homogeneous solutions. The first one is $(\rho,w) = (1,0,0)$ and it represents spatially homogeneous disordered motion of particles. The second solution represents partially synchronized collective motion $(\rho,w) = (1, \Vert w^*\Vert\cos\varphi_0, \Vert w^*\Vert\sin\varphi_0)$, where the degree of polarization is found to be
\begin{equation}
\label{eq:solution_hydrodynamic_polarization_traveling_wave}
	\Vert w^*\Vert = \sqrt{\frac{1}{\hat{D}_\varphi}(4\hat{D}_\varphi^2 + v^2)(\cos\alpha - 2\hat{D}_\varphi)}
\end{equation}
and $\varphi_0\in\mathbb{T}$ is an arbitrary direction subject to initial conditions. In this regime, the macroscopic fraction of particles synchronizes in phase and rotates steadily with frequency $v$. One of the assumptions that we have used to derive the hydrodynamic equations is that diffusion is strong enough to guarantee the negligence of higher order Fourier modes, i.e., $n\geq3$. 
The limitations are that \EqRef{eq:solution_hydrodynamic_polarization_traveling_wave} is valid only close to the order-disorder transition line $\hat{D}_\varphi=\frac{1}{2}\cos\alpha$ up to $\hat{D}_\varphi=\frac{1}{4}\cos\alpha$, where it reaches its maximum. From Fig.~\ref{fig:selfconsistent_equations_nonzero_alpha}(a), we see that the polarization must actually increase further with $\hat{D}_\varphi\rightarrow0_+$ for fixed $\alpha$. 
Note that in a linear regime $\alpha=0$, particles do not rotate, i.e., $v=0$, and we retrieve the well-known polarization level for the VM and the KM as $\Vert w^*\Vert = 2\sqrt{\hat{D}_\varphi(1 - 2\hat{D}_\varphi)}$.

The linear stability analysis from the point of view of the hydrodynamic theory of the disordered state as well as the partially synchronized state for $\alpha=0$ does not reveal any additional instabilities.
The latter result appears as a contradiction to the one obtained for the standard VM, which was shown to exhibit longitudinal long wavelength instabilities leading to the emergence of traveling bands. The explanation for this lies in the type of the continuum limit we derived, and the subsequent requirement to have the normalization in the alignment term. For many time continuous modifications of the VM, in the limit $N/L^2 = \text{const}$ for $N,L\rightarrow\infty$, one does not use the normalization by the number of particles to handle the alignment term during the transition $N\rightarrow\infty$. In our case, we do not assume $N/L^2 = \text{const}$. Therefore, in order to keep the alignment term finite in the transition $N\rightarrow\infty$, we have to have the normalization by the number of particles $\vert B_{\varrho}^i\vert$. We conclude that the presence of the normalization term in the continuum limit PDE makes spatially homogeneous partially synchronized motion more stable against spatially nonhomogeneous perturbations compared to continuum limit PDEs without such normalization.


\begin{figure}[t]
	\includegraphics[width=0.5\textwidth]{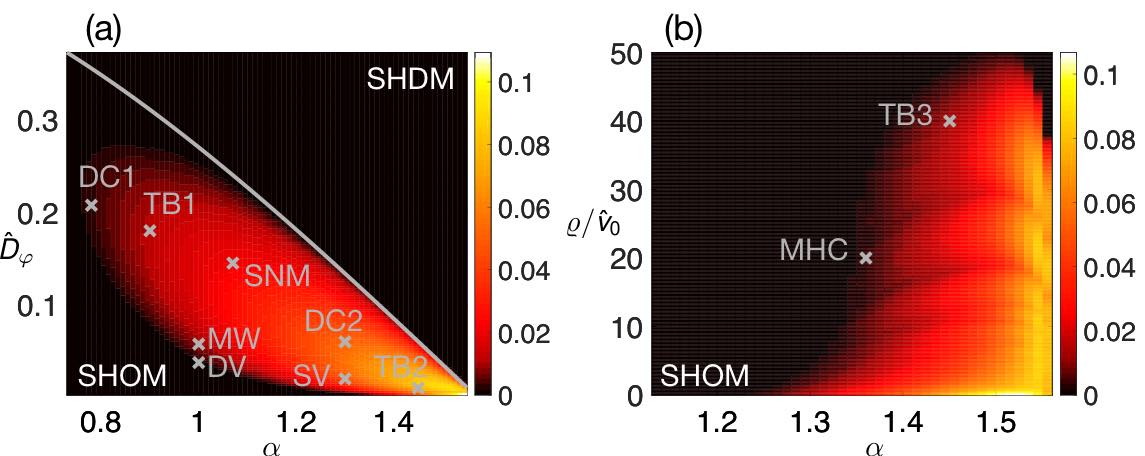}
	\caption
	{
		\label{fig:phase_diagram}
		Phase diagrams in the parameter space of (a) the noise strength $\hat{D}_\varphi$ and the phase lag $\alpha$, and (b) the rescaled radius of interaction $\varrho/\hat{v}_0$ and the phase lag, as predicted by the kinetic theory (see \ApRef{sec:stability_analysis_via_kinetic_theory}). Color shows the maximal real part of the strongest unstable mode. 
		The gray line is the order-disorder transition line $\hat{D}_\varphi=\frac{1}{2}\cos\alpha$.
		Above the line in (a), spatially homogeneous disordered motion (SHDM) is stable; below the line in (a) and to the left in (b), spatially homogeneous ordered motion (SHOM) given by \EqRef{eq:solution_profile_uniform_nonzero_lag} is stable in the black region. 
		Gray crosses indicate parameter values, selected to exemplify particle dynamics in Fig.~\ref{fig:particle_dynamics}.
	}
\end{figure}

For $\alpha\neq0$, the linear stability analysis \cite{bertin:jpa,mishra:pre,grossmann:iop} of \EqRef{eq:solution_hydrodynamic_polarization_traveling_wave} shows that a parameter regime where instabilities could occur lies on the margins of validity of the hydrodynamic equations.
Therefore, we turn to the kinetic theory \cite{degond:arma} (see Appendices \ref{sec:stability_analysis_via_kinetic_theory} and \ref{sec:solutions_of_eigenvalue_problems}). The solution \EqRef{eq:solution_profile_uniform_nonzero_lag} is stable against spatially homogeneous perturbations for $D_\varphi<\frac{1}{2}\cos\alpha$.
Regarding spatially nonhomogeneous perturbations, the linear stability analysis is summarized in the phase diagrams in Fig.~\ref{fig:phase_diagram}. All spatially dependent instabilities occur for $\alpha$ sufficiently large. As one approaches $\alpha\rightarrow\pi/2$, the number of unstable wave vectors and corresponding maximal real parts of dispersion relations increase. The phase diagrams were obtained by considering perturbations of any direction. Note that since we consider periodic boundary conditions, wave vectors are discrete $(k_x,k_y)\in\mathbb{Z}^2$. As we wanted to emphasize from the very beginning, varying $\varrho$ may lead to new system behavior. Such results are summarized in Fig.~\ref{fig:phase_diagram}(b). Both phase diagrams demonstrate regions where the spatially homogeneous solution \EqRef{eq:solution_profile_uniform_nonzero_lag} becomes unstable subject to spatially dependent perturbations whose Fourier transforms contain concrete unstable modes (cf. Fig.~\ref{fig:dispersion_relations_unstable} for examples of such modes). However, neither does it mean that spatially nonhomogeneous solutions exclusively exist inside such instability regions nor does it mean that spatially homogeneous solutions exist only outside them. For an example of such a conclusion for the Vicsek model with nematic alignment see, e.g., \cite{peshkov:prl}.
 
Exemplary particle dynamics can be found in Fig.~\ref{fig:particle_dynamics} and respective movies can be found in \cite{supplemental_material,bcs_youtube_channel,*figshare}. We do not go into the details of analyzing each of those states because it extends beyond the scope of the paper. We only comment on their key features. One of the states is a cloud of high density (DC1 and DC2). In both cases, particles self-organize into circular shapes of high density (cf. Fig.~\ref{fig:coarse_grained_fields}), which we call clouds. While a momentum field is quite homogeneous for DC1, it has a clear radial structure for DC2. The same holds true for traveling bands TB1 and TB2. The dense part of TB1 is characterized with a uniform momentum field while TB2 has points with the radial change of a momentum field. Moreover, we have found a traveling band of low density TB3 for large $\tilde{\varrho}$ values only. The other dynamics include (i) a multiheaded chimera state (cf. Fig.~\ref{fig:particle_dynamics}(f)) characterized by the formation of several synchronized and spatially localized groups that rotate with constant frequency. This state is the generalization of a localized chimera state reported in \cite{kruk:aps2018}. By decreasing $\tilde{\varrho}$, one increases the number of chimeric heads. By increasing $\alpha$, the chaotic background becomes more pronounced until the heads become unstable and one observes giant number fluctuations in the density field. (ii) There are vortical structures where each one is either static (cf. Fig.~\ref{fig:particle_dynamics}(g)) in shape or periodically expands and shrinks (cf. Fig.~\ref{fig:particle_dynamics}(h)). By changing $\tilde{\varrho}$, one can control the number of vortices appearing. (iv) Particles may organize in structures of uniform density but with the direction of a momentum field uniformly distributed horizontally or vertically (cf. Fig.~\ref{fig:particle_dynamics}(i)). (v) We also find a configuration with a spatially homogeneous density but a nonhomogeneous momentum field (cf. Fig.~\ref{fig:particle_dynamics}(j)).

\section{Conclusions}

In this paper, we have considered the ABP model with alignment interactions subject to phase lag $\alpha$. Such interactions facilitate chirality of particle motion which is manifested only as a collective phenomenon as opposed to other chiral ABP models with explicit rotational frequencies. We showed that in the continuum limit, there are two spatially homogeneous system states, i.e., with particles moving chaotically or self-organizing into uniformly rotating polar clusters. The transition between these two states is of second order and depends not only on the interplay between coupling and noise coefficients but also on the phase lag. When the lag is zero, our model becomes a continuous time variation of the Vicsek model. Moreover, for sufficiently large phase lags, the spatially homogeneous ordered motion becomes linearly unstable against spatially dependent perturbations, and we observe a wide range of spatially nonhomogeneous patters, e.g., traveling bands, dense clouds, vortical motion, irregular momentum fields, and multiheaded chimera states.


\begin{figure*}[]
	\includegraphics[width=1.0\textwidth]{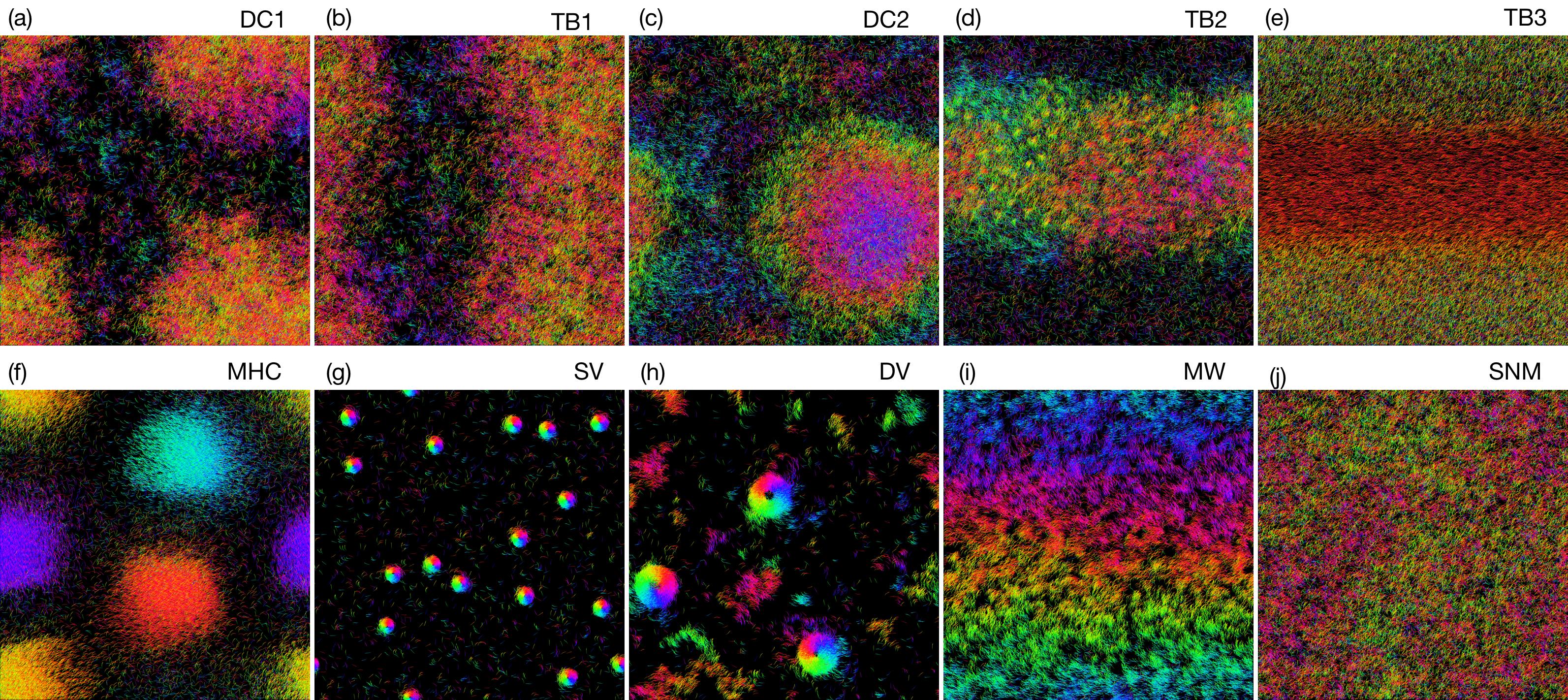}
	\caption
	{
		\label{fig:particle_dynamics} Particle dynamics according to \EqRef{eq:chimera_sde} (see the Supplemental Material \cite{supplemental_material,bcs_youtube_channel,*figshare} for corresponding movies). Abbreviations stand for a dense cloud (DC), a traveling band (TB), a multiheaded chimera (MHC), a stationary vortex (SV), a dynamic vortex (DV), a momentum wave (MW), and a spatially nonhomogeneous momentum (SNM). Color indicates the direction of motion in the HSV color map. Each particle is represented as an elongated object over several time points. Parameter values correspond to those marked on the phase diagrams in Fig.~\ref{fig:phase_diagram}. Parameters: $N=5\cdot10^4$, $L=1$, $\tilde{\varrho}=1$, $\hat{v}_0=0.01$, (a) $\varrho=0.01,\alpha=0.78,\hat{D}_\varphi=0.2075$, (b) $\varrho=0.01,\alpha=0.9,\hat{D}_\varphi=0.18$, (c) $\varrho=0.01,\alpha=1.3,\hat{D}_\varphi=0.06$, (d) $\varrho=0.01,\alpha=1.45,\hat{D}_\varphi=0.01$, (e) $\varrho=0.4,\alpha=1.45,\hat{D}_\varphi=0.005$, (f) $\varrho=0.2,\alpha=1.36,\hat{D}_\varphi=0.005$, (g) $\varrho=0.01,\alpha=1.3,\hat{D}_\varphi=0.02$, (h) $\varrho=0.01,\alpha=1.0,\hat{D}_\varphi=0.0375$, (i) $\varrho=0.01,\alpha=1.0,\hat{D}_\varphi=0.0575$, and (j) $\varrho=0.01,\alpha=1.07,\hat{D}_\varphi=0.145$.
	}
\end{figure*}

We have illustrated that nonlocalized interactions in the large $N$ limit play a significant role as well. Namely, we have found that the length scale of each presented pattern inversely depends on $\varrho/\hat{v}_0$ meaning that the microscopic particle velocity alone is not enough to characterize the dynamics. Moreover, the presence of both $\varrho$ and $\hat{v}_0$ allows us to build a connection between the KM for the stationary phase oscillators and the time continuous variations of the VM known so far. By keeping interactions normalized by a neighborhood cardinality, we reach a conclusion that the particle behavior qualitatively differs from the one where interactions are not normalized. Namely, the presence of normalization makes dynamics more robust against spatial perturbations. 

We have reached the point where we have discovered a wide range of spatially nonhomogeneous patterns, many of which not described for chiral active matter systems yet. Thereupon, the detailed analysis of each of them is needed, as well as the study of related phase transitions. As we have mentioned, some of those patterns bear resemblance to solutions of other models, which needs to be thoroughly investigated. Another interesting question is how the presented analysis compares to the previous approaches where the continuum limit is taken under hydrodynamic scaling, and to investigate which of the reported patterns would survive such a transition. However, the answers to these questions go beyond the scope of this paper and would be subject to future research. The present work does not claim to give a universal model of collective chiral behavior in the large $N$ limit but invites further studies to characterize various kinds of related continuum dynamics.


\begin{acknowledgments}
JAC was partially supported by EPSRC grant number EP/P031587/1 and the Advanced Grant Nonlocal-CPD (Nonlocal PDEs for Complex Particle Dynamics: 
	Phase Transitions, Patterns and Synchronization) of the European Research Council Executive Agency (ERC) under the European Union's Horizon 2020 research and innovation programme (grant agreement No. 883363).
\end{acknowledgments}

\appendix
\section{Spatially Nonhomogeneous Particle Dynamics\label{sec:spatially_nonhomogeneous_particle_dynamics}}

In this appendix, we provide the snapshots of exemplary particle dynamics mentioned in the main text and briefly describe the differences in collective behavior for pairs of solutions having qualitatively similar macroscopic structure by looking at their coarse grained hydrodynamic description in terms of $\rho(r,t)$ and $w(r,t)$.

\begin{figure*}[t]
	\includegraphics[width=1.0\textwidth]{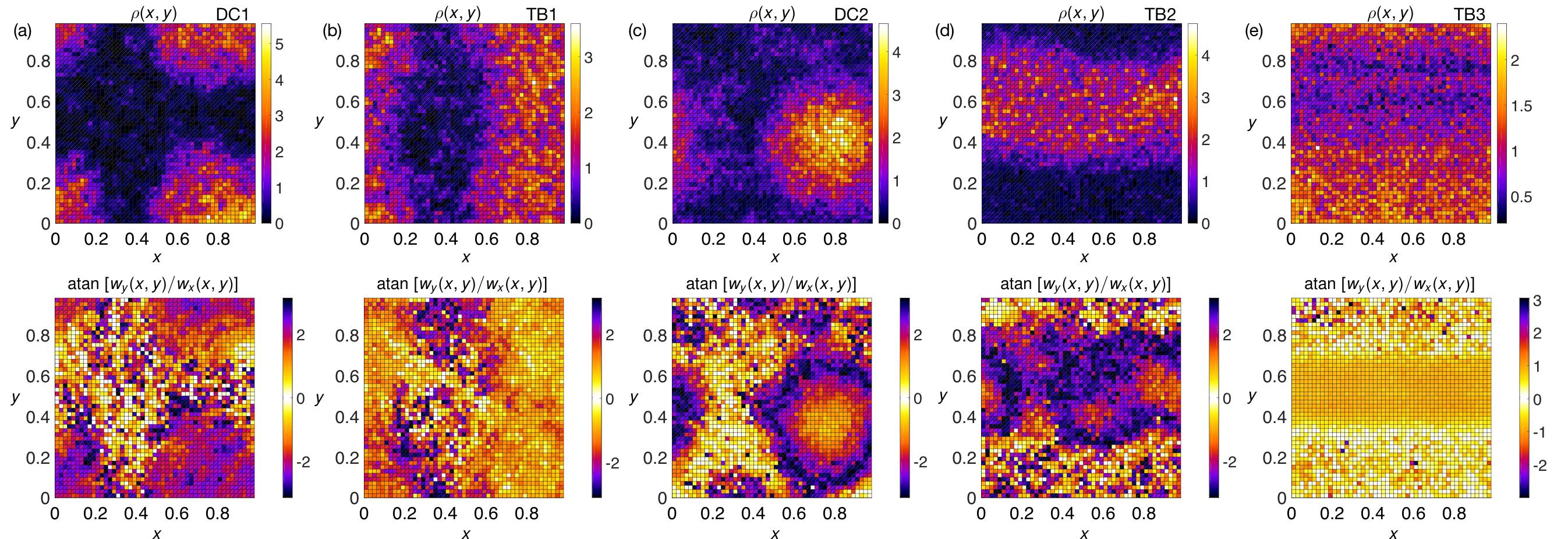}
	\caption
	{
		\label{fig:coarse_grained_fields} Coarse grained marginal density function $\rho(r,t)$ (the upper row) and the direction of a momentum field $w(r,t)$ (the lower row). Parameters: $N=5\cdot10^4$, $\tilde{\varrho}=1$, $v_0=0.01$, (a) $\varrho=0.01,\alpha=0.78,D_\varphi=0.2075$, (b) $\varrho=0.01,\alpha=0.9,D_\varphi=0.18$, (c) $\varrho=0.01,\alpha=1.3,D_\varphi=0.06$, (d) $\varrho=0.01,\alpha=1.45,D_\varphi=0.01$, and (e) $\varrho=0.4,\alpha=1.45,D_\varphi=0.005$.
	}
\end{figure*}

First, we find two types of solutions where particles accumulate into clouds of high density (DC1 and DC2 in Figs.~\ref{fig:particle_dynamics}(a) and (c), respectively). Inside both such clouds, particles are distributed quite uniformly with respect to $r$ but the momentum fields structurally differ (cf. Fig.~\ref{fig:coarse_grained_fields}(a) and (c), respectively). For DC1, the mean direction $\arg(w(r,t))$ is also quite uniform. Therefore, on average, particles inside the cloud are oriented similarly but due to the small microscopic velocity $v_0$ they stay in the cloud for a long time. For DC2, the momentum field clearly possesses a radial structure. During such motion, central particles first define the orientation which later (in time) is assumed by particles further away from the center.

Second, we observe three types of traveling bands (TB1, TB2, and TB3 in Figs.~\ref{fig:particle_dynamics}(b),(d), and (e), respectively). TB1 and TB2 are characterized by the formation of bands of high density, which align horizontally or vertically depending on initial conditions. The hydrodynamic structure inside these bands follows the description of DC1 and DC2 with the hydrodynamic field illustrated in Figs.~\ref{fig:coarse_grained_fields}(b) and (d), respectively. We also observe TB3 where particles, which become synchronized, form a band that does not comprise most of the population. However, the other particles not inside this band do not become completely disordered. Due to the large interaction radius $\varrho=0.4$, they are significantly influenced by the synchronized group and follow their orientation with some lag in time (cf. Fig.~\ref{fig:coarse_grained_fields}(e)).

Third, we find two vortical structures (SV and DV in Figs.~\ref{fig:particle_dynamics}(g) and (h), respectively). For SV, when particles are entrained into one of the vortexes, they begin to rotate on average around a common center and do not deviate from it much. For DV, particles periodically approach the center of a vortex but then rotate away from it. Thus, these vortexes remind a 'breathing' shape.

Last, we would like to comment on localized (self-propelled) chimera structures, introduced in \cite{kruk:aps2018}, and generalized here to multiple 'heads' (cf. Fig.~\ref{fig:particle_dynamics}(f)). A localized chimera state is a solution of \EqRef{eq:chimera_sde} in which a particle system splits into two distinct populations. Particles in the first population synchronize and additionally gather into a compact rotating cloud (sometimes called a 'head'). The rest of the particles remain disordered and are uniformly distributed across the domain. In \cite{kruk:aps2018}, we reported the existence of a localized chimera state with one 'head'. In Fig.~\ref{fig:particle_dynamics}(f), one can observe a four-headed localized chimera state. Moreover, by changing $\tilde{\varrho}$, one can obtain such chimera states with a different number of compact clouds.

The integration of the SDEs \EqRef{eq:chimera_sde} was performed using the strong order 1.5 Taylor scheme \cite{platen}. The movies representing these exemplary particle dynamics can be found in the supplemental material \cite{supplemental_material} as well as in \cite{bcs_youtube_channel,*figshare}.

\section{Continuum Limit Derivation\label{sec:continuum_limit_derivation}}

In this section, we present how to obtain the equations that describe the dynamics of an ensemble of particles in the continuum limit $N\rightarrow\infty$ within the framework of Fokker-Planck equations. The continuum limit is understood in such a way that for each fixed $N$, a two-dimensional system domain is divided into $\sqrt{N}$ units \cite{kipnis1998scaling}. The approach we follow here is different from the one discussed in \cite{kruk:aps2018}, and it eventually provides us with a hierarchy of evolution equations for density functions that incorporate inter-particle interactions of any order. In the main text, we nondimensionalized the particle model by introducing dimensionless quantities such as a particle velocity $\hat{v}_0$ and a phase diffusion intensity $\hat{D}_\varphi$. In this Supplemental Material, we use the same variables but omit the $\hat{}$ symbol hereafter for the sake of simplicity.

Let $\mathbb{U} \coloneqq \mathbb{R}/(L\mathbb{Z})$ and $\mathbb{T} \coloneqq \mathbb{R}/(2\pi\mathbb{Z})$ be one-dimensional spaces with periodic boundaries extending from $[0,L]$ and $[0,2\pi]$, respectively. In the main text, the system size $L$ is assumed to be equal to one, but throughout this Supplemental Material we keep it arbitrary but constant. We will denote a three-dimensional state space of each particle by $\Omega = \mathbb{U}\!\times\!\mathbb{U}\!\times\!\mathbb{T}$. We introduce new variables to keep notation more compact. We will denote a spatial position of each particle with index $i\in\{1,\dots,N\}$ by $r_i = (x_i, y_i) \in \mathbb{U}^2$. We will also denote a state of each particle by $p_i = (x_i,y_i,\varphi_i) \in \Omega$. First, let's define a microscopic density function as
\begin{equation}
\label{sm:microscopic_density_function}
\begin{aligned}
\hat{f}(p&,t;p_1,\dots,p_N) = \frac{1}{N}\sum_{i=1}^N \delta(p_i(t) - p) \\
&= \frac{1}{N}\sum_{i=1}^{N} \delta(x_i(t) - x)\delta(y_i(t) - y)\delta(\varphi_i(t) - \varphi).
\end{aligned}
\end{equation}
This function should be treated the following way. For a given solution of the particle SDE (see the main text), we compute the value of an integral of $\hat{f}$ against some sufficiently smooth test function $\phi$.
As the next step, we consider a probability of finding particles with coordinates $\{p_i\}_{i=1,\dots,N}$ at time $t$, and denote its probability density function by $w = w(p_1,\dots,p_N,t)$. The time evolution of such a probability density function is given by the Fokker-Planck equation \cite{risken} and it reads
	\begin{equation}
	\label{sm:chimera_fpe}
	\begin{aligned}
	\partial_t w &= -\sum_{i=1}^N \Biggl( v_0\cos\varphi_i \partial_{x_i}w + v_0\sin\varphi_i \partial_{y_i}w \\
	&\qquad+  \partial_{\varphi_i} \biggl( \frac{1}{\vert B_\varrho^i \vert} \sum_{j \in B_\varrho^i} \sin(\varphi_j - \varphi_i - \alpha)w \biggr) \\
	&\qquad- D_\varphi \partial_{\varphi_i\varphi_i}w \Biggr).
	\end{aligned}
	\end{equation}
This equation requires information about each particle, the fact of which is prohibitive in practice. Therefore, we define an ensemble averaged probability density function $f = f(p,t)$ as the microscopic density $\hat{f}$ averaged with respect to the configuration probability $w$:
\begin{equation}
\begin{aligned}
	f(p,t) = \int_{\Omega^N} &\hat{f}(p,t;p_1,\dots,p_N) \times\\
	&\times w(p_1,\dots,p_N,t) \mathrm{d}p_1\dots\mathrm{d}p_N,
\end{aligned}
\end{equation}
where $\mathrm{d}p_i = \mathrm{d}x_i\mathrm{d}y_i\mathrm{d}\varphi_i$ is a three dimensional volume element in $\Omega$.
Since the particles are considered to be identical, the probability of the system configuration $w$ is symmetric with respect to permutations of particles. We can rewrite the ensemble averaged microscopic density as \cite{archer:jpa}
\begin{equation}
f(p,t) = \int_{\Omega^{N-1}} w(p, p_2,\dots, p_N) \mathrm{d}p_2\dots\mathrm{d}p_N.
\end{equation}

We can now use the Fokker-Planck equation \EqRef{sm:chimera_fpe} to obtain the time evolution for the one-particle density function. Namely, we integrate out $N-1$ particles in \EqRef{sm:chimera_fpe}. The first terms are transformed as follows
\begin{equation}
	\int_{\Omega^{N-1}} \partial_t w\; \mathrm{d}p_2\dots\mathrm{d}p_N \\
	= \partial_t f(p_1,t),
\end{equation}
\begin{equation}
\begin{aligned}
	\int_{\Omega^{N-1}} \sum_{i=1}^N v_0\cos\varphi_i \partial_{x_i}w\; \mathrm{d}p_2\dots\mathrm{d}p_N \\
	= v_0\cos\varphi_1 \partial_{x_1}f(p_1,t),
\end{aligned}
\end{equation}
\begin{equation}
\begin{aligned}
	\int_{\Omega^{N-1}} \sum_{i=1}^N v_0\sin\varphi_i \partial_{y_i}w\; \mathrm{d}p_2\dots\mathrm{d}p_N \\
	= v_0\sin\varphi_1 \partial_{y_1}f(p_1,t).
\end{aligned}
\end{equation}
We have used the fact that surface terms, which appear in integrals with partial derivatives, vanish due to the periodic boundary conditions. For the interaction term, we have
	\begin{equation}
	\label{sm:interaction_term_infinite_limit_1}
	\begin{aligned}
		&\int\limits_{\Omega^{N-1}} \sum_{i=1}^N \partial_{\varphi_i} \Biggl( \frac{1}{\vert B_\varrho^i \vert} \times\\
		&\qquad\times \sum_{j \in B_\varrho^i} \sin(\varphi_j - \varphi_i - \alpha)w \Biggr)\mathrm{d}p_2\dots\mathrm{d}p_N \\
		&= \int\limits_{\Omega^{N-1}} \sum_{i=1}^N \partial_{\varphi_i} \Biggl( \frac{1}{\sum_{\substack{j=1 \\ j \neq i}}^N H(\varrho - \Vert r_{ji} \Vert)} \times\\
		&\qquad\times \sum_{\substack{j=1 \\ j \neq i}}^N \sin(\varphi_j - \varphi_i - \alpha)H(\varrho - \Vert r_{ji} \Vert)w \Biggr) \mathrm{d}p_2\dots\mathrm{d}p_N,
	\end{aligned}
	\end{equation}
where $H$ is a Heaviside step function. We have rewritten the summation term and the neighborhood cardinality using the definition of particle's neighborhood $B_\varrho^i$, defined in the main text. The interparticle distance is computed with respect to $L^2(\mathbb{U}^2)$-norm. All the integrand terms except for the first particle cancel out because of the periodic boundaries, and we write \EqRef{sm:interaction_term_infinite_limit_1} as
	\begin{equation}
	\label{sm:interaction_term_infinite_limit_2}
	\begin{aligned}
		&\partial_{\varphi_1} \int_{\Omega^{N-1}} \frac{1}{\sum_{j=2}^N H(\varrho - \Vert r_{j1} \Vert)} \times\\
		&\qquad\times \sum_{j=2}^{N} \sin(\varphi_j - \varphi_1 - \alpha)H(\varrho - \Vert r_{j1} \Vert) w\; \mathrm{d}p_2\dots\mathrm{d}p_N \\
		&= (N-1) \partial_{\varphi_1} \int_{\Omega} \sin(\varphi_2 - \varphi_1 - \alpha) H(\varrho - \Vert r_{21} \Vert) \times\\
		&\qquad\times \left( \int_{\Omega^{N-2}} \frac{w\; \mathrm{d}p_3\dots\mathrm{d}p_N}{\sum_{j=2}^N H(\varrho - \Vert r_{j1} \Vert)} \right) \mathrm{d}p_2.
	\end{aligned}
	\end{equation}
The denominator in the last expression does not allow to integrate $w$ out straightforwardly. However, we are interested in the continuum limit $N\rightarrow\infty$. In this limit, by the law of large numbers
\begin{equation}
\begin{aligned}
\lim\limits_{N\rightarrow\infty} &\frac{1}{N-1} \sum_{j=2}^N H(\varrho - \Vert r_{j1} \Vert) \\
&= \iint_{\mathbb{U}^2} H(\varrho - \Vert r_2 - r_1 \Vert)\tilde{f}(x_2,y_2,t) \mathrm{d}x_2\mathrm{d}y_2 \\
&= \iiint_{\Omega} H(\varrho - \Vert r_2 - r_1 \Vert) f(x_2,y_2,\varphi_2,t) \mathrm{d}x_2\mathrm{d}y_2\mathrm{d}\varphi_2,
\end{aligned}
\end{equation}
where in the intermediate step, $\tilde{f}$ denotes a marginal density function of spatial variables. We use this fact to rewrite the right hand side of \EqRef{sm:interaction_term_infinite_limit_2} further as
\begin{align}
\begin{aligned}
	\partial_{\varphi_1} &\int_{\Omega} \sin(\varphi_2 - \varphi_1 - \alpha)H(\varrho - \Vert r_2 - r_1 \Vert) \times\\
	&\times \left( \frac{f^{(2)}(p_1,p_2,t)}{\int_{\Omega} H(\varrho - \Vert r_2 - r_1 \Vert) f(p_2,t)\; \mathrm{d}p_2} \right) \mathrm{d}p_2,
\end{aligned}
\end{align}
where 
$
	f^{(2)}(p_1,p_2,t)=\lim\limits_{N\rightarrow\infty}\int_{\Omega^{N-2}} w\; \mathrm{d}p_3\dots\mathrm{d}p_N
$
is a two-particle density function. Similarly to how we expressed the one-particle density function by averaging the microscopic density function, we can obtain the two-particle density function in the limit of infinitely many particles \cite{gupta:pre} in the following way:
	\begin{equation}
	\begin{aligned}
		&\lim\limits_{N\rightarrow\infty} \int_{\Omega^N} \hat{f}(p,t;p_1,\dots,p_N)\hat{f}(p',t;p_1,\dots,p_N) \times \\
		&\qquad\times w(p_1,\dots,p_N,t)\; \mathrm{d}p_1\dots\mathrm{d}p_N \\
		&= \lim\limits_{N\rightarrow\infty} \frac{N-1}{N} \int_{\Omega^{N-2}} w(p,p',p_3,\dots,p_N) \mathrm{d}p_3\dots\mathrm{d}p_N \\
		&\qquad+ \frac{1}{N}\delta(p\!-\!p')f(p,t) = f^{(2)}(p,p',t),
	\end{aligned}
	\end{equation}
where the first transition has been performed using the definition \EqRef{sm:microscopic_density_function} and the symmetry of $w$ under permutations.

Lastly, we integrate the diffusion term in the Fokker-Planck equation \EqRef{sm:chimera_fpe} and obtain
\begin{equation}
\begin{aligned}
\int_{\Omega^{N-1}} D_\varphi\sum_{i=1}^N \partial_{\varphi_i\varphi_i} w\; \mathrm{d}p_2\dots\mathrm{d}p_N = D_\varphi \partial_{\varphi_1\varphi_1}f(p_1,t).
\end{aligned}
\end{equation}

As a result, the time evolution of the one-particle density function $f=f(p_1,t)=f(r_1,\varphi_1,t)$ reads
\begin{widetext}
	\begin{equation}
	\begin{aligned}
		\partial_t f = -v_0e(\varphi_1)\cdot\nabla_{r_1}f - \frac{1}{\vert C(r_1;\varrho) \vert} \partial_{\varphi_1} \int_{C(r_1;\varrho)} \sin(\varphi_2 - \varphi_1 - \alpha)f^{(2)}(p_1,p_2,t)\; \mathrm{d}p_2 + D_\varphi \partial_{\varphi_1\varphi_1} f,
	\end{aligned}
	\end{equation}
\end{widetext}
where $e(\varphi_1) = (\cos\varphi_1,\sin\varphi_1)\in\mathbb{S}^1\subset\mathbb{R}^2$ is a unit velocity vector, $\nabla_{r_1}=(\partial_{x_1},\partial_{y_1})$ is a spatial gradient, and 
\begin{equation}
\begin{aligned}
	\vert C(r_1;&\varrho) \vert = \int_{C(r_1;\varrho)} f(p_2,t)\; \mathrm{d}p_2 \\
	&= \int_\Omega f(r_2,\varphi_2,t) H(\varrho - \Vert r_2 - r_1 \Vert) \mathrm{d}r_2\mathrm{d}\varphi_2
\end{aligned}
\end{equation}
is a neighborhood mass. The neighborhood domain $C(r;\varrho)$ is itself defined as
\begin{equation}
\label{sm:continuum_neighborhood_domain_3d}
	C(r;\varrho) = \left\{ (r',\varphi') \in \mathbb{U}^2\times\mathbb{T} \mid \Vert r'-r \Vert \leq \varrho \right\}.
\end{equation}
We hereafter use $C(r)$ instead of $C(r;\varrho)$ for shorter notation. We see that for an interacting particle system, the time evolution of a one-particle density function is not a closed equation since it depends on a two-particle density function. In order to obtain a closure, one often admits the simplest mean field approximation known as a molecular chaos assumption \cite{laney:comp_gas_dyn}. It postulates that particle correlations are negligible and the following factorization of the two-particle density function is possible:
\begin{equation}
	f^{(2)}(p_1,p_2,t) \approx f(p_1,t)f(p_2,t).
\end{equation}
Under that assumption, the time evolution of $f$ is given by
	\begin{equation}
	\label{sm:chimera_pde_global}
	\begin{aligned}
		\partial&_t f = -v_0e(\varphi_1)\cdot\nabla_{r_1}f + D_\varphi \partial_{\varphi_1\varphi_1} f \\
		&- \frac{\partial_{\varphi_1}}{\vert C(r_1) \vert} \left(f \!\int_{C(r_1)}\! \sin(\varphi_2 \!-\! \varphi_1 \!-\! \alpha)f(p_2,t)\; \mathrm{d}p_2 \right).
	\end{aligned}
	\end{equation}
This is the main equation that we will work with in the next chapters.

\subsection{Two-particle Density Function}

In the case that the closure at the first order is not sufficient, we may proceed in the same manner and next define a three-particle density function. In the limit of infinitely many particles, we have

\begin{equation}
\begin{aligned}
	f&^{(3)}(p,p',p'',t) \\
	&= \lim\limits_{N\rightarrow\infty} \int_{\Omega^{N-3}} w(p,p',p'',p_4,\dots,p_N,t)\; \mathrm{d}p_4\dots\mathrm{d}p_N.
\end{aligned}
\end{equation}

If we integrate out $N-2$ particles from the Fokker-Planck equation \EqRef{sm:chimera_fpe}, we derive the equation for the time evolution of the two-particle density function $f^{(2)}=f^{(2)}(p_1,p_2,t)=f^{(2)}(r_1,\varphi_1,r_2,\varphi_2,t)$. It reads
	\begin{equation}
	\begin{aligned}
		\partial&_t f^{(2)} = -v_0e(\varphi_1)\cdot\nabla_{r_1}f^{(2)} + D_\varphi \partial_{\varphi_1\varphi_1}f^{(2)} \\
		&- \frac{\partial_{\varphi_1}}{\vert C(r_1) \vert} \int_{C(r_1)} \sin(\varphi_3 \!-\! \varphi_1 \!-\! \alpha) f^{(3)}(p_1,p_2,p_3,t)\; \mathrm{d}p_3 \\
		&- v_0e(\varphi_2)\cdot\nabla_{r_2}f^{(2)} + D_\varphi \partial_{\varphi_2\varphi_2}f^{(2)} \\
		&- \frac{\partial_{\varphi_2}}{\vert C(r_2) \vert} \int_{C(r_2)} \sin(\varphi_3 \!-\! \varphi_2 \!-\! \alpha) f^{(3)}(p_1,p_2,p_3,t)\; \mathrm{d}p_3,
	\end{aligned}
	\end{equation}
where the neighborhood domain $C(r)$ is defined in \EqRef{sm:continuum_neighborhood_domain_3d}.
The time evolution of the two-particle density function $f^{(2)}$ now depends on the three-particle density function $f^{(3)}$. If we continue further, we can derive a corresponding equation for an $n$-particle density function which will further depend on an $n+1$-particle density function. This infinite hierarchy of integro-differential equations is similar to the Born-Bogolubov-Green-Kirkwood-Yvon (BBGKY) or Vlasov hierarchies in statistical physics \cite{spohn:mmas,braun:cmp}. We could theoretically close the hierarchy at any level provided that we have a required closure. Besides the molecular chaos assumption that closes it at the first order, the so-called Kirkwood superposition approximation can be used to obtain the second order closure. It assumes that the three-particle density function is factorized as a product of two-particle density functions as
\begin{equation}
\begin{aligned}
	f^{(3)}&(p_1,p_2,p_3,t) \\
	&\approx \frac{f^{(2)}(p_1,p_2,t) f^{(2)}(p_1,p_3,t) f^{(2)}(p_2,p_3,t)}{f(p_1,t)f(p_2,t)f(p_3,t)}.
\end{aligned}
\end{equation}
The time evolution of the two-particle density function under this approximation reads
\begin{widetext}
	\begin{equation}
	\begin{aligned}
		\partial_t &f^{(2)} = -v_0e(\varphi_1)\cdot\nabla_{r_1}f^{(2)} - v_0e(\varphi_2)\cdot\nabla_{r_2}f^{(2)} + D_\varphi \partial_{\varphi_1\varphi_1}f^{(2)} + D_\varphi \partial_{\varphi_2\varphi_2}f^{(2)} \\
		&- \frac{1}{\vert C(r_1) \vert} \partial_{\varphi_1} \left[ \frac{f^{(2)}(p_1,p_2,t)}{f(p_1,t)f(p_2,t)} \int_{C(r_1)} \sin(\varphi_3 - \varphi_1 - \alpha) \frac{f^{(2)}(p_1,p_3,t)f^{(2)}(p_2,p_3,t)}{f(p_3,t)} \mathrm{d}p_3 \right] \\
		&- \frac{1}{\vert C(r_2) \vert} \partial_{\varphi_2} \left[ \frac{f^{(2)}(p_1,p_2,t)}{f(p_1,t)f(p_2,t)} \int_{C(r_2)} \sin(\varphi_3 - \varphi_2 - \alpha) \frac{f^{(2)}(p_1,p_3,t)f^{(2)}(p_2,p_3,t)}{f(p_3,t)} \mathrm{d}p_3 \right].
	\end{aligned}
	\end{equation}
\end{widetext}
The  approximation is usually required when the dynamics due to a self-propelled particle system involves hard-core repulsion interactions \cite{marconi:jcp}. Since our model does not contain such terms, we limit ourselves to subsequently work with the one particle density function, whose dynamics is described by \EqRef{sm:chimera_pde_global}.

\section{Solutions for the Continuum Limit PDE\label{sec:solutions_for_the_continuum_limit_pde}}

From now on, we do not use $p$ as particle's state variable. Instead, we split it up into position $r=(x,y)\in\mathbb{U}^2$ and phase $\varphi$ variables. The easiest solution to \EqRef{sm:chimera_pde_global} (and to all of its variations) is the uniform probability density function, i.e.,

\begin{equation}
	f(r,\varphi,t) = \frac{1}{2\pi}.
\end{equation}
It corresponds to the chaotic behavior of the particle system, for which the continuum limit has been derived. One also says that this solution represents a globally disordered state.


The model \EqRef{sm:chimera_pde_global} admits a major simplification if we assume that solutions are spatially homogeneous. Based on the results from \cite{kruk:aps2018}, we know that a subset of chimeric solutions are of such a form. Under such an assumption of spatial homogeneity, we obtain a 1+1-dimensional PDE, which we can also consider as the continuum Kuramoto-Sakaguchi model:
\begin{equation}
\begin{aligned}
\partial_t &f(\varphi,t) = D_\varphi\partial_{\varphi\varphi} f(\varphi,t) \\
&-\partial_\varphi \left( f(\varphi,t) \;\frac{\int_\mathbb{T} f(\varphi',t)\sin(\varphi' - \varphi - \alpha) \mathrm{d}\varphi'}{\int_\mathbb{T} f(\varphi',t) \mathrm{d}\varphi'} \right). \\
\end{aligned}
\end{equation}
Since we treat the function $f(\varphi,t)$ as a probability density function, we have that $\int_\mathbb{T} f(\varphi,t) \mathrm{d}\varphi = 1$. The time evolution of the density function then becomes
\begin{equation}
\label{sm:chimera_pde_local}
\begin{aligned}
\partial_t &f(\varphi,t) = D_\varphi\partial_{\varphi\varphi} f(\varphi,t) \\
&-\partial_\varphi \left( f(\varphi,t) \;\int_\mathbb{T} f(\varphi',t)\sin(\varphi' - \varphi - \alpha) \mathrm{d}\varphi' \right).
\end{aligned}
\end{equation}
This model have certain symmetries. Generally, we could rescale $f(\varphi,t) \mapsto cf(\varphi,t), c\in\mathbb{R}$ but since we treat $f$ as a probability density function, this symmetry is of no importance to us. The equation is also invariant under the phase translation $f(\varphi,t) \mapsto f(\varphi+\varphi_0,t) \quad \forall\varphi_0\in\mathbb{T}$. This means that we can shift the distribution by any $\varphi_0$ and obtain another solution. This is particularly important in the stability analysis conducted later. The third symmetry is due to the invariance under the reflection of phase $f(\varphi,t) \mapsto f(-\varphi,t)$ if $\alpha \mapsto -\alpha$. Thus, we see that we can obtain another solution by flipping the signs of phases and the parameter $\alpha$ simultaneously. Given that, we will subsequently consider $\alpha\in[0,\pi/2]$ only. We also see that the solution will be symmetric with respect to some $\varphi_0$ only if $\alpha=0$. Thus, we will first look for solutions of \EqRef{sm:chimera_pde_local} without a phase lag.

\subsection{Stationary Solutions}

To find a nontrivial stationary solution to \EqRef{sm:chimera_pde_local}, we put $\partial_t f(\varphi,t) = 0$. This gives us a second order ordinary differential equation (ODE) of the form:
\begin{equation}
D_\varphi \frac{\mathrm{d}^2 f(\varphi)}{\mathrm{d}\varphi^2} - \frac{\mathrm{d}}{\mathrm{d}\varphi} \left( f(\varphi) \;\int_\mathbb{T} f(\varphi')\sin(\varphi' - \varphi - \alpha) \mathrm{d}\varphi' \right) = 0.
\end{equation}
In terms of the order parameter, we can write it as
\begin{equation}
D_\varphi \frac{\mathrm{d}^2 f(\varphi)}{\mathrm{d}\varphi^2} - \frac{\mathrm{d}}{\mathrm{d}\varphi}\left[ f(\varphi)R\sin(\Theta - \varphi - \alpha) \right] = 0.
\end{equation}
To solve this equation, we integrate it once and then look for the solution of the form
\begin{equation}
f(\varphi) = c(\varphi) e^{\gamma\cos(\Theta - \varphi - \alpha)},
\end{equation}
where the function $c(\varphi)$ is to be determined from the ODE, and $\gamma = R/D_\varphi$. It can be shown that eventually the solution is of the form
\begin{equation}
f(\varphi) = c_1 e^{\gamma\cos(\Theta - \varphi - \alpha)} \left( 1 + c_2 \int e^{-\gamma\cos(\Theta - \varphi - \alpha)} \mathrm{d}\varphi \right),
\end{equation}
where $c_1,c_2$ are the constants to be determined. We are interested in smooth solutions to \EqRef{sm:chimera_pde_local}, so we require that $f\in C(\mathbb{T})$. This implies that $f(0) = f(2\pi)$, which can be shown to hold if and only if $c_2=0$. From the normalization condition, we find that $c_1 = (2\pi I_0(\gamma))^{-1}$, where $I_0(\gamma) = 1/(2\pi) \int_\mathbb{T} \exp(\gamma\cos\varphi) \mathrm{d}\varphi$ denotes the modified Bessel function of the first kind \cite{olver:nist_handbook}. As a result, we have the nontrivial stationary solution of the form:
\begin{equation}
f(\varphi) = \frac{e^{\gamma\cos(\Theta - \varphi - \alpha)}}{2\pi I_0(\gamma)}.
\end{equation}
Due to the translational invariance of \EqRef{sm:chimera_pde_local} with respect to the phase $\varphi$, we can put $\Theta = 0$ without loss of generality. This simplifies the solution to
\begin{equation}
\label{sm:solution_uniform_zero_lag_1}
f(\varphi) = \frac{e^{\gamma\cos(\varphi + \alpha)}}{2\pi I_0(\gamma)}.
\end{equation}
In this form, the solution is not particularly useful since the density function is recursively contained in the definition of the order parameter. However, we are able to determine the latter the other way. If we multiply \EqRef{sm:solution_uniform_zero_lag_1} by $\cos\varphi$ and integrate over the domain $\mathbb{T}$, we find that
\begin{equation}
R = \frac{1}{2\pi I_0(\gamma)} \int_\mathbb{T} e^{\gamma\cos(\varphi + \alpha)} e^{i\varphi} \mathrm{d}\varphi = e^{-i\alpha} \frac{I_1(\gamma)}{I_0(\gamma)}.
\end{equation}
From the equation for the imaginary part, we have that either $\alpha=0$ or $I_1(\gamma)=0$. The latter case is true for $\gamma=R/D_\varphi=0$. But if the order parameter magnitude is zero, the density function \EqRef{sm:solution_uniform_zero_lag_1} becomes just a constant and the phase lag $\alpha$ does not play any role. Therefore, we conclude that the system \EqRef{sm:chimera_pde_local} is solved by \EqRef{sm:solution_uniform_zero_lag_1} only when $\alpha=0$:
\begin{equation}
\label{sm:solution_uniform_zero_lag_2}
f(\varphi) = \frac{e^{\gamma\cos\varphi}}{2\pi I_0(\gamma)},
\end{equation}
for which the order parameter magnitude is determined from
\begin{equation}
\label{sm:order_parameter_uniform_zero_lag}
R = \frac{I_1\left(\frac{R}{D_\varphi}\right)}{I_0\left(\frac{R}{D_\varphi}\right)}.
\end{equation}

\subsubsection{The Onset of Orientational Order}

Even though we do not have a closed form solution for the density function $f$ that solves \EqRef{sm:chimera_pde_local}, we can extract the information on what relation the model parameters should satisfy in order to allow the existence of this nontrivial solution. It was shown in \cite{pearce:jsp} that for positive values of $R$, the relation \EqRef{sm:order_parameter_uniform_zero_lag} has a unique solution. That solution allows us to find a condition where the nonconstant density function of the form \EqRef{sm:solution_uniform_zero_lag_2} appears. We should search for parameters for which the slope on the right hand side of \EqRef{sm:order_parameter_uniform_zero_lag} is greater than the slope of the left hand side at $R=0$. Namely, we consider
\(
\frac{\mathrm{d}}{\mathrm{d}R} [I_1(\gamma) / I_0(\gamma)] \geq 1,
\)
where we denote $\gamma = R / D_\varphi$. Using the properties \cite{olver:nist_handbook} of the Bessel function $\frac{\mathrm{d}}{\mathrm{d}R} I_0(\gamma) = \frac{\sigma}{D_\varphi}I_1(\gamma)$ and $\frac{\mathrm{d}}{\mathrm{d}R} I_1(\gamma) = \frac{\sigma}{2D_\varphi}[I_0(\gamma) + I_2(\gamma)]$, we rewrite the above inequality as
\begin{equation}
\frac{1}{2D_\varphi} + \frac{1}{2D_\varphi}\frac{I_2(\gamma)}{I_0(\gamma)} - \frac{1}{D_\varphi}\frac{I_1^2(\gamma)}{I_0^2(\gamma)} \geq 1.
\end{equation}
Using the property \cite{joshi:australian_math_soc} $I_0(\gamma) - I_2(\gamma) = \frac{2}{\gamma}I_1(\gamma)$, we obtain the desired inequality for the order parameter magnitude
\begin{equation}
\frac{1}{D_\varphi}R^2 - \frac{1}{D_\varphi} + 2 \leq 0.
\end{equation}
Since we are interested in the value of the slope at $R=0$, we derive the following condition for the existence of the nontrivial stationary solution to \EqRef{sm:chimera_pde_local}:
\begin{equation}
\label{sm:critical_coupling_zero_lag}
D_\varphi \leq \frac{1}{2}.
\end{equation}
We also see from \EqRef{sm:solution_uniform_zero_lag_2} and \EqRef{sm:order_parameter_uniform_zero_lag} that when $D_\varphi\rightarrow0$, the order parameter approaches $1$ and we have the completely synchronous stationary state, namely,
\begin{equation}
f(\varphi,t) = \delta(\varphi - \varphi_0),
\end{equation}
where the phase $\varphi_0$ is determined from the initial condition.

\subsection{Traveling Wave Solutions}

Our next step is to investigate the solution to \EqRef{sm:chimera_pde_local} in the presence of a nonzero phase lag $\alpha$. We know, when $\alpha\neq0$ the density $f(\varphi,t)$ is no longer a symmetric function. It moves to the left if $\alpha>0$ and to the right if $\alpha<0$ with some constant speed $v$. This fact allows us to look for a solution in the form of a traveling wave. Thus, we introduce an ansatz $f(\varphi,t) = g(\varphi-vt) = g(\omega)$, where $v$ is the speed of the traveling wave, which is also to be determined. After the substitution, we obtain the following second order ODE:
\begin{equation}
\label{sm:chimera_pde_local_traveling_wave}
\begin{aligned}
	D_\varphi &\frac{\mathrm{d}^2}{\mathrm{d}\omega^2}g(\omega) \\
	&+ \frac{\mathrm{d}}{\mathrm{d}\omega}\left\{\left[ v - R\sin(\Theta-\omega-\alpha) \right] g(\omega)\right\} = 0,
\end{aligned}
\end{equation}
where $R$ and $\Theta$ are now constants. We trivially integrate it with respect to $\omega$ and get
\begin{equation}
	D_\varphi \frac{\mathrm{d}}{\mathrm{d}\omega} g(\omega) + \left[ v - R\sin(\Theta - \omega - \alpha) \right] g(\omega) = c_1,
\end{equation}
where $c_1\in\mathbb{R}$ is some constant. The method to solve such an equation is again to look for a solution of the form 
\begin{equation}
	g(\omega) = c(\omega) \exp\left( -\int \left[ \frac{v}{D_\varphi} - \frac{R}{D_\varphi} \sin(\Theta - \omega - \alpha) \right] \mathrm{d}\omega \right),
\end{equation}
where the function $c(\omega)$ is to be determined. After we substitute the function of this form into the above differential equation, we find the following solution:
\begin{equation}
	g(\omega) = E_1(\omega) \left( c_1 \int E_1^{-1}(\omega') \mathrm{d}\omega' + c_2 \right),
\end{equation}
where we have denoted $E_1(\omega) = \exp\left[ -\frac{v}{D_\varphi}\omega + \frac{R}{D_\varphi} \cos(\Theta - \omega - \alpha) \right]$ and $c_2\in\mathbb{R}$ is some constant. One of the constants can be found by recalling that we are looking for a periodic and continuous solution, i.e., $g(0) = g(2\pi)$. The other constant is determined from the normalization condition of the probability density function. Eventually, we arrive at the following solution (compare to the form of the solution in the case of the Kuramoto model with frequency distribution and zero phase lag \cite{gupta:jsm}):
\begin{equation}
	g(\omega) = c_0 E_1(\omega) 
	\left( 1 + c_1 \frac{\int_{0}^{\omega} E_1^{-1}(\omega') \mathrm{d}\omega'}{\int_\mathbb{T} E_1^{-1}(\omega') \mathrm{d}\omega'} \right),
\end{equation}
where $c_0\in\mathbb{R}$ is a normalization constant and $c_1 = \left( e^{2\pi\frac{v}{D_\varphi}} - 1 \right)$ comes form a periodicity constraint.

One may notice that due to the translational invariance of \EqRef{sm:chimera_pde_local}, by the suitable shift of $\omega - \Theta \mapsto \omega$, we have the system where the order parameter phase can be put equal to zero without loss of generality. In other words, we could initially have introduced an ansatz $f(\varphi,t) = g(\varphi - vt - \Theta_0) = g(\omega)$, with $v$ to be determined and where $\Theta_0 = \Theta(0)$. The density function, expressed in terms of new $\omega$, can be shown to read
\begin{equation}
\label{sm:solution_profile_uniform_nonzero_lag}
	g(\omega) = c_0 E_2(\omega) 
	\left( 1 + c_1 \frac{\int_{0}^\omega E_2^{-1}(\omega') \mathrm{d}\omega'}{\int_\mathbb{T} E_2^{-1}(\omega') \mathrm{d}\omega'} \right),
\end{equation}
where the exponential function is redefined as $E_2(\omega) = \exp\left[ -\frac{v}{D_\varphi}\omega + \frac{R}{D_\varphi} \cos(\omega + \alpha) \right]$. 

The above solution \EqRef{sm:solution_profile_uniform_nonzero_lag} is expressed in terms of the traveling wave variable $\omega$, and we now want to return to the original variables $\varphi$ and $t$. Inserting them back, we find the solution of the traveling wave form to be
\begin{equation}
\begin{aligned}
	&f(\varphi,t) = c_0 E(\varphi,t) \left( 1 + c_1 \frac{\int_{vt}^{\varphi_0+vt} E^{-1}(\varphi',t) \mathrm{d}\varphi'}{\int_{vt}^{2\pi+vt} E^{-1}(\varphi',t) \mathrm{d}\varphi'} \right)
\end{aligned}
\end{equation}
with $\varphi\in[vt,2\pi+vt)$, $E(\varphi,t) = \exp\left[ -\frac{v}{D_\varphi}\varphi + \frac{R}{D_\varphi} \cos(\varphi - vt + \alpha) \right]$, and $\varphi_0 = \left.\varphi\right|_{t=0} = \varphi-vt$ is the initial reference frame. For the computational purposes, it is better to perform the change of variables $\varphi_0 = \varphi - vt$ in the integrals so that they are independent of time, giving
\begin{equation}
\label{sm:solution_uniform_nonzero_lag}
\begin{aligned}
	&f(\varphi,t) = c_0 E(\varphi,t) \left( 1 + c_1 \frac{\int_{0}^{\varphi_0} E^{-1}(\varphi_0,0) \mathrm{d}\varphi_0}{\int_{\mathbb{T}} E^{-1}(\varphi_0,0) \mathrm{d}\varphi_0} \right)\\
	&\text{with } \varphi\in[vt,2\pi+vt).
\end{aligned}
\end{equation}

\subsubsection{The Onset of Orientational Order}

To learn the behavior of the order parameter, corresponding to \EqRef{sm:solution_uniform_nonzero_lag}, it is enough to study its profile $g(\omega)$. The global order parameter is defined as
\begin{equation}
\label{sm:order_parameter_uniform_nonzero_lag}
	R = \int_{\mathbb{T}} e^{i\omega} g(\omega) \mathrm{d}\omega,
\end{equation}
where the average direction is shifted to the origin so that $\Theta\equiv0$ without loss of generality. Thus, the order parameter must satisfy the following set of self-consistency equations:
\begin{align}
\label{sm:order_parameter_self_consistent_real}
	R = \int_\mathbb{T} g(\omega)\cos\omega \mathrm{d}\omega, \\
\label{sm:order_parameter_self_consistent_imaginary}
	0 = \int_\mathbb{T} g(\omega)\sin\omega \mathrm{d}\omega,
\end{align}
where $g$ is given by \EqRef{sm:solution_profile_uniform_nonzero_lag}. This system does not have an analytical solution but can be solved numerically for $R$ and $v$, assuming that $\Theta\equiv0$. The numerical results are presented in the main text. Note that in order to obtain them, one has to use multiprecision arithmetic. For parameter values away from the order-disorder transition line, the exponents in \EqRef{sm:solution_profile_uniform_nonzero_lag} assume values not valid for the double precision format.

As before, we can use the set of self-consistent equations to determine conditions on the model parameters that lead to the existence of the nontrivial solution $g$. If we substitute \EqRef{sm:solution_profile_uniform_nonzero_lag} into \EqRef{sm:order_parameter_self_consistent_real}, expand the right hand side of \EqRef{sm:order_parameter_self_consistent_real} in powers of $\gamma\coloneqq R / D_\varphi$, and take the limit $\gamma\rightarrow0+0$ \cite{gupta:jsm}, we find the following equality:
\begin{equation}
\begin{aligned}
	R &= -\frac{\cos\alpha}{2} \frac{\gamma}{\frac{v^2}{D_\varphi^2} + 1}\frac{v^2}{D_\varphi^2} \\
	&- \frac{\sin\alpha}{2}  \frac{\gamma}{\frac{v^2}{D_\varphi^2} + 1}\frac{v}{D_\varphi} + \frac{\cos\alpha}{2} \gamma + O\left(\gamma^2\right).
\end{aligned}
\end{equation}
If we divide both sides by the order parameter magnitude $R$, we obtain the relation between the critical coupling strength, the diffusion constant $D_\varphi$, and the phase lag $\alpha$ expressed as
\begin{equation}
	1 = \frac{2(D_\varphi^2 + v^2)}{D_\varphi \cos\alpha - v \sin\alpha}.
\end{equation}
The drawback is that it also involves the unknown parameter $v$. But fortunately as we have mentioned, we must simultaneously satisfy \EqRef{sm:order_parameter_self_consistent_imaginary}. Thus, we substitute \EqRef{sm:solution_profile_uniform_nonzero_lag} into \EqRef{sm:order_parameter_self_consistent_imaginary} and perform the expansion again. We find the condition for the critical velocity as $v = -D_\varphi \tan\alpha$.
Combining it with the last expression, we obtain the value for the critical coupling strength in terms of the known system parameters as
\begin{equation}
	\frac{1}{2}\cos\alpha = D_\varphi.
\end{equation}
We thus deduce that the condition to completely desynchronize the system is $D_\varphi \geq \frac{1}{2}\cos\alpha$. Note that if we let $\alpha\rightarrow0$, we obtain the same condition as given by \EqRef{sm:critical_coupling_zero_lag} for the case of zero phase lag.

\section{Stability Analysis via Kinetic Theory\label{sec:stability_analysis_via_kinetic_theory}}

Now that we have derived several spatially homogeneous solutions for the original problem \EqRef{sm:chimera_pde_global}, we want to know parameter regions, where these solutions become unstable and spatially nonhomogeneous structures appear. In order to do that, we perform linear stability analysis of the solutions in Fourier space from the point of view of the kinetic theory first.

\subsection{Stationary Solutions}

First, let us provide the version of the nonhomogeneous continuum limit PDE, we will build our further analysis upon. We start with \EqRef{sm:chimera_pde_global} for a one-particle density function $f=f(p,t)=f(r,\varphi,t)$, which we state here one more time for the easier reference:
	\begin{equation}
	\label{sm:chimera_pde_global_2}
	\begin{aligned}
		\partial&_t f = - v_0e(\varphi)\cdot\nabla_r f + D_\varphi \partial_{\varphi\varphi}f \\
		&- \partial_{\varphi}\left( f \frac{\int_{C(r)} f(r',\varphi',t)\sin(\varphi' \!-\! \varphi \!-\! \alpha) \mathrm{d}r'\mathrm{d}\varphi'}{\int_{C(r)} f(r',\varphi',t) \mathrm{d}r'\mathrm{d}\varphi'} \right),
	\end{aligned}
	\end{equation}
where we have explicitly separated the combined variable $p$ into the position vector $r$ and the phase $\varphi$. Note that the neighborhood domain $C(r)$ implicitly depends on the radius of interaction $\varrho$ according to \EqRef{sm:continuum_neighborhood_domain_3d}. We are interested in the solutions that are periodic in spatial and phase variables. We can generally represent it in a Fourier series as
\begin{equation}
	f(r,\varphi,t) = \sum_{k\in\mathbb{Z}^2} \sum_{n\in\mathbb{Z}} f_{n,k}(t) e^{-in\varphi - i\frac{2\pi}{L}k\cdot r}.
\end{equation}
Since spatial and phase scales have different periodicities, we will perform the two corresponding transforms separately.

Our first step is to transform \EqRef{sm:chimera_pde_global_2} into the Fourier space with respect to the phase variable $\varphi$. The density function can be represented as
\begin{equation}
\label{sm:fourier_series_wrt_phase}
	f(r,\varphi,t) = \sum_{n\in\mathbb{Z}} f_n e^{-in\varphi},
\end{equation}
where each Fourier mode is defined as
\begin{equation}
\label{sm:fourier_mode_wrt_phase}
	f_n(r,t) = \frac{1}{2\pi}\int_{\mathbb{T}} f(r,\varphi,t) e^{in\varphi} \mathrm{d}\varphi.
\end{equation}
Using the above decomposition, each term of \EqRef{sm:chimera_pde_global_2} can be rewritten as
\begin{equation}
	\partial_t f(r,\varphi,t) = \sum_{n\in\mathbb{Z}} \partial_t f_n(r,t) e^{-in\varphi},
\end{equation}
\begin{equation}
\begin{aligned}
	v_0e&(\varphi)\cdot \nabla_r f(r,\varphi,t) \\
	&= -\frac{v_0\partial_x}{2} \sum_{n\in\mathbb{Z}} (f_ne^{-i(n-1)\varphi} + f_ne^{-i(n+1)\varphi}) \\
	&+ \frac{iv_0\partial_y}{2} \sum_{n\in\mathbb{Z}} (f_ne^{-i(n-1)\varphi} - f_ne^{-i(n+1)\varphi}),
\end{aligned}
\end{equation}
\begin{equation}
\begin{aligned}
\partial_{\varphi}&\left( f(r,\varphi,t)\frac{\iiint_{C(r)} f(r',\varphi',t)\sin(\varphi' - \varphi - \alpha) \mathrm{d}r'\mathrm{d}\varphi'}{\iiint_{C(r)} f(r',\varphi',t) \mathrm{d}r'\mathrm{d}\varphi'} \right) \\ 
&= \frac{1}{2} \sum_{n\in\mathbb{Z}} f_n \left[ (n-1)e^{-i(n-1)\varphi} \frac{\iint_{B(r)} f_{-1}(r',t)\mathrm{d}r'}{\iint_{B(r)} f_0(r',t)\mathrm{d}r'} e^{i\alpha} \right.\\
&\left.\qquad\qquad - (n+1)e^{-i(n+1)\varphi} \frac{\iint_{B(r)} f_1(r',t)\mathrm{d}r'}{\iint_{B(r)} f_0(r',t)\mathrm{d}r'} e^{-i\alpha} \right],
\end{aligned}
\end{equation}
\begin{equation}
\partial_{\varphi\varphi} f(r,\varphi,t) = -\sum_{n\in\mathbb{Z}} n^2 f_ne^{-in\varphi}.
\end{equation}
Gathering the terms corresponding to each Fourier mode $e^{-in\varphi}$, the evolution equation for each mode $f_n = f_n(r,t), n\in\mathbb{Z}$ becomes
	\begin{equation}
	\label{sm:fourier_mode_pde_wrt_phase}
	\begin{aligned}
		\partial_t f_n &= - n^2 D_\varphi f_n \\
		&-\frac{v_0}{2}\partial_x(f_{n+1} + f_{n-1}) + \frac{iv_0}{2}\partial_y(f_{n+1} - f_{n-1}) \\
		&+ \frac{n}{2} \Biggl( f_{n-1} \frac{\iint_{B(r)} f_1(r',t) \mathrm{d}r'}{\iint_{B(r)} f_0(r',t) \mathrm{d}r'} e^{-i\alpha} \\
		&\qquad\qquad- f_{n+1} \frac{\iint_{B(r)} f_{-1}(r',t) \mathrm{d}r'}{\iint_{B(r)} f_0(r',t) \mathrm{d}r'} e^{i\alpha} \Biggr).
	\end{aligned}
	\end{equation}
The integration over the cylinder $C(r) = C(r;\varrho)$ becomes the integration over the disk $B(r) = B(r;\varrho)$, which is defined as
\begin{equation}
\label{sm:continuum_neighborhood_domain_2d}
	B(r;\varrho) = \left\{ r' \in \mathbb{U}^2 \mid \Vert r' - r \Vert \leq \varrho \right\}.
\end{equation}
As before, we usually suppress the explicit dependence of the circular neighborhood $B(r)$ on the parameter $\varrho$ for the sake of brevity.

Next, we perform the Fourier transform of \EqRef{sm:fourier_mode_pde_wrt_phase} with respect to spatial coordinates $r$. Each Fourier mode $f_n(r,t)$ can be decomposed into a series as
\begin{equation}
f_n(r,t) = \sum_{k\in\mathbb{Z}^2} g_n(k,t) e^{-i\frac{2\pi}{L}k\cdot r},
\end{equation}
where its coefficients are defined as
\begin{equation}
\label{sm:forward_fourier_transform_wrt_xy}
\begin{aligned}
	g_n(k,t) &= \mathcal{F}\{f_n(r,t)\}(k,t) \\
	&= \frac{1}{L^2} \iint_{\mathbb{U}^2} f_n(r,t) e^{i\frac{2\pi}{L}k\cdot r} \mathrm{d}r.
\end{aligned}
\end{equation}

Our next goal is to obtain the differential equations for each mode $g_n(k,t)$ in the Fourier space with respect to the spatial variables, with the subsequent goal of deriving its linearized dynamics. With that regard, all terms in \EqRef{sm:fourier_mode_pde_wrt_phase} except for the nonlinear interaction one are easily transformed as
\begin{equation}
	\mathcal{F}\left\{ \partial_t f_n(r,t) \right\}(k,t) = \partial_t g_n(k,t),
\end{equation}
\begin{equation}
\begin{aligned}
	\mathcal{F}&\left\{ \partial_x [f_{n+1}(r,t) + f_{n-1}(r,t)] \right\}(k,t) \\
	&= -i\frac{2\pi}{L}k_x[g_{n+1}(k,t) + g_{n-1}(k,t)],
\end{aligned}
\end{equation}
\begin{equation}
\begin{aligned}
	\mathcal{F}&\left\{ \partial_y [f_{n+1}(r,t) - f_{n-1}(r,t)] \right\}(k,t) \\
	&= -i\frac{2\pi}{L}k_y[g_{n+1}(k,t) - g_{n-1}(k,t)].
\end{aligned}
\end{equation}
As to the interaction terms (the ones with the integrals), we will consider the derivation only for the first one of them. In order to find its transformation, we first represent it as
	\begin{equation}
	\begin{aligned}
		\frac{1}{L^2} &\iint_{\mathbb{U}^2} f_{n-1}(r,t) \frac{\iint_{B(r)} f_1(r',t) \mathrm{d}r'}{\iint_{B(r)} f_0(r',t) \mathrm{d}r'} e^{i\frac{2\pi}{L}k\cdot r} \mathrm{d}r \\
		&= \sum_{q\in\mathbb{Z}^2} g_{n-1}(q,t) \frac{1}{L^2} \iint_{\mathbb{U}^2} \frac{\iint_{B(r)} f_1(r',t) \mathrm{d}r'}{\iint_{B(r)} f_0(r',t) \mathrm{d}r'} \times\\
		&\qquad\qquad\qquad\qquad\qquad\qquad\times e^{i\frac{2\pi}{L}(k-q)\cdot r} \mathrm{d}r \\
		&= \sum_{q\in\mathbb{Z}^2} g_{n-1}(q,t) K_1(k-q,t),
	\end{aligned}
	\end{equation}
where $K_1$ denotes the Fourier transform \EqRef{sm:forward_fourier_transform_wrt_xy} of one of the interaction force terms. The other appears as $K_{-1}$ and inherently depends on $f_{-1}$. To obtain the representation of the interaction kernel $K_1$ in the Fourier space solely, i.e., via the wave vectors, we need to transform the integrals involving the primed variables \cite{grossmann:iop}. Generally, we consider
	\begin{equation}
	\begin{aligned}
		\iint_{B(r)} &f_n(r',t) \mathrm{d}r' = \sum_{k\in\mathbb{Z}^2} \iint_{B(r)} g_n(k,t) e^{-i\frac{2\pi}{L}k\cdot r'} \mathrm{d}r' \\
		&= \sum_{k\in\mathbb{Z}^2} g_n(k,t) e^{-i\frac{2\pi}{L}k\cdot r} \iint_{B(0)} e^{-i\frac{2\pi}{L}k\cdot z} \mathrm{d}z \\
		&= \sum_{k\in\mathbb{Z}^2} g_n(k,t) e^{-i\frac{2\pi}{L}k\cdot r} \times\\
		&\qquad\times \int_{0}^\varrho\int_{\mathbb{T}} \Vert z\Vert e^{-i\frac{2\pi}{L}\Vert z\Vert\Vert k\Vert\cos(\zeta - \chi)} \mathrm{d}\Vert z\Vert\mathrm{d}\zeta \\
		&= 2\pi \sum_{k\in\mathbb{Z}^2} g_n(k,t)e^{-i\frac{2\pi}{L}k\cdot r} \times\\
		&\qquad\times \int_{0}^\varrho \Vert z\Vert J_0\left(\frac{2\pi}{L}\Vert z\Vert\Vert k\Vert\right) \mathrm{d}\Vert z\Vert \\
		&= L\varrho \sum_{k\in\mathbb{Z}^2} g_n(k,t)e^{-i\frac{2\pi}{L}k\cdot r} \frac{J_1\left(\frac{2\pi}{L} \varrho\Vert k\Vert\right)}{\Vert k\Vert}.
	\end{aligned}
	\end{equation}
In the above derivation, we have made use of the polar representation of $z=\Vert z\Vert(\cos\zeta,\sin\zeta)^T$ and $k=\Vert k\Vert(\cos\chi,\sin\chi)^T$. The functions $J_0$ and $J_1$ denote the Bessel functions of the first kind. The transition between these functions was performed using the identity $\int_{0}^{R} sJ_0(s) \mathrm{d}s = RJ_1(R), R\in\mathbb{R}_+$. Note that the last transition of the above chain is valid as long as $k\neq(0,0)^T$. Otherwise, in the zero wave number regime, one has $\int_{0}^\varrho \Vert z\Vert J_0(0) \mathrm{d}\Vert z\Vert = \varrho^2/2$.

The last sum of the above chain of equations is the Fourier series with the coefficients containing the Bessel function. To simplify the further notation, we will denote

\begin{equation}
j_\varrho(k) = \frac{J_1\left(\frac{2\pi}{L}\varrho\Vert k\Vert\right)}{\Vert k\Vert}.
\end{equation}
Note that despite of the division by the norm of a wave vector, it is possible to consider the dynamics in the hydrodynamic limit, since $\lim\limits_{x\rightarrow0+0} \frac{1}{x}J_1(x) = \frac{1}{2}$. With the new notation, we can write the interaction kernel as

\begin{equation}
\label{sm:interaction_kernel_in_fourier_space}
\begin{aligned}
&K_{\pm1}(k-q,t) \\
&= \mathcal{F}\left\{ \frac{\sum_{p\in\mathbb{Z}^2}  g_{\pm1}(p,t)j_\varrho(p) e^{-i\frac{2\pi}{L}p\cdot r}}{\sum_{p\in\mathbb{Z}^2} g_0(p,t)j_\varrho(p) e^{-i\frac{2\pi}{L}p\cdot r}} \right\}(k-q,t).
\end{aligned}
\end{equation}
The kernel in this representation still depends on the Fourier coefficients in a nonlinear way but further decomposition of the kernel in a linear combination requires the knowledge of $g_n$.

We see that after the Fourier transform with respect to spatial variables, the coefficient $\frac{2\pi}{L}$ exclusively appears in front of $v_0$ and $\varrho$. Thus, we introduce $v_0^*=\frac{2\pi}{L}v_0$ and $\varrho^*=\frac{2\pi}{L}\varrho$ to shorten the further notation. The auxiliary function $j_\varrho(k)$ becomes $j_\varrho(k)=\frac{J_1(\varrho^* \Vert k\Vert)}{\Vert k\Vert}$.

With the current representation of the interaction kernels, the time evolution of the Fourier coefficients $g_n = g_n(k,t)$ reads
	\begin{equation}
	\label{sm:fourier_mode_pde_wrt_xy}
	\begin{aligned}
		\partial_t &g_n(k,t) = - n^2D_\varphi g_n \\
		&+\frac{iv_0^*}{2}k_x (g_{n+1} + g_{n-1}) + \frac{v_0^*}{2}k_y (g_{n+1} - g_{n-1}) \\
		&+ \frac{n}{2} \Biggl( e^{-i\alpha} \sum_{q\in\mathbb{Z}^2} g_{n-1}(q,t)K_1(k-q,t) \\
		&\qquad- e^{i\alpha} \sum_{q\in\mathbb{Z}^2} g_{n+1}(q,t)K_{-1}(k-q,t) \Biggr),
	\end{aligned}
	\end{equation}
where the Fourier transforms $K_{\pm1}$ of the interaction kernels are defined in \EqRef{sm:interaction_kernel_in_fourier_space}. As an outline, we mention that since we are going to test only the spatially homogeneous solutions on the matter of stability, their Fourier transforms with respect to $x,y$ will contain Kronecker delta functions. Upon the substitution of such transforms into the above expression, the sums over the wave numbers $q$ will be resolved. 

For convenience, we denote
\begin{equation}
	\mathcal{L}[g_0,g_{\pm1},g_{n\mp1}](q,k,t) = g_{n\mp1}(q,t) K_{\pm1}(k-q,t),
\end{equation}
where the dependence on $g_0$ and $g_{\pm1}$ comes through $K_{\pm1}$. We will use this expression in the linearization procedure described next.

Let the stationary spatially homogeneous solution of \EqRef{sm:chimera_pde_global_2}, transformed to the Fourier space with respect to the spatial variables, be $g_n^*(k)$. We denote the components of a small perturbation to the solution as 
\begin{equation}
\label{sm:perturbation_component}
	\delta g_n(k,t) = g_n(k,t) - g_n^*(k).
\end{equation}
To see how those perturbations behave over time, we need to derive differential equations for $\delta g_n$. To do that, we linearize \EqRef{sm:fourier_mode_pde_wrt_xy} around $g_n^*(k)$. Since the right hand side of \EqRef{sm:fourier_mode_pde_wrt_xy} depends on several $\delta g_n$, by Taylor series expansion we find
\begin{widetext}
	\begin{equation}
	\label{sm:fourier_mode_pde_wrt_xy_taylor}
	\begin{aligned}
	\partial_t &\delta g_n(k,t) = \frac{iv_0^*}{2}k_x [\delta g_{n+1}(k,t) + \delta g_{n-1}(k,t)] + \frac{v_0^*}{2}k_y [\delta g_{n+1}(k,t) - \delta g_{n-1}(k,t)] - n^2D_\varphi \delta g_{n}(k,t) \\
	&+ \frac{n}{2} e^{-i\alpha} \sum_{m\in\{0,1,n-1\}} \sum_{q\in\mathbb{Z}^2} \frac{\partial\mathcal{L}[g_0,g_1,g_{n-1}]}{\partial g_m}(q,k,t) \Biggr|_{g_0^*,g_1^*,g_{n-1}^*} \delta g_m(q,t) \\
	&- \frac{n}{2} e^{i\alpha} \sum_{m\in\{0,-1,n+1\}} \sum_{q\in\mathbb{Z}^2} \frac{\partial\mathcal{L}[g_0,g_{-1},g_{n+1}]}{\partial g_m}(q,k,t) \Biggr|_{g_0^*,g_{-1}^*,g_{n+1}^*} \delta g_m(q,t).
	\end{aligned}
	\end{equation}
\end{widetext}
One can show that the derivatives of the functional $\mathcal{L}$ are
	\begin{equation}
	\begin{aligned}
		&\frac{\partial\mathcal{L}[g_0,g_1,g_{n-1}]}{\partial g_0}(q,k,t) = -j_\varrho(q) \sum_{p\in\mathbb{Z}^2} g_{n-1}(p,t) \times\\
		&\quad\times \mathcal{F}\left\{ \frac{\sum_{s\in\mathbb{Z}^2} g_1(s,t)j_\varrho(s) e^{-i\frac{2\pi}{L}s\cdot r} }{[\sum_{s\in\mathbb{Z}^2} g_0(s,t)j_\varrho(s) e^{-i\frac{2\pi}{L}s\cdot r} ]^2} \right\}(k-p-q),
	\end{aligned}
	\end{equation}
	
	\begin{equation}
	\begin{aligned}
		&\frac{\partial\mathcal{L}[g_0,g_1,g_{n-1}]}{\partial g_1}(q,k,t) = j_\varrho(q) \sum_{p\in\mathbb{Z}^2} g_{n-1}(p,t) \times\\
		&\quad\times \mathcal{F}\left\{ \frac{1}{\sum_{s\in\mathbb{Z}^2} g_0(s,t)j_\varrho(s) e^{-i\frac{2\pi}{L}s\cdot r}} \right\}(k-p-q),
	\end{aligned}
	\end{equation}
	
	\begin{equation}
	\frac{\partial\mathcal{L}[g_0,g_1,g_{n-1}]}{\partial g_{n-1}}(q,k,t) = K_1(k-q,t),
	\end{equation}
and the other three derivatives are determined similarly. In the derivation of the above derivatives, one needs to take particular care so as to ensure that the perturbed variables depend on the same variables as the functions, with respect to which the differentiation is performed.

As we have already mentioned, we are interested here in stationary spatially homogeneous solutions of \EqRef{sm:chimera_pde_global_2}, i.e., $f^*(r,\varphi,t) = f^*(\varphi)$. This property results in the fact that its Fourier coefficients $f_n^*$ in Fourier space with respect to the spatial variables are factorized as $g_n^*(k) = \delta_{k,0} f_n^*$ with $\delta_{k,0}=\delta_{k_x,0}\delta_{k_y,0}$ as a product of two Kronecker delta functions. Plugging the found expressions for all the functional derivatives into \EqRef{sm:fourier_mode_pde_wrt_xy_taylor} and evaluating them at the fixed points by using the factorization property, we obtain
	\begin{equation}
	\begin{aligned}
	\partial_t \delta g_n(k,t) &= - n^2D_\varphi \delta g_{n}(k,t) \\
	&+\frac{iv_0^*}{2}k_x [\delta g_{n+1}(k,t) + \delta g_{n-1}(k,t)] \\
	&+ \frac{v_0^*}{2}k_y [\delta g_{n+1}(k,t) - \delta g_{n-1}(k,t)] \\
	&- \frac{n}{2f_0^*} \biggl[ \frac{f_1^* f_{n-1}^*}{f_0^*} j_1(k) \delta g_0(k,t) \\
	&\qquad\qquad- f_{n-1}^* j_1(k) \delta g_1(k,t) \\
	&\qquad\qquad- f_1^* \delta g_{n-1}(k,t) \biggr] e^{-i\alpha} \\
	&+ \frac{n}{2f_0^*} \biggl[ \frac{f_{-1}^* f_{n+1}^*}{f_0^*} j_1(k) \delta g_0(k,t) \\
	&\qquad\qquad- f_{n+1}^* j_1(k) \delta g_{-1}(k,t) \\
	&\qquad\qquad- f_{-1}^* \delta g_{n+1}(k,t) \biggr] e^{i\alpha},
	\end{aligned}
	\end{equation}
where we have denoted $j_1(k) = j_\varrho(k) / j_\varrho(0) = 2J_1\left(\varrho^*\Vert k\Vert\right) / \left(\varrho^*\Vert k\Vert\right)$.

Since $n\in\mathbb{Z}$, we have obtained an infinite linear system of ODEs, which we can write more compactly as
\begin{equation}
\label{sm:linearized_dynamics_in_matrix_form}
	\partial_t \delta g_n(k,t) = \sum_{m=-\infty}^\infty M_{n,m}(k)\delta g_m(k,t),
\end{equation}
where the matrix coefficients are given by
	\begin{equation}
	\label{sm:stability_matrix_for_stationary_solutions}
	\begin{aligned}
	M&_{n,m} = - n^2D_\varphi\delta_{n,m} \\
	&+\frac{v_0^*}{2}(ik_x - k_y) \delta_{n-1,m} + \frac{v_0^*}{2}(ik_x + k_y) \delta_{n+1,m} \\
	&- \frac{n}{2f_0^*} \biggl( \frac{f_1^* f_{n-1}^*}{f_0^*} j_1(k) \delta_{0,m} \\
	&\qquad\qquad- f_{n-1}^* j_1(k) \delta_{1,m} - f_1^* \delta_{n-1,m} \biggr) e^{-i\alpha} \\
	&+ \frac{n}{2f_0^*} \biggl( \frac{f_{-1}^* f_{n+1}^*}{f_0^*} j_1(k) \delta_{0,m} \\
	&\qquad\qquad- f_{n+1}^* j_1(k) \delta_{-1,m} - f_{-1}^* \delta_{n+1,m} \biggr) e^{i\alpha},
	\end{aligned}
	\end{equation}
where $\delta_{n,m}$ is the Kronecker delta symbol for $n,m\in\mathbb{Z}$. This system is the linearization of \EqRef{sm:fourier_mode_pde_wrt_xy} around a stationary spatially homogeneous solution as postulated by \EqRef{sm:perturbation_component}.

The further stability analysis for the spatially homogeneous solutions proceeds as follows. One needs to calculate the eigenvalues $\lambda$ of the stability matrix $M=(M_{n,m})_{n,m\in\mathbb{Z}}$, each of which is a function of the wave vector $k$, from the characteristic equation
\begin{equation}
\det(M - \lambda(k)I) = 0,
\end{equation}
where $I$ is the identity matrix. The relationship $\lambda = \lambda(k)$ is known as a dispersion relation and it defines the stability of a solution with respect to a given wave vector $k$. If the real part of all eigenvalues $\lambda_n, n\in\mathbb{Z}$ for all values of the wave vector is negative, then the solution is stable. If there exists an eigenvalue such that for a range of $k$ its real part becomes positive, the solution is unstable. Additionally, in the latter case, if the imaginary part of the eigenvalue is zero, we should expect another stationary pattern for that parameter set. Otherwise, we expect a nonstationary behavior of the system.

In the following, we apply the developed stability analysis framework to the known stationary spatially homogeneous solutions. Namely, a uniform density function and a von Mis\'{e}s density function \EqRef{sm:solution_uniform_zero_lag_2}.

\subsubsection{The Uniform Solution}

We are first interested in the stability analysis of the simplest solution that satisfies \EqRef{sm:chimera_pde_global}, i.e., a uniform density function $f^*(r,\varphi,t) = \frac{1}{2\pi}$. Its Fourier transform with respect to the phase variable is $f_n^*=\frac{1}{2\pi}\delta_{n,0}$. Subsequently, its Fourier transform with respect to spatial $r$ and angular $\varphi$ variables is $g_n(k,t) = \delta_{n,0}\delta_{k,0}$. We have constructed everything we need for the linear stability analysis of this solution so far. We substitute the Fourier modes of this fixed point into \EqRef{sm:stability_matrix_for_stationary_solutions} and obtain the linearized system of ODEs described by the following matrix:
\begin{equation}
\begin{aligned}
	M&_{n,m} = - n^2D_\varphi \delta_{n,m} \\
	&+ \frac{v_0^*}{2}(ik_x - k_y) \delta_{n-1,m} + \frac{v_0^*}{2}(ik_x + k_y) \delta_{n+1,m} \\
	& + \frac{n}{2}j_1(k) \left(\delta_{n,1}e^{-i\alpha} - \delta_{n,-1}e^{i\alpha}\right) \delta_{n,m}.
\end{aligned}
\end{equation}

If we consider the spatially homogeneous system \EqRef{sm:chimera_pde_local}, the above matrix simplifies by setting $k=(0,0)^T$:
\begin{equation}
\begin{aligned}
	M_{n,m} &= - n^2D_\varphi \delta_{n,m} \\
	&+\frac{n}{2} \left(\delta_{n,1}e^{-i\alpha} - \delta_{n,-1}e^{i\alpha}\right) \delta_{n,m} .
\end{aligned}
\end{equation}
Since this is a diagonal matrix, we find the eigenvalues straightforwardly. They are
\begin{equation}
\begin{aligned}
	\lambda_0 &= 0, \\
	\lambda_{\pm 1} &= \frac{1}{2}\cos\alpha - D_\varphi \mp \frac{i}{2}\sin\alpha, \\
	\lambda_{n} &= -n^2 D_\varphi, |n| \geq 2.
\end{aligned}
\end{equation}
The zeroth eigenvalue is always neutrally stable and the eigenvalues with $|n| \geq 2$ are always stable since $\text{Re}(\lambda_n)<0$. The only instability may arise for $\lambda_{\pm 1}$. The uniform solution $f(r,\varphi,t)=\frac{1}{2\pi}$ is stable, if $\text{Re}(\lambda_{\pm 1}) < 0$, i.e., if $D_\varphi > \frac{1}{2}\cos\alpha$. As a result, the line 
\begin{equation}
D_\varphi = \frac{1}{2}\cos\alpha
\end{equation}
is the transition line for the onset of polar order. This is in accordance with the result that we had when we analyzed the traveling wave solution \EqRef{sm:solution_profile_uniform_nonzero_lag}. The numerical investigations of this solution against both the parallel and transversal perturbations does not reveal any new instability mechanisms for $D_\varphi > \frac{1}{2}\cos\alpha$.

\subsubsection{The Zero Phase Lag Case}

We have showed that the other stationary solution to the spatially homogeneous system \EqRef{sm:chimera_pde_local} in the absence of phase lag $\alpha$ is
\begin{equation}
f^*(\varphi) = \frac{e^{\gamma\cos\varphi}}{2\pi I_0(\gamma)},
\end{equation}
where $\gamma = R/D_\varphi$. The Fourier modes in the series expansion of this solution with respect to the phase variable $\varphi$ read
\begin{equation}
f_n^* = \frac{I_n(\gamma)}{2\pi I_0(\gamma)}.
\end{equation}
For completeness, we here provide the stability matrix \EqRef{sm:stability_matrix_for_stationary_solutions} for the linearized dynamics of the perturbations to this solution:
	\begin{equation}
	\begin{aligned}
		M&_{n,m} = - n^2D_\varphi \delta_{n,m} \\
		&+ \frac{v_0^*}{2}(ik_x - k_y) \delta_{n-1,m} + \frac{v_0^*}{2}(ik_x + k_y) \delta_{n+1,m}  \\
		&- \frac{n}{2} \biggl( j_1(k) \frac{I_{1}(\gamma)I_{n-1}(\gamma)}{I_0^2(\gamma)} \delta_{0,m} \\
		&\qquad - j_1(k) \frac{I_{n-1}(\gamma)}{I_0(\gamma)} \delta_{1,m} - \frac{I_1(\gamma)}{I_0(\gamma)} \delta_{n-1,m} \biggr) \\
		&+ \frac{n}{2} \biggl( j_1(k) \frac{I_{-1}(\gamma)I_{n+1}(\gamma)}{I_0^2(\gamma)} \delta_{0,m} \\
		&\qquad - j_1(k) \frac{I_{n+1}(\gamma)}{I_0(\gamma)} \delta_{-1,m} - \frac{I_{-1}(\gamma)}{I_0(\gamma)} \delta_{n+1,m} \biggr).
	\end{aligned}
	\end{equation}
Because of its form, we cannot solve an eigenvalue problem for this stability matrix in the spatially nonhomogeneous case analytically. Thus, we solve it numerically. As a result, it appears that the von Mis\'{e}s density function is always stable for $D_\varphi < \frac{1}{2}$ (see its further analysis in the next section on the hydrodynamic theory approach).

\subsection{Traveling Wave Solutions}

Inside the region of partial polar order, we have shown that the solution to \EqRef{sm:chimera_pde_local} in the presence of the phase lag is given by \EqRef{sm:solution_profile_uniform_nonzero_lag}. If we use the ansatz $\omega = \varphi - vt$, we find that \EqRef{sm:solution_uniform_nonzero_lag} solves \EqRef{sm:chimera_pde_local} as well as \EqRef{sm:chimera_pde_global_2}. The stability analysis framework, developed so far, is valid only for the stationary solutions of \EqRef{sm:chimera_pde_global_2}, which was stated by \EqRef{sm:perturbation_component}. In order to make the same framework be applicable to \EqRef{sm:solution_uniform_nonzero_lag}, we introduce the following ansatz being an extension to the traveling wave ansatz used before:

\begin{equation}
\label{sm:helical_ansatz}
\begin{aligned}
\xi &= \cos(vt)x + \sin(vt)y, \\
\eta &= -\sin(vt)x + \cos(vt)y, \\
\omega &= \varphi - vt,\quad h(\xi,\eta,\omega,t) = f(x,y,\varphi,t).
\end{aligned}
\end{equation}
Leaving alone the spatial variables for the moment, the previous substitution of the form $f(\varphi,t)=g(\omega)$ transforms the PDE of two variables into the ODE. Instead, if we consider the substitution like $f(\varphi,t)=h(\omega,t)$, we transform the PDE to the moving frame $\varphi-vt$, which has the solution \EqRef{sm:solution_profile_uniform_nonzero_lag} as its stationary solution. Now, in order to perform the stability analysis of \EqRef{sm:chimera_pde_global_2} instead of \EqRef{sm:chimera_pde_local}, one also needs to take into account that with the substitution $\omega=\varphi-vt$, the spatial advection terms become dependent on time and the stability analysis is again not applicable. In order to circumvent that, we introduce the ansatz \EqRef{sm:helical_ansatz}. After its application, the PDE becomes
	\begin{equation}
	\label{sm:chimera_pde_helical_ansatz}
	\begin{aligned}
		\partial_t &h = - (v_0\cos\omega \!+\! v\eta) \partial_\xi h - (v_0\sin\omega \!-\! v\xi) \partial_\eta h \\
		&+ D_\varphi\partial_{\omega\omega}h + v\partial_\omega h \\
		&- \partial_\omega \Biggl( h \frac{\int_{C(r)} h(r',\omega',t)\sin(\omega' \!-\! \omega \!-\! \alpha) \mathrm{d}r'\mathrm{d}\omega'}{\int_{C(r)} h(r',\omega',t) \mathrm{d}r'\mathrm{d}\omega'} \Biggr),
	\end{aligned}
	\end{equation}
where we now denote the spatial vector as $r=(\xi,\eta)\in\mathbb{U}^2$.

This equation admits the profile of the spatially homogeneous traveling wave solution \EqRef{sm:solution_profile_uniform_nonzero_lag} as its stationary solution:

\begin{equation}
\label{sm:solution_profile_uniform_nonzero_lag_2}
	h(\omega) = c_0 E(\omega)
	\left( 1 + c_1 \frac{\int_{0}^{\omega} E^{-1}(\omega') \mathrm{d}\omega'}{\int_{\mathbb{T}} E^{-1}(\omega') \mathrm{d}\omega'} \right).
\end{equation}
where $E(\omega) = \exp\left[ -\frac{v}{D_\varphi}\omega + \frac{R}{D_\varphi} \cos(\omega + \alpha) \right]$, $c_0$ is a normalization constant, and $c_1 = \left( e^{2\pi\frac{v}{D_\varphi}} - 1 \right)$ comes from a periodicity constraint. 

Due to that fact, we can proceed in the same manner as we did for the stationary solutions. First, we expand the new density function $h$ into a Fourier series with respect to the phase $\omega$ and substitute the expansion into \EqRef{sm:chimera_pde_helical_ansatz}. If we gather the terms of each Fourier mode together, we obtain
	\begin{equation}
	\label{sm:fourier_mode_pde_wrt_phase_traveling_wave}
	\begin{aligned}
		\partial_t &h_n(r,t) = - n^2D_\varphi h_n + v(r\times\nabla)h_n - invh_n \\
		&-\frac{v_0}{2}\partial_\xi(h_{n+1} + h_{n-1}) + \frac{iv_0}{2}\partial_\eta(h_{n+1} - h_{n-1})  \\
		&+ \frac{n}{2}\Biggl( h_{n-1} \frac{\iint_{B(r)} h_1(r',t) \mathrm{d}r'}{\iint_{B(r)} h_0(r',t) \mathrm{d}r'} e^{-i\alpha} \\
		&\qquad\qquad - h_{n+1} \frac{\iint_{B(r)} h_{-1}(r',t) \mathrm{d}r'}{\iint_{B(r)} h_0(r',t) \mathrm{d}r'} e^{i\alpha} \Biggr),
	\end{aligned}
	\end{equation}
where $\times$ denotes the third component of a cross product, i.e., $r\times\nabla = \xi\partial_\eta - \eta\partial_\xi$.

Next, we need to transform this hierarchy of position dependent Fourier modes into the Fourier space with respect to the spatial variables using \EqRef{sm:forward_fourier_transform_wrt_xy}. Most of the terms are transformed as described in the previous section, except for

\begin{equation}
\begin{aligned}
	\mathcal{F}\{ v(r\times\nabla)h_n \}&(k,t) \\
	= iv\frac{2\pi}{L} \mathcal{F}&\left\{ (k\times r) h_n(r,t) \right\}(k,t) \\
	= iv\pi(k_x&-k_y) g_n(k_x,k_y,t) \\
	&+ \sum_{\substack{q_y\in\mathbb{Z} \\ k_y\neq q_y}} \frac{v k_x}{k_y-q_y}g_n(k_x,q_y,t) \\
	&- \sum_{\substack{q_x\in\mathbb{Z} \\ k_x\neq q_x}} \frac{v k_y}{k_x-q_x}g_n(q_x,k_y,t).
\end{aligned}
\end{equation}
The appearance of the couplings between $g_n(k)$ and $g_n(k')$ with $k\neq k'$, i.e., to the function values at the wave vectors other than $k$, makes the subsequent linear stability analysis convoluted and computationally unfeasible. We could map three Fourier indices $n$, $k_x$, and $k_y$ into one index and perform the stability analysis of all the perturbations together but the solution to the eigenvalue problem would lose the spatial dependence, and we would not be able to obtain the dispersion relations and draw necessary conclusions. But note that if we rescale the spatial dimension in \EqRef{sm:fourier_mode_helical_pde_wrt_xy} as $k'=v_0k$ (see the detailed example about the rescaling of the marginal density function in the section S4.4), we see that the above Fourier transform is mainly determined by the first term in the limit of small particle velocities $v_0$ so that we may assume
\begin{equation}
	\mathcal{F}\{ v(r\times\nabla)h_n \}(k,t) \approx iv\pi(k_x-k_y) g_n(k_x,k_y,t).
\end{equation}

As a result, the PDE \EqRef{sm:fourier_mode_pde_wrt_phase_traveling_wave} for the Fourier modes of the traveling wave solution in the Fourier space with respect to the spatial variables becomes
	\begin{equation}
	\label{sm:fourier_mode_helical_pde_wrt_xy}
	\begin{aligned}
		\partial_t &g_n(k,t) = - (n^2D_\varphi + inv) g_n \\
		&+ iv\pi(k_x-k_y) g_n(k,t) \\
		&+ \frac{iv_0^*}{2}k_x (g_{n+1} + g_{n-1}) + \frac{v_0^*}{2}k_y (g_{n+1} - g_{n-1}) \\
		&+ \frac{n}{2} \Biggl[ e^{-i\alpha} \sum_{q\in\mathbb{Z}^2} g_{n-1}(q,t)K_1(k-q,t) \\
		&\qquad\qquad- e^{i\alpha} \sum_{q\in\mathbb{Z}^2} g_{n+1}(q,t)K_{-1}(k-q,t) \Biggr].
	\end{aligned}
	\end{equation}
where we have again denoted $v_0^* = \frac{2\pi}{L} v_0$ and we will also use $\varrho^* = \frac{2\pi}{L} \varrho$.

At this point, we postulate again that in order to proceed further in the derivation of the linearized differential equations for the perturbations, we use the fact that the solutions, we are interested in, are stationary solutions of \EqRef{sm:chimera_pde_helical_ansatz}, i.e., their Fourier transforms with respect to the spatial variables are $g_n^*(k)=\delta_{k,0}h_n^*$. Under such assumptions, the perturbations have the form \EqRef{sm:perturbation_component} and we need to linearize \EqRef{sm:fourier_mode_helical_pde_wrt_xy} around $g_n^*,n\in\mathbb{Z}$.

Following the same procedure as for the stationary solutions, we find the time evolution of the linearized dynamics of small spatially dependent perturbations to be
	\begin{equation}
	\label{sm:fourier_mode_helical_pde_wrt_xy_2}
	\begin{aligned}
		\partial_t \delta &g_n(k,t) = - (n^2D_\varphi + inv) \delta g_{n} \\
		&+ iv\pi(k_x-k_y) \delta g_n \\
		&+\frac{iv_0^*}{2}k_x (\delta g_{n+1} + \delta g_{n-1}) + \frac{v_0^*}{2}k_y (\delta g_{n+1} - \delta g_{n-1}) \\ 
		&- \frac{n}{2h_0^*} \biggl[ \frac{h_1^* h_{n-1}^*}{h_0^*} j_1(k) \delta g_0(k,t) \\
		&\qquad- h_{n-1}^* j_1(k) \delta g_1(k,t) - h_1^* \delta g_{n-1}(k,t) \biggr] e^{-i\alpha} \\
		&+ \frac{n}{2h_0^*} \biggl[ \frac{h_{-1}^* h_{n+1}^*}{h_0^*} j_1(k) \delta g_0(k,t) \\
		&\qquad- h_{n+1}^* j_1(k) \delta g_{-1}(k,t) - h_{-1}^* \delta g_{n+1}(k,t) \biggr] e^{i\alpha}.
	\end{aligned}
	\end{equation}
The corresponding stability matrix coefficients \EqRef{sm:linearized_dynamics_in_matrix_form} read
	\begin{equation}
	\label{sm:infinite_stability_matrix_for_traveling_wave}
	\begin{aligned}
		M&_{n,m} = - (n^2D_\varphi + inv) \delta_{n,m} \\
		&+ iv\pi(k_x-k_y)\delta_{n,m} \\
		&+\frac{v_0^*}{2} (ik_x - k_y) \delta_{n-1,m} + \frac{v_0^*}{2} (ik_x + k_y) \delta_{n+1,m} \\
		&- \frac{n}{2h_0^*} \biggl( \frac{h_1^*h_{n-1}^*}{h_0^*} j_1(k) \delta_{0,m} \\
		&\qquad- h_{n-1}^* j_1(k) \delta_{1,m} - h_1^* \delta_{n-1,m} \biggr) e^{-i\alpha} \\
		&+ \frac{n}{2h_0^*} \biggl( \frac{h_{-1}^*h_{n+1}^*}{h_0^*} j_1(k) \delta_{0,m} \\
		&\qquad- h_{n+1}^* j_1(k) \delta_{-1,m} - h_{-1}^* \delta_{n+1,m} \biggr) e^{i\alpha}.
	\end{aligned}
	\end{equation}

\subsubsection{The Nonzero Phase Lag Case}

In order to perform the stability analysis of the traveling wave solution, we need to transform it first into the Fourier space with respect to the phase variable $\varphi$. However, one cannot straightforwardly integrate it using \EqRef{sm:fourier_mode_wrt_phase} as required by the definition of Fourier modes. We first make use of the following decomposition of an exponential function into a series containing the modified Bessel functions of the first kind:
\begin{equation}
	e^{\gamma\cos\varphi} = I_0(\gamma) + 2\sum_{\nu=1}^\infty I_\nu(\gamma)\cos(\nu\varphi),\quad \gamma\in\mathbb{R},\varphi\in\mathbb{T}.
\end{equation}
in order to rewrite the corresponding terms in the density function. Such a decomposition makes it possible to integrate \EqRef{sm:solution_profile_uniform_nonzero_lag_2} when applying \EqRef{sm:fourier_mode_wrt_phase}. Performing the lengthy integration, one can show that the Fourier modes take the following form:
\begin{equation}
\label{sm:fourier_mode_of_homogeneous_solution}
\begin{aligned}
	f_n(t) = c_0 \biggl\{ c_1(&n) I_0(\gamma)I_n(-\gamma) \\
	+ &\sum_{\nu=1}^\infty I_\nu(\gamma) \bigl[ c_1(n+\nu)I_{n+\nu}(-\gamma) \\
	&+ c_1(n-\nu)I_{n-\nu}(-\gamma) \bigr] \biggr\} e^{-in\alpha},
\end{aligned}
\end{equation}
where as previous $\gamma = R/D_\varphi$ and the normalization constant can be shown to be
\begin{equation}
	c_0 = \frac{1}{2\pi} \left\{ c_1(0)I_0^2(\gamma) + \sum_{\nu=1}^\infty I_\nu(\gamma)I_\nu(-\gamma) \left[ c_1(\nu) + \bar{c}_1(\nu) \right] \right\} ^ {-1},
\end{equation}
where $\bar{c}_1$ denotes a complex conjugate. We have also denoted for brevity
\begin{align}
	c_1(\nu) = \frac{v/D_\varphi + i\nu}{(v/D_\varphi)^2 + \nu^2}.
\end{align}

Given such a representation of the solution \EqRef{sm:solution_profile_uniform_nonzero_lag_2} and even more complicated form of the stability matrix than it was for the zero phase lag case, the only way to study stability properties here is using the numerical methods. For spatially homogeneous perturbations, we find that \EqRef{sm:solution_profile_uniform_nonzero_lag_2} is always stable for $D_\varphi < \frac{1}{2}\cos\alpha$. However, we are interested in the development of small spatially dependent perturbations to the solution \EqRef{sm:solution_profile_uniform_nonzero_lag_2}. The dynamics of such perturbations is governed by the linearized system \EqRef{sm:fourier_mode_helical_pde_wrt_xy_2}. It depends on the Fourier modes $f_n,n\in\mathbb{Z}$ of the solution, which are given by \EqRef{sm:fourier_mode_of_homogeneous_solution}.

\section{Stability Analysis via Hydrodynamic Theory\label{sec:stability_analysis_via_hydrodynamic_theory}}

The stability analysis from the point of view of the kinetic theory is effective when using the numerical methods, thus, providing us with the quantitative information about the instability mechanisms that act on a solution. If we want to have the qualitative description of the system behavior with respect to the microscopic parameters of the model, we must restrict the consideration of the infinite hierarchy of the Fourier modes \EqRef{sm:fourier_mode_pde_wrt_phase} to the first several ones. The common strategy to pursue is the following. The expansion of the density function that solves the original spatially nonhomogeneous PDE \EqRef{sm:chimera_pde_global_2} into Fourier series transforms the problem of solving the temporal dynamics of the $3+1$-dimensional density function into the problem of solving the temporal dynamics of the infinite system of $2+1$-dimensional density functions. The first modes from that hierarchy can be given a reasonable interpretation. Namely, the integration of $f(r,\varphi,t)$ over the phase variable gives a marginal density function of spatial coordinates $r=(x,y)\in\mathbb{U}^2$:
\begin{equation}
	\rho(r,t) = \int_{\mathbb{T}} f(r,\varphi,t) \mathrm{d}\varphi = 2\pi f_0(r,t).
\end{equation}
This definition also establishes the connection of the marginal density function to the zeroth Fourier mode defined in \EqRef{sm:fourier_series_wrt_phase}. Furthermore, we consider an arbitrary unit velocity vector $e(\varphi)=(\cos\varphi,\sin\varphi)\in\mathbb{S}^1\subset\mathbb{R}^2$ and find its expectation with respect to the one particle density function $f(r,\varphi,t)$. The result of this operation is known to be a momentum field $w(r,t)=(w_x(r,t),w_y(r,t))\in\mathbb{R}^2$, which is defined in the relation to the Fourier modes as
\begin{equation}
\begin{aligned}
	w_x(r,t) &= \int_{\mathbb{T}} \cos\varphi f(r,\varphi,t) \mathrm{d}\varphi \\
	&= \pi [f_1(r,t) + f_{-1}(r,t)], \\
	w_y(r,t) &= \int_{\mathbb{T}} \sin\varphi f(r,\varphi,t) \mathrm{d}\varphi \\
	&= -i\pi [f_1(r,t) - f_{-1}(r,t)].
\end{aligned}
\end{equation}
The marginal density function and the momentum field constitute a hydrodynamic description of a system of interacting particles.

The momentum field, divided by the marginal density function, is isomorphic to the order parameter we introduced earlier \EqRef{sm:order_parameter_uniform_nonzero_lag}, i.e., we could associate $\dfrac{w_x(r,t)}{\rho(r,t)} = R(r,t)\cos\Theta(r,t)$ and $\dfrac{w_y(r,t)}{\rho(r,t)} = R(r,t)\sin\Theta(r,t)$. Thus, the knowledge of the hydrodynamic variables $\rho(r,t)$ and $w(r,t)$ automatically allows us to evaluate the degree of polarization in the particle flow. Examples of coarse grained hydrodynamic variables can be found in Fig.~\ref{fig:coarse_grained_fields}.

\subsection{Stationary Solutions\label{subsec:stability_analysis_via_hydrodynamic_theory_stationary_solutions}}

We are interested in the time evolution of the marginal density function and the momenta. They are obtained directly from the above definitions using the differential equations of the respective Fourier coefficients. Since the temporal dynamics of each Fourier mode is coupled to the neighboring modes, we cannot derive the required equations immediately. Namely, we see that the first Fourier mode couples to the second one through the convective terms and through the nonlinear alignment terms \EqRef{sm:fourier_mode_pde_wrt_phase}. In order to obtain the closure, we adopt the approach of \cite{bertin:jpa}. We assume that the temporal evolution of the nematic order field, which is related to the second Fourier modes, is a small quantity, giving $\partial_t f_{\pm2} \approx 0$. Furthermore, we assume that the higher order fields are negligible $f_{n}\approx0,|n|\geq3$. This is appropriate for sufficiently high diffusion levels since $\partial_t f_n \propto -n^2D_\varphi$. As a result, the second Fourier mode is approximated by
\begin{equation}
\label{sm:closure_relation}
\begin{aligned}
	f_2(r,t) &\approx -\frac{v_0}{8D_\varphi} [\partial_x f_1(r,t) + i\partial_y f_1(r,t)] \\
	&+ \frac{f_1(r,t)}{4D_\varphi} \frac{\iint_{B(r)} f_1(r',t) \mathrm{d}r'}{\iint_{B(r)} f_0(r',t) \mathrm{d}r'} e^{-i\alpha}.
\end{aligned}
\end{equation}
Using the closure relation \EqRef{sm:closure_relation}, we find the following system of differential equations for the marginal density $\rho=\rho(r,t)$ and the momentum field $w=w(r,t)$:
\begin{equation}
\label{sm:hydrodynamic_equations}
\begin{aligned}
	\partial_t \rho &= -v_0 \nabla\cdot w, \\
	\partial_t w &= -\frac{v_0}{2} \nabla\rho - D_\varphi w + \frac{v_0^2}{16D_\varphi} \Delta w + \frac{\rho}{2} Q_{-\alpha} W \\
	&+ \frac{1}{8D_\varphi} \biggl\{ \frac{v_0}{2} Q_\alpha \left[ (W\cdot\nabla)w + (W_\perp\cdot\nabla)w_\perp \right] \\
	& - w\Vert W\Vert^2 + v_0 Q_{-\alpha} \Bigl[ \nabla(w\cdot W) - (W\cdot\nabla)w \\
	& - (\nabla\cdot W)w - W(\nabla\cdot w) - (w\cdot\nabla)W \Bigr] \biggr\}.
\end{aligned}
\end{equation}
where $w_\perp=(-w_y,w_x)^T$ and $W_\perp=(-W_y,W_x)^T$. We have denoted the spatially averaged momentum field as $W = W(r,t) = \iint_{B(r)} w(r',t) \mathrm{d}r' / \iint_{B(r)} \rho(r',t) \mathrm{d}r'$ by analogy with the interaction terms in \EqRef{sm:chimera_pde_global}. Note that the neighborhood domain $B(r)$ implicitly depends on the radius of interaction $\varrho$, as defined previously by \EqRef{sm:continuum_neighborhood_domain_2d}. The matrix $Q_\alpha = \begin{pmatrix} \cos\alpha & -\sin\alpha \\ \sin\alpha & \cos\alpha \end{pmatrix}$ represents anticlockwise rotation by $\alpha$ radians.


We begin our analysis by looking for the solutions of \EqRef{sm:hydrodynamic_equations}. As usual for such models, \EqRef{sm:hydrodynamic_equations} has two stationary spatially homogeneous solutions. The first one $(\rho,w) = (1,0,0)$ represents a spatially uniform disordered state of the system. The second one is best found using the polar representation for the momentum field. Note that due to the approach we used to derive the continuum limit, the particle density $\rho_0\equiv1$ since we are bound to work only with probability DFs, and as a result the marginal density function $\rho=1$. Assuming the temporal and spatial independence of the solutions, we have $W = (\pi\varrho^2 w_x,\pi\varrho^2 w_y)^T$. We find the second solution, which represents the partially synchronized flocking, to be $(\rho,w) = (1, \Vert w^*\Vert\cos\varphi_0, \Vert w^*\Vert\sin\varphi_0)$, where the degree of polarization is
\begin{equation}
\label{sm:solution_hydrodynamic_polarization}
	\Vert w^*\Vert = 2\sqrt{D_\varphi(\cos\alpha - 2D_\varphi)}
\end{equation}
and $\varphi_0\in\mathbb{T}$ is an arbitrary direction subject to initial conditions. One of the assumptions that we have used to obtain the closure relation \EqRef{sm:closure_relation} is that the diffusion level is high enough to justify the negligence of the higher order Fourier modes. We can now see from \EqRef{sm:solution_hydrodynamic_polarization} the limitations of those assumptions. Namely, it shows that the polarization level goes to zero with the diffusion value going to zero which is definitely not correct. From our previous study we know that the polarization level goes up to one with the decrease of $D_\varphi$. But at the onset of the flocking $D_\varphi = 1/2$, the answer is correct. By analyzing \EqRef{sm:solution_hydrodynamic_polarization}, we find that the maximum polarization level could be $\frac{\sqrt{2}}{2} \cos\alpha$ and it is attained at $D_\varphi = \frac{1}{4}\cos\alpha$. Thus, the hydrodynamic equations \EqRef{sm:hydrodynamic_equations} are correct for the range $\frac{1}{4}\cos\alpha \leq D_\varphi \leq \frac{1}{2}\cos\alpha$, the later of which is the order-disorder transition line as we already know and it is also the existence condition for \EqRef{sm:solution_hydrodynamic_polarization}.

Now that we have the solutions of the model, we proceed with the analysis of their stability. As previously, we first transform the equations to the Fourier space with respect to the spatial variables. Since we work in the space with periodic boundaries, we may expand the marginal density function as
\begin{equation}
	\rho(r,t) = \sum_{k\in\mathbb{Z}^2} \hat{\rho}(k,t) e^{-i\frac{2\pi}{L}k\cdot r},
\end{equation}
where the coefficients of each mode are defined as
\begin{equation}
\begin{aligned}
	\hat{\rho}(k,t) &= \mathcal{F}\{\rho(r,t)\}(k,t) \\
	&= \frac{1}{L^2}\iint_{\mathbb{U}^2} \rho(r,t) e^{i\frac{2\pi}{L}k\cdot r} \mathrm{d}r.
\end{aligned}
\end{equation}
The expansion for the momentum field is defined similarly.

The transformation of \EqRef{sm:hydrodynamic_equations} into the Fourier space with respect to the spatial variables is then performed the same way as we did it in the kinetic theory. Thus, we do not delve into all the details here but mention several key points. Namely, the transforms of all functionally different terms can be found to be
\begin{equation}
	\mathcal{F}\{\nabla\cdot w(r,t)\}(k,t) = -i\frac{2\pi}{L} (\hat{w}(k,t)\cdot k),
\end{equation}
\begin{equation}
	\mathcal{F}\{\nabla\rho(r,t)\}(k,t) = -i\frac{2\pi}{L}\hat{\rho}(k,t)k,
\end{equation}
\begin{equation}
	\mathcal{F}\{\Delta w(r,t)\}(k,t) = -\left(\frac{2\pi}{L}\right)^2 |k|^2 \hat{w}(k,t),
\end{equation}
\begin{equation}
	\mathcal{F}\{\rho(r,t)W(r,t)\}(k,t) = \sum_{q\in\mathbb{Z}^2} \hat{\rho}(q,t) K_1(k-q,t),
\end{equation}
\begin{equation}
\begin{aligned}
	\mathcal{F}&\{\nabla[w(r,t)\cdot W(r,t)]\}(k,t) \\
	&= -i\frac{2\pi}{L} k\sum_{q\in\mathbb{Z}^2} [\hat{w}(q,t)\cdot K_1(k-q,t)],
\end{aligned}
\end{equation}
\begin{equation}
\begin{aligned}
	\mathcal{F}&\{w(r,t)[\nabla\cdot W(r,t)]\}(k,t) \\
	&= -i\frac{2\pi}{L} \sum_{q\in\mathbb{Z}^2} \hat{w}(q,t) [(k-q)\cdot K_1(k-q,t)],
\end{aligned}
\end{equation}
\begin{equation}
\begin{aligned}
	\mathcal{F}&\{[w(r,t)\cdot\nabla]W(r,t)\}(k,t) \\
	&= -i\frac{2\pi}{L} \sum_{q\in\mathbb{Z}^2} [(k-q)\cdot\hat{w}(q,t)] K_1(k-q,t),
\end{aligned}
\end{equation}
\begin{equation}
\begin{aligned}
	\mathcal{F}&\{W(r,t)[\nabla\cdot w(r,t)]\}(k,t) \\
	&= -i\frac{2\pi}{L} \sum_{q\in\mathbb{Z}^2} [q\cdot\hat{w}(q,t)] K_1(k-q,t),
\end{aligned}
\end{equation}
\begin{equation}
\begin{aligned}
	\mathcal{F}&\{[W(r,t)\cdot\nabla]w(r,t)\}(k,t) \\
	&= -i\frac{2\pi}{L} \sum_{q\in\mathbb{Z}^2} \hat{w}(q,t) [q\cdot K_1(k-q,t)],
\end{aligned}
\end{equation}
\begin{equation}
\begin{aligned}
	\mathcal{F}&\{w(r,t)\Vert W(r,t)\Vert^2\}(k,t) \\
	&= \sum_{q\in\mathbb{Z}^2} \hat{w}(q,t) K_2(k-q,t),
\end{aligned}
\end{equation}
where we have introduced two kernel functions in comparison with the ones from the previous chapter:
\begin{equation}
\begin{aligned}
K_1(k-q,t) &= \frac{1}{L^2} \iint_{\mathbb{U}^2} W(r,t) e^{i\frac{2\pi}{L}(k-q)\cdot r} \mathrm{d}r, \\
K_2(k-q,t) &= \frac{1}{L^2} \iint_{\mathbb{U}^2} \Vert W(r,t)\Vert^2 e^{i\frac{2\pi}{L}(k-q)\cdot r} \mathrm{d}r.
\end{aligned}
\end{equation}
Note that the spatially averaged momentum field can be written in terms of the wave vectors as
\begin{equation}
	W(r,t) = \frac{\sum_{p\in\mathbb{Z}^2} \hat{w}(p,t) j_\varrho(p) e^{-i\frac{2\pi}{L} p\cdot r}}{\sum_{p\in\mathbb{Z}^2} \hat{\rho}(p) j_\varrho(p) e^{-i\frac{2\pi}{L} p\cdot r}},
\end{equation}
where $j_\varrho(p) = J_1\left(\frac{2\pi}{L}\varrho\Vert p\Vert\right) / \Vert p\Vert$ as before.

The hydrodynamic equations in the Fourier space with respect to the spatial variables read
\begin{widetext}
	\begin{equation}
	\label{sm:hydrodynamic_equations_fourier_wrt_xy}
	\begin{aligned}
		\partial_t \hat{\rho}&(k) = iv_0^* [k\cdot\hat{w}(k)], \\
		\partial_t \hat{w}&(k) = \frac{iv_0^*}{2} k\hat{\rho}(k) - D_\varphi \hat{w}(k) - \frac{(v_0^*)^2}{16D_\varphi} \vert k\vert^2 \hat{w}(k) \\
		&+ \sum_{q\in\mathbb{Z}^2}\Biggl\{ \frac{1}{2}Q_{-\alpha} \hat{\rho}(q)K_1(k\!-\!q) - \frac{1}{8D_\varphi}\hat{w}(q)K_2(k\!-\!q) - \frac{iv_0^*}{16D_\varphi} Q_\alpha \left\{ \hat{w}(q) [q\!\cdot\! K_1(k\!-\!q)] + \hat{w}_\perp(q) [q\!\cdot\! K_{1,\perp}(k\!-\!q)] \right\} \\
		&- \frac{iv_0^*}{8D_\varphi} Q_{-\alpha} \{ k[\hat{w}(q)\!\cdot\! K_1(k\!-\!q)] - \hat{w}(q)[k\!\cdot\! K_1(k\!-\!q)] - [k\!\cdot\!\hat{w}(q)] K_1(k\!-\!q) \} \Biggr\},
	\end{aligned}
	\end{equation}
\end{widetext}
where $K_{1,\perp}=(-K_{1,y},K_{1,x})^T$ and we use $v_0^*=\frac{2\pi}{L}v_0$ and $\varrho^*=\frac{2\pi}{L}\varrho$ as before. Note that we suppressed the explicit time dependence of $\hat{\rho}$, $\hat{w}$, $K_1$, and $K_2$ for compactness.

Let a stationary spatially homogeneous solution to \EqRef{sm:hydrodynamic_equations} be $\rho(r,t)=1$ (since we are allowed to work only with probability density functions) and $w(r,t) = w^*$. Due to its spatial homogeneity, the Fourier transform of such a solution is $\hat{\rho}(k,t)=\hat{\rho}^*(k) = \delta_{k,0}$, $\hat{w}(k,t) = \hat{w}^*(k) = w^*\delta_{k,0}$ with $\delta_{k,0}=\delta_{k_x,0}\delta_{k_y,0}$. We consider the infinitesimal deviations from such a solution as

\begin{equation}
\begin{aligned}
	\delta \hat{\rho}(k,t) &= \hat{\rho}(k,t) - \hat{\rho}^*(k), \\
	\delta \hat{w}(k,t) &= \hat{w}(k,t) - \hat{w}^*(k)
\end{aligned}
\end{equation}
and we want to derive the linearized dynamics for these perturbations if they evolve according to \EqRef{sm:hydrodynamic_equations_fourier_wrt_xy}. Note that the kernels $K_{1,2}$ implicitly depend on the marginal density $\hat{\rho}$ and the momentum field $\hat{w}$. 

The complete procedure how the linearization is done is the same as it was previously for the kinetic equations. For that reason, we do not go into the details here. One can show that the linearized dynamics of the perturbations around a stationary solution follow
\begin{widetext}
	\begin{equation}
	\begin{aligned}
		\partial_t \delta\hat{\rho}&(k,t) = iv_0^* [k\cdot\delta\hat{w}(k,t)], \\
		\partial_t \delta\hat{w}&(k,t) = \left\{ \frac{iv_0^*}{2}k + \frac{1}{2}[1 - j_1(k)] Q_{-\alpha} w^* + \frac{1}{4D_\varphi} j_1(k) \Vert w^*\Vert^2 w^* + \frac{iv_0^*}{8D_\varphi} Q_{-\alpha} j_1(k) \left( \Vert w^* \Vert^2 k - 2(k\cdot w^*)w^* \right) \right\} \delta\hat{\rho}(k,t) \\
		&+ \left\{ -D_\varphi - \frac{(v_0^*)^2}{16D_\varphi} \vert k\vert^2 + \frac{1}{2}j_1(k) Q_{-\alpha} - \frac{1}{8D_\varphi}\Vert w^*\Vert^2 I - \frac{1}{4D_\varphi} j_1(k) (w^*\otimes w^*) \right.\\
		&\left. - \frac{iv_0^*}{16D_\varphi} Q_\alpha \left[ (k\cdot w^*)I + (k\cdot w_\perp^*) Q_{\frac{\pi}{2}} \right] - \frac{iv_0^*}{8D_\varphi} [1+j_1(k)] Q_{-\alpha} \left[  (k\otimes w^*) - (w^*\otimes k) - (k\cdot w^*)I \right] \right\} \delta\hat{w}(k,t),
	\end{aligned}
	\end{equation}
\end{widetext}
where $\otimes$ denotes the outer product. As before, we have denoted $j_1(k) = 2J_1\left(\varrho^* \vert k\vert\right) / \left(\varrho^* \vert k\vert\right)$. 

For the further analysis, it would be helpful to rewrite this linearized system in a matrix form as

\begin{equation}
	\partial_t \begin{pmatrix} \delta\hat{\rho} \\ \delta\hat{w}_x \\ \delta\hat{w}_y \end{pmatrix} = 
	M \begin{pmatrix} \delta\hat{\rho} \\ \delta\hat{w}_x \\ \delta\hat{w}_y \end{pmatrix},
\end{equation}
where $M=(M_{n,m})_{n,m=1,2,3}$ is the stability matrix with the coefficients

\begin{equation}
	M_{11} = 0,\quad M_{12} = iv_0^*k_x,\quad M_{13} = iv_0^*k_y,
\end{equation}
\begin{equation}
\begin{aligned}
	M_{21} &= \frac{iv_0^*}{2}k_x + \frac{1}{2} [1-j_1(k)] (w^*\cdot n_\alpha) \\
	&+ \frac{1}{4D_\varphi}j_1(k) \Vert w^*\Vert^2 w_x^* \\
	&+ \frac{iv_0^*}{8D_\varphi} j_1(k) \left[ \Vert w^*\Vert^2 (k\cdot n_\alpha) - 2(k\cdot w^*) (w^*\cdot n_\alpha) \right],
\end{aligned}
\end{equation}
\begin{equation}
\begin{aligned}
	M_{22} &= \frac{iv_0^*}{8D_\varphi} \biggl\{ [1\!+\!j_1(k)] [(w^*\!\cdot\! n_\alpha) k_x - w_x^* (k\!\cdot\! n_\alpha)] \\
	&+ \left[\frac{1}{2}\!+\!j_1(k)\right] (k\!\cdot\! w^*) \cos\alpha \\
	&+ \frac{1}{2} (k\!\cdot\! w_\perp^*) \sin\alpha \biggr\} - D_\varphi - \frac{(v_0^*)^2}{16D_\varphi} \vert k\vert^2 \\
	&+ \frac{1}{2}j_1(k)\left( \cos\alpha - \frac{(w_x^*)^2}{2D_\varphi} \right) - \frac{\Vert w^*\Vert^2}{8D_\varphi},
\end{aligned}
\end{equation}
\begin{equation}
\begin{aligned}
	M_{23} &= \frac{iv_0^*}{8D_\varphi} \biggl\{ [1 + j_1(k)] [(w^*\cdot n_\alpha) k_y - w_y^* (k\cdot n_\alpha)] \\
	&+ \left[\frac{3}{2} + j_1(k)\right] (k\cdot w^*) \sin\alpha \\
	&+ \frac{1}{2} (k\cdot w_\perp^*) \cos\alpha \biggr\} + \frac{1}{2}j_1(k)\left( \sin\alpha - \frac{w_x^*w_y^*}{2D_\varphi} \right),
\end{aligned}
\end{equation}
\begin{equation}
\begin{aligned}
	M_{31} &= \frac{iv_0^* k_y}{2} - \frac{1}{2} [1-j_1(k)] (n_\alpha\cdot w_\perp^*) \\
	&+ \frac{1}{4D_\varphi}j_1(k) \Vert w^*\Vert^2 w_y^* \\
	&+ \frac{iv_0^*}{8D_\varphi} j_1(k) \left[ 2(k\cdot w^*) (n_\alpha\cdot w_\perp^*) - \Vert w^*\Vert^2 (n_\alpha\cdot k_\perp) \right],
\end{aligned}
\end{equation}
\begin{equation}
\begin{aligned}
	M_{32} &= \frac{iv_0^*}{8D_\varphi} \biggl\{ [1+j_1(k)] [w_x^* (k_\perp\cdot n_\alpha) - (w_\perp^*\cdot n_\alpha) k_x] \\
	&- \left[\frac{3}{2} + j_1(k)\right] (k\cdot w^*) \sin\alpha \\
	&- \frac{1}{2} (k\cdot w_\perp^*) \cos\alpha \biggr\} -\frac{1}{2} j_1(k) \left( \sin\alpha + \frac{w_x^*w_y^*}{2D_\varphi} \right),
\end{aligned}
\end{equation}
\begin{equation}
\begin{aligned}
	M_{33} &= \frac{iv_0^*}{8D_\varphi} \biggl\{ [1+j_1(k)] [w_y^* (k_\perp\cdot n_\alpha) - (w_\perp^*\cdot n_\alpha) k_y] \\
	&+ \left[\frac{1}{2} + j_1(k)\right] (k\cdot w^*) \cos\alpha \\
	&+ \frac{1}{2} (k\cdot w_\perp^*) \sin\alpha \biggr\} - D_\varphi - \frac{(v_0^*)^2}{16D_\varphi} \vert k\vert^2 \\
	&+ \frac{1}{2} j_1(k) \left( \cos\alpha - \frac{(w_y^*)^2}{2D_\varphi} \right) - \frac{\Vert w^*\Vert^2}{8D_\varphi},
\end{aligned}
\end{equation}
where $n_\alpha=(\cos\alpha,\sin\alpha)^T$ and $k_\perp=(-k_y,k_x)^T$.

The general form of the linearized matrix is not particularly informative and we need to instantiate the solutions to \EqRef{sm:hydrodynamic_equations} as well as impose various assumptions on the parameters in order to simplify the above expressions and draw any conclusions. Still, there are some useful observations from the form of the linearized dynamics. 
First, the phase lag $\alpha$ and the noise strength $D_\varphi$ appear nontrivially in most of the matrix coefficients, so they are definitely the parameters that determine the linear stability of the system. Second, the particle velocity $v_0$ and the interaction range $\varrho$ both appear only as multipliers of the wave vector $k$. Thus, they both influence the spatial scale of the perturbations. But they do so separately, i.e., $\varrho$ only appears inside $j_1(k)$. As a result, by rescaling $\tilde{k}=v_0^*k$, we have $j_1(k) = 2J_1\left(\frac{\varrho^*}{v_0^*} \vert\tilde{k}\vert\right) / \left(\frac{\varrho^*}{v_0^*} \vert\tilde{k}\vert\right)$, which signifies that the third independent parameter that is important for the stability of the solutions is $\varrho^*/v_0^*$ ratio. For this reason, the length scale of the patterns that result from the instability of the partially synchronized solution, scales proportionally to $\varrho^*/v_0^*$. However, since the radius of interaction is restricted to be in $[0,\frac{1}{2}]$ range, we will keep these two parameters separately in the subsequent derivations. Note also that in the view of the fact that the wave numbers are integers, $v_0^*$ and $\varrho^*$ would determine whether the instabilities are detected at all.

\subsection{The Uniform Solution}

The uniform solution, which signifies the disordered motion of particles, is

\begin{equation}
(\rho, w_x, w_y) = (1, 0, 0).
\end{equation}
The stability matrix $M$ for this solution simplifies to

\begin{widetext}
	\begin{equation}
	\begin{pmatrix}
	0 & iv_0^*k_x & iv_0^*k_y \\
	\frac{iv_0^*}{2}k_x & \frac{1}{2} j_1(k) \cos\alpha - D_\varphi - \frac{(v_0^*)^2}{16D_\varphi}\vert k\vert^2 & \frac{1}{2} j_1(k) \sin\alpha \\
	\frac{iv_0^*}{2}k_y & -\frac{1}{2} j_1(k) \sin\alpha & \frac{1}{2} j_1(k) \cos\alpha - D_\varphi - \frac{(v_0^*)^2}{16D_\varphi}\vert k\vert^2
	\end{pmatrix}
	\end{equation}
\end{widetext}

In the presence of a phase lag, the characteristic equation to the above matrix is a third degree polynomial and the explicit solutions of it are not particularly informative. In order to gain insight whether the long wavelength instability is possible for this solution, we expand the eigenvalues up to the second order both in the wave number $k_x$ around $k_x=0$ and in the diffusion constant around the order-disorder transition line $D_\varphi=\frac{1}{2}$. The resulting expansions read
\begin{equation}
\begin{aligned}
	\lambda_1&(\vert k\vert,D_\varphi) \approx -\frac{4\sqrt{3}}{9|\sin\alpha|} \left(D_\varphi - \frac{1}{2}\cos\alpha\right)^2, \\
	\lambda_{2,3}&(\vert k\vert,D_\varphi) \approx \pm\frac{i}{2}|\sin\alpha| - \frac{4}{3} \left(D_\varphi \!-\! \frac{1}{2}\cos\alpha \right) \\
	&-\! \frac{1}{2} \left( \frac{(v_0^*)^2}{4\cos\alpha} \!+\! \frac{(\varrho^*)^2\cos\alpha}{8} \!\mp\! \frac{i(v_0^*)^2}{|\sin\alpha|} \right) \vert k\vert^2 \\
	&\!+\! \frac{1}{|\sin\alpha|} \left(\frac{2\sqrt{3}}{9} \!\pm\! 6i \right) \left(D_\varphi \!-\! \frac{1}{2}\cos\alpha \right)^2.
\end{aligned}
\end{equation}
The first eigenvalue is negative for all the parameters and wave numbers. The other two eigenvalues are oscillating quantities for $\alpha\neq0$ and they are stable for $D_\varphi > \frac{1}{2}\cos\alpha$ which is the order-disorder transition line we have encountered earlier. As a result, along a transition line that characterizes the onset of the synchronized motion, there are no long wavelength instability mechanisms leading to the formation of a new behavior. The scanning of the regions farther from the transition line with the help of numerical methods shows that there are no other instabilities for this solution too.

\subsection{The Synchronized Solution in the Zero Phase Lag Case}

The solution that signifies the onset of collective motion may point to an arbitrary direction. Thus, we write it as $w^* = \Vert w^*\Vert e(\varphi)$, where as previous $e(\varphi) = (\cos\varphi,\sin\varphi)\in\mathbb{S}^1\subset\mathbb{R}^2$ is a unit vector in the direction of $\varphi\in\mathbb{T}$. Without loss of generality, we put $e(\varphi)=(1,0)$ henceforth.

\subsubsection{Longitudinal Perturbations}

We consider the longitudinal perturbations of the form $k=(k_x,0)^T$, $\delta \hat{w} = (\delta\hat{w}_x,\delta\hat{w}_y)^T$. The matrix coefficients thus read

\begin{equation}
	M_{11} = 0, M_{12} = iv_0^*k_x, M_{13} = 0,
\end{equation}
\begin{equation}
\begin{aligned}
	M_{21} &= \Vert w^*\Vert\left( \frac{1}{2} [1 + j_1(k)] \cos\alpha - 2j_1(k) D_\varphi \right) \\
	&+ \frac{iv_0^*}{2}k_x \left[1 - j_1(k) (\cos\alpha - 2D_\varphi) \cos\alpha\right],
\end{aligned}
\end{equation}
\begin{equation}
\begin{aligned}
	M_{22} &= -\frac{(v_0^*)^2}{16D_\varphi} k_x^2 - \left( \frac{1}{2} [1 + j_1(k)] \cos\alpha - 2j_1(k) D_\varphi \right) \\
	&+ \frac{iv_0^*}{8D_\varphi} k_x \left[\frac{1}{2} + j_1(k)\right] \Vert w\Vert \cos\alpha,
\end{aligned}
\end{equation}
\begin{equation}
\begin{aligned}
	M_{23} &= \frac{1}{2} j_1(k) \sin\alpha \\
	&+ \frac{iv_0^*}{8D_\varphi} k_x \left[\frac{3}{2} + j_1(k)\right] \Vert w\Vert \sin\alpha,
\end{aligned}
\end{equation}
\begin{equation}
\begin{aligned}
	M_{31} &= -\frac{1}{2}[1 - j_1(k)] \Vert w^*\Vert \sin\alpha \\
	&+ \frac{iv_0^*}{2} k_x j_1(k) (\cos\alpha - 2D_\varphi) \sin\alpha,
\end{aligned}
\end{equation}
\begin{equation}
	M_{32} = -M_{23},
\end{equation}
\begin{equation}
\begin{aligned}
	M_{33} &= -\frac{(v_0^*)^2}{16D_\varphi} k_x^2 + \frac{1}{2} (j_1(k) - 1) \cos\alpha \\
	&+ \frac{iv_0^*}{8D_\varphi} k_x \left[ \frac{1}{2} + j_1(k) \right] \Vert w^*\Vert \cos\alpha.
\end{aligned}
\end{equation}

The general form of the dispersion relations given by solving the third order polynomial of the characteristic equation is a complex and uninformative expression. In order to get the insights about the behavior of the perturbations near the order-disorder transition, we consider some simplified cases \cite{mishra:pre}.

We first analyze the Vicsek model, which is obtained by setting $\alpha=0$. We see that the dynamics of the perturbations towards the marginal density function decouples from the one towards the momenta, and we have
\begin{equation}
\begin{aligned}
	\lambda_1(k) &= M_{33},\\
	\lambda_{2,3}(k) &= \frac{1}{2}\left( M_{22} \pm \sqrt{M_{22}^2 + 4M_{12}M_{21}} \right).
\end{aligned}
\end{equation}
In the long wavelength limit $k_x\rightarrow0+0$, two of the above dispersion relations tend to zero and the third one is always negative. The first one up to the fourth order in $k_x$ reads
\begin{equation}
\begin{aligned}
	\lambda_1(k_x) &= \frac{3iv_0^*}{16D_\varphi}\Vert w^*\Vert k_x - \left(\frac{(v_0^*)^2}{D_\varphi} + (\varrho^*)^2\right) \frac{k_x^2}{16} \\
	&- \frac{iv_0^*(\varrho^*)^2}{64D_\varphi}\Vert w^*\Vert k_x^3 + \frac{(\varrho^*)^4}{384} k_x^4 + \mathcal{O}(k_x^5).
\end{aligned}
\end{equation}
The real part of the dispersion relation is negative for small $k_x$. One could find the condition for the instability as $k_x^2>\frac{24}{(\varrho^*)^4} \left(\frac{(v_0^*)^2}{D_\varphi}+(\varrho^*)^2\right)$. This condition is satisfied provided $k_x\gg0$, which is however out of the validity of the approximation. The expansion of the other hydrodynamic mode reads
\begin{equation}
\begin{aligned}
	\lambda_2(k_x) &= iv_0^*\Vert w^*\Vert k_x - \frac{(v_0^*)^2}{2}\Biggl( \frac{9}{128D_\varphi} + \frac{1}{1-2D_\varphi} \\
	&- 1 - \frac{D_\varphi}{2} \left(4 - \frac{3}{8D_\varphi}\right)^2 \Biggr) k_x^2 + \mathcal{O}(k_x^3).
\end{aligned}
\end{equation}
One can show that the real part of this dispersion relation is always negative for $D_\varphi<1/2$. Thus, the second dispersion relation is always stable. As a result, for the Vicsek model with $\alpha=0$, the synchronized homogeneous solution is always stable against long wavelength perturbations. 

This result appears as a contradiction to the one obtained for the classical Vicsek model that was shown to exhibit longitudinal long wavelength instability leading to the emergence of the traveling waves. The explanation for this is the type of the continuum limit we derived and the subsequent requirement to have a normalization in the alignment term. In the limit $\frac{N}{L^2} = \text{const}$ for $N,L\rightarrow\infty$, they do not use the normalization by the number of particles in time continuous modifications for the Vicsek model. This is not required during the derivation of the continuum limit because of the assumption $\frac{N}{L^2} = \text{const}$, which allows to handle the alignment term. In our case, we do not use such an assumption. Therefore, in order to keep the alignment term finite in the transition $N\rightarrow\infty$, we introduce the normalization by the number of particles.

\subsubsection{Transversal Perturbations}

We consider the transversal perturbations of the form $k=(0,k_y)^T$, $\delta \hat{w} = (0,\delta\hat{w}_y)$, i.e., orthogonal to the direction of collective motion. The matrix coefficients the read

\begin{equation}
	M_{11} = 0, M_{12} = 0, M_{13} = iv_0^*k_y,
\end{equation}
\begin{equation}
\begin{aligned}
	M_{21} &= \Vert w^*\Vert \left( \frac{1}{2} [1 + j_1(k)] \cos\alpha - 2j_1(k)D_\varphi \right) \\
	&+ \frac{iv_0^*}{2} k_y j_1(k)(\cos\alpha - 2D_\varphi) \sin\alpha,
\end{aligned}
\end{equation}
\begin{equation}
\begin{aligned}
	M_{22} &= - \frac{(v_0^*)^2}{16D_\varphi} k_y^2 - \left( \frac{1}{2} (j_1(k) + 1) \cos\alpha - 2j_1(k)D_\varphi \right) \\
	&- \frac{iv_0^*}{8D_\varphi} k_y \left[\frac{1}{2} + j_1(k)\right] \Vert w^*\Vert \sin\alpha,
\end{aligned}
\end{equation}
\begin{equation}
\begin{aligned}
	M_{23} &= \frac{1}{2}j_1(k) \sin\alpha \\
	&+ \frac{iv_0^*}{8D_\varphi} k_y \left[\frac{3}{2} + j_1(k)\right] \Vert w^*\Vert \cos\alpha,
\end{aligned}
\end{equation}
\begin{equation}
\begin{aligned}
	M_{31} &= \frac{iv_0^*}{2} k_y - \frac{1}{2}[1 - j_1(k)] \Vert w^*\Vert \sin\alpha \\
	&+ \frac{iv_0^*}{2} k_y j_1(k) (\cos\alpha - 2D_\varphi) \cos\alpha,
\end{aligned}
\end{equation}
\begin{equation}
	M_{32} = -M_{23},
\end{equation}
\begin{equation}
\begin{aligned}
	M_{33} &= - \frac{(v_0^*)^2}{16D_\varphi} k_y^2 + \frac{1}{2} [j_1(k) - 1] \cos\alpha \\
	&- \frac{iv_0^*}{8D_\varphi} k_y \left[\frac{1}{2} + j_1(k)\right] \Vert w^*\Vert \sin\alpha.
\end{aligned}
\end{equation}

First, we consider the simplified case of the zero phase lag $\alpha = 0$, which is again equivalent to the linear Vicsek regime.

The two dispersion relations are then given by
\begin{equation}
	\lambda_{1,2}(k_y) = \frac{1}{2}\left( M_{33} \pm \sqrt{M_{33}^2 + 4M_{13}M_{31}} \right).
\end{equation}
Expanding them to the second order in $k_y$ around $k_y=0$ reveals
\begin{equation}
\begin{aligned}
	\lambda_{1,2}(k_y) &= \pm iv_0^*\sqrt{1-D_\varphi} k_y \\
	&- \left( \frac{(v_0^*)^2}{D_\varphi} + (\varrho^*)^2 \right) \frac{k_y^2}{32} + \mathcal{O}(k_y^3).
\end{aligned}
\end{equation}
We see that their real part is always negative. Thus, the solution is also stable against transversal perturbations in the linear Vicsek regime $\alpha=0$.

\subsection{The Traveling Wave Solution}

As it was the case for the kinetic theory, the stability analysis of the traveling wave solutions cannot be carried out using \EqRef{sm:hydrodynamic_equations} because it has been developed under the assumption of the stationarity of the solutions. Therefore, we need to rederive the corresponding hydrodynamic equations for the traveling wave solutions anew, starting from the Fourier modes \EqRef{sm:fourier_mode_pde_wrt_phase_traveling_wave}. Since the most of the derivations are straightforward generalizations of the ones from the previous section, we provide only the key steps here.

The assumptions made to obtain the closure relation \EqRef{sm:closure_relation} are the same as for the stationary solutions, except that we additionally assume that the spatial variation of the nematic order field is negligible $\nabla h_{\pm2}\approx0$. This is appropriate for sufficiently high diffusion levels, which we have also assumed previously in the context of the hydrodynamic theory. The hydrodynamic equations describing the evolution of the marginal density function $\rho=\rho(r,t)$ and the momentum field $w=w(r,t)$ read
\begin{equation}
\label{sm:hydrodynamic_equations_traveling_wave}
\begin{aligned}
\partial_t \rho &= -v_0 \nabla\cdot w + v(r\times\nabla)\rho, \\
\partial_t w &= -\frac{v_0}{2} \nabla\rho - C_1 w + v(r\times\nabla)w + \frac{v_0^2}{8} C_2 \Delta w \\
&+ \frac{\rho}{2} Q_{-\alpha} W + \frac{1}{4} C_2 \Biggl\{ \frac{v_0}{2} Q_\alpha \left[ (W\cdot\nabla)w + (W_\perp\cdot\nabla)w_\perp \right] \\
&- w\Vert W\Vert^2 + v_0 Q_{-\alpha} \Bigl[ \nabla(w\cdot W) - (W\cdot\nabla)w \\
&- (\nabla\cdot W)w - W(\nabla\cdot w) - (w\cdot\nabla)W \Bigr] \Biggr\},
\end{aligned}
\end{equation}
where the matrix $Q_\alpha = \begin{pmatrix} \cos\alpha & -\sin\alpha \\ \sin\alpha & \cos\alpha \end{pmatrix}$ represents anticlockwise rotation by $\alpha$ radians; the matrix $C_1 = \begin{pmatrix} D_\varphi & -v \\ v & D_\varphi \end{pmatrix}$ arises due to the coupling between the longitudinal and transversal length scales; the matrix $C_2 = \dfrac{1}{4D_\varphi^2 + v^2} \begin{pmatrix} 2D_\varphi & v \\ -v & 2D_\varphi \end{pmatrix}$ arises due to the coupling between the first and second Fourier modes.

The system has two stationary spatially homogeneous solutions. The first one $(\rho,w) = (1,0,0)$ represents a spatially homogeneous disordered state of the system. The second one, which represents the partially synchronized flocking, is $(\rho,w) = (1, \Vert w^*\Vert\cos\varphi_0, \Vert w^*\Vert\sin\varphi_0)$, where the degree of polarization is
\begin{equation}
\label{sm:solution_hydrodynamic_polarization_traveling_wave}
	\Vert w^*\Vert = \sqrt{\frac{1}{D_\varphi}(4D_\varphi^2+v^2)(\cos\alpha - 2D_\varphi)}
\end{equation}
and $\varphi_0\in\mathbb{T}$ is an arbitrary direction subject to initial conditions. The group velocity $v$ is not a parameter of choice here but implicitly depends on other microscopic parameters of the Langevin dynamics (see the main text). Generally, it can be estimated from the self-consistent system of equations \EqRef{sm:order_parameter_self_consistent_real}. But we have shown that near the order-disorder transition line the group velocity is equal $v=-\frac{1}{2}\sin\alpha$. Thus, the degree of polarization \EqRef{sm:solution_hydrodynamic_polarization_traveling_wave} next to that transition line is
\begin{equation}
	\Vert w^*\Vert = \sqrt{\left( 4D_\varphi + \frac{\sin^2\alpha}{4D_\varphi} \right)(\cos\alpha - 2D_\varphi)},
\end{equation}
which agrees well with the result shown in Fig.~2(a) of the main text.

Next, we are going to test the solutions on the matter of stability. We transform \EqRef{sm:hydrodynamic_equations_traveling_wave} into the Fourier space with respect to the spatial variables. Most of the terms are transformed as it was described in the previous section, except for the following term arising after the application of the ansatz \EqRef{sm:helical_ansatz}:
\begin{equation}
\begin{aligned}
	\mathcal{F}&\{ v(r\times\nabla) \rho \}(k,t) = iv\frac{2\pi}{L} \mathcal{F}\left\{ (k\times r) \rho(r,t) \right\}(k,t) \\
	&= iv\pi(k_x-k_y) \hat{\rho}(k_x,k_y,t) \\
	&+ \sum_{\substack{q_y\in\mathbb{Z} \\ k_y\neq q_y}} \frac{v k_x}{k_y-q_y}\hat{\rho}(k_x,q_y,t) - \sum_{\substack{q_x\in\mathbb{Z} \\ k_x\neq q_x}} \frac{v k_y}{k_x-q_x}\hat{\rho}(q_x,k_y,t).
\end{aligned}
\end{equation}
The corresponding term of the momentum equation is treated similarly.

The appearance of the couplings to the function values at other wave vectors except for $k$ hinders the subsequent linear stability analysis we have been developing so far. We will not be able to represent the linearized dynamics of the perturbations using the stability matrix $M=(M_{n,m})_{n,m=1,2,3}$ because the time dynamics of $\hat{\rho}(k)$ and $\hat{w}(k)$ is not a closed system anymore. Theoretically, we could map three Fourier indices $n$, $k_x$, and $k_y$ into one index and write down the linearized dynamics of all the perturbations with respect to $n$, $k_x$, and $k_y$ together. By doing so, first, we would end up with an infinite hierarchy equations again, which we wanted to circumvent on the first place by using the hydrodynamic theory. Second, the solution of the eigenvalue problem would lose the spatial dependence and we would not be able to obtain the results in the form of the dispersion relations. The hydrodynamic equation for the marginal density function would then become
\begin{equation}
\begin{aligned}
	\partial_t \hat{\rho}(k) &= iv_0\frac{2\pi}{L}[k\cdot\hat{w}(k)] + iv\pi(k_x-k_y) \hat{\rho}(k_x,k_y) \\
	&+ \sum_{\substack{q_y\in\mathbb{Z} \\ k_y\neq q_y}} \frac{v k_x}{k_y-q_y}\hat{\rho}(k_x,q_y) - \sum_{\substack{q_x\in\mathbb{Z} \\ k_x\neq q_x}} \frac{v k_y}{k_x-q_x}\hat{\rho}(q_x,k_y).
\end{aligned}
\end{equation}
We see that the spatial scale is influenced by $v_0$. If we introduce the change of variables $k'=v_0k$, $\hat{\rho}'(k',t)=\hat{\rho}(k,t)$, $\hat{w}'(k',t)=\hat{w}(k,t)$, we rewrite the equation for the marginal density function as
\begin{equation}
\begin{aligned}
	\partial_t \hat{\rho}'(k') &= i\frac{2\pi}{L}[k'\cdot\hat{w}'(k')] + \frac{iv\pi}{v_0}(k_x'-k_y') \hat{\rho}'(k_x',k_y') \\
	&+ \sum_{\substack{q_y\in\mathbb{Z} \\ k_y'\neq v_0q_y}} \frac{v k_x'}{k_y'-v_0q_y}\hat{\rho}'(k_x',v_0q_y) \\
	&- \sum_{\substack{q_x\in\mathbb{Z} \\ k_x'\neq v_0q_x}} \frac{v k_y'}{k_x'-v_0q_x}\hat{\rho}'(v_0q_x,k_y').
\end{aligned}
\end{equation}
If we restrict ourselves only to small values of the particle velocity $v_0\ll1$, we see that the first term of the Fourier transform of $v(r\times\nabla)\rho$ would make the major impact. Therefore, we assume that in the limit of small $v_0$, that Fourier transform is approximated by
\begin{equation}
\label{sm:independence_of_wavevectors_approximation}
	\mathcal{F}\{ v(r\times\nabla) \rho \}(k,t) \approx iv\pi(k_x-k_y) \hat{\rho}(k_x,k_y,t)
\end{equation}
and the respective Fourier transform for the momentum field is approximated similarly. In the rest of the discussion, we follow this assumption.

Finally, the hydrodynamic equations \EqRef{sm:hydrodynamic_equations_traveling_wave} in the Fourier space with respect to the spatial variables read
\begin{widetext}
	\begin{equation}
	\begin{aligned}
	\partial_t &\hat{\rho}(k) = iv_0^* [k\cdot\hat{w}(k)] + iv\pi(k_x-k_y) \hat{\rho}(k_x,k_y), \\
	\partial_t &\hat{w}(k) = \frac{iv_0^*} {2} k\hat{\rho}(k) - C_1 \hat{w}(k) - \frac{(v_0^*)^2}{8} C_2 \vert k\vert^2 \hat{w}(k) + iv\pi(k_x-k_y) \hat{w}(k_x,k_y) \\
	&+ \sum_{q\in\mathbb{Z}^2}\Biggl\{ \frac{\hat{\rho}(q)}{2} Q_{-\alpha} K_1(k-q) - \frac{1}{4}C_2 \biggl[ \hat{w}(q)K_2(k-q) + \frac{iv_0^*}{2} Q_\alpha \Bigl\{\hat{w}(q)[q\cdot K_1(k-q)] + \hat{w}_\perp(q) [q\cdot K_{1,\perp}(k-q)]\Bigr\} \\
	&+ iv_0^* Q_{-\alpha} \Bigl\{k[\hat{w}(q)\cdot K_1(k-q)] - \hat{w}(q)[k\cdot K_1(k-q)] - [k\cdot\hat{w}(q)]K_1(k-q)\Bigr\} \biggr]\Biggr\},
	\end{aligned}
	\end{equation}
\end{widetext}
where $K_{1,\perp}=(-K_{1,y},K_{1,x})^T$ and we denote $v_0^* = \frac{2\pi}{L} v_0$ and $\varrho^* = \frac{2\pi}{L} \varrho$ as previous. Note that we suppressed the explicit time dependence of $\hat{\rho}$, $\hat{w}$, $K_1$, and $K_2$ for compactness.

If we consider the infinitesimal deviations from a stationary (here, in a moving reference frame) spatially homogeneous solution as

\begin{equation}
\begin{aligned}
	\delta \hat{\rho}(k,t) &= \hat{\rho}(k,t) - \hat{\rho}^*(k), \\
	\delta \hat{w}(k,t) &= \hat{w}(k,t) - \hat{w}^*(k),
\end{aligned}
\end{equation}
their linearized dynamics read
\begin{widetext}
	\begin{equation}
	\label{sm:linearized_equations_for_perturbations_of_hydrodynamaic_variables}
	\begin{aligned}
	\partial_t \delta\hat{\rho}&(k) = iv\pi(k_x-k_y) \delta\hat{\rho}(k) + iv_0^* [k\cdot\delta\hat{w}(k)], \\
	\partial_t \delta\hat{w}&(k) = \biggl\{ \frac{iv_0^*}{2} k + \frac{1}{2}[1 - j_1(k)] Q_{-\alpha}w^* + \frac{1}{4}j_1(k) C_2 \left[ 2\Vert w^* \Vert^2 w^* + iv_0^* Q_{-\alpha} \left( \Vert w^* \Vert^2 k - 2(k\cdot w^*)w^* \right) \right] \biggr\} \delta\hat{\rho}(k) \\
	&+ \biggl\{ -C_1 - \frac{(v_0^*)^2}{8} C_2 \vert k\vert^2 + \frac{1}{2}j_1(k) Q_{-\alpha} + iv\pi(k_x-k_y)I - \frac{1}{4} C_2 \biggl[ \Vert w^*\Vert^2I + 2j_1(k) (w^*\otimes w^*) \\
	&+ \frac{iv_0^*}{2} Q_\alpha \left[ (k\cdot w^*)I + (k\cdot w_\perp^*) Q_{\frac{\pi}{2}} \right] + iv_0^* [1+j_1(k)] Q_{-\alpha} \left[ (k\otimes w^*) - (w^*\otimes k) - (k\cdot w^*)I \right] \biggr] \biggr\} \delta\hat{w}(k), \\
	\end{aligned}
	\end{equation}
\end{widetext}
where $\otimes$ denotes the outer product, $I$ is the identity matrix, $Q_{\frac{\pi}{2}} = \begin{pmatrix}0&-1\\1&0\end{pmatrix}$, and $j_1(k) = 2J_1\left(\varrho^* \vert k\vert\right) / \left(\varrho^* \vert k\vert\right)$.

Since the direction of collective motion may be arbitrary, we put $\varphi_0=0$ without the loss of generality. To solve the eigenvalue problem for the linearized dynamics, we first need to rewrite these equations in the matrix form:
\begin{equation}
	\partial_t \begin{pmatrix} \delta\hat{\rho} \\ \delta\hat{w}_x \\ \delta\hat{w}_y \end{pmatrix} = 
	M \begin{pmatrix} \delta\hat{\rho} \\ \delta\hat{w}_x \\ \delta\hat{w}_y \end{pmatrix}.
\end{equation}
The matrix coefficients are found to be
\begin{equation}
	M_{11} = iv\pi (k_x-k_y),\quad M_{12} = iv_0^* k_x,\quad M_{13} = iv_0^* k_y,
\end{equation}
\begin{equation}
\begin{aligned}
	M_{21} &= \frac{iv_0^*}{2} k_x + \frac{1}{2}[1-j_1(k)]\Vert w^*\Vert \cos\alpha \\
	&+ (\cos\alpha - 2D_\varphi)j_1(k)\Vert w^*\Vert \\
	&+ \frac{iv_0^*}{4D_\varphi} j_1(k)(\cos\alpha - 2D_\varphi) \bigl[k_x(2D_\varphi\cos\alpha - v\sin\alpha) \\
	&\qquad+ k_y(2D_\varphi\sin\alpha + v\cos\alpha)\bigr] \\
	&- \frac{iv_0^*}{2D_\varphi} (\cos\alpha - 2D_\varphi) j_1(k)k_x (2D_\varphi\cos\alpha - v\sin\alpha),
\end{aligned}
\end{equation}
\begin{equation}
\begin{aligned}
	M&_{22} = -D_\varphi - \frac{(v_0^*)^2}{4(4D_\varphi^2+v^2)} \vert k\vert^2 D_\varphi \\
	&+ \frac{1}{2}j_1(k)\cos\alpha + iv\pi(k_x-k_y) \\
	&- \frac{\cos\alpha-2D_\varphi}{2} - \frac{j_1(k)}{2(4D_\varphi^2+v^2)} (2D_\varphi w_x^2 + vw_xw_y) \\
	&- \frac{iv_0^*}{8(4D_\varphi^2+v^2)} \bigl[(2D_\varphi\cos\alpha + v\sin\alpha) (k\cdot w^*) \\
	&\qquad+ (-2D_\varphi\sin\alpha + v\cos\alpha) (k\cdot w_\perp^*)\bigr] \\
	&+ \frac{iv_0^* [1+j_1(k)]}{4(4D_\varphi^2+v^2)} \bigl[(2D_\varphi\cos\alpha - v\sin\alpha) (k\cdot w^*) \\
	&\qquad- (2D_\varphi\sin\alpha + v\cos\alpha) (k\cdot w_\perp^*)\bigr],
\end{aligned}
\end{equation}
\begin{equation}
\begin{aligned}
	M&_{23} = v - \frac{(v_0^*)^2}{8(4D_\varphi^2+v^2)} \vert k\vert^2 v + \frac{1}{2}j_1(k)\sin\alpha \\
	&- \frac{\cos\alpha - 2D_\varphi}{4D_\varphi}v - \frac{j_1(k)}{2(4D_\varphi^2+v^2)} (2D_\varphi w_xw_y + vw_y^2) \\
	&+ \frac{iv_0^*}{8(4D_\varphi^2+v^2)} \bigl[(2D_\varphi\cos\alpha + v\sin\alpha) (k\cdot w_\perp^*) \\
	&\qquad- (-2D_\varphi\sin\alpha + v\cos\alpha) (k\cdot w^*)\bigr] \\
	&+ \frac{iv_0^* [1+j_1(k)]}{4(4D_\varphi^2+v^2)} \bigl[(2D_\varphi\cos\alpha - v\sin\alpha) (k\cdot w_\perp^*) \\
	&\qquad+ (2D_\varphi\sin\alpha + v\cos\alpha) (k\cdot w^*)\bigr],
\end{aligned}
\end{equation}
\begin{equation}
\begin{aligned}
	M_{31} &= \frac{iv_0^*}{2} k_y - \frac{1}{2}[1-j_1(k)]\Vert w^*\Vert \sin\alpha \\
	&- \frac{\cos\alpha-2D_\varphi}{2D_\varphi}vj_1(k)\Vert w^*\Vert \\
	&- \frac{iv_0^*}{4D_\varphi} j_1(k)(\cos\alpha - 2D_\varphi) \bigl[k_x(v\cos\alpha + 2D_\varphi\sin\alpha) \\
	&\qquad+ k_y(v\sin\alpha - 2D_\varphi\cos\alpha)\bigr] \\
	&+ \frac{iv_0^*}{2D_\varphi} (\cos\alpha - 2D_\varphi) j_1(k)k_x (v\cos\alpha + 2D_\varphi\sin\alpha),
\end{aligned}
\end{equation}
\begin{equation}
\begin{aligned}
	M&_{32} = -v + \frac{(v_0^*)^2}{8(4D_\varphi^2+v^2)} \vert k\vert^2 v - \frac{1}{2}j_1(k)\sin\alpha \\
	&+ \frac{\cos\alpha - 2D_\varphi}{4D_\varphi}v - \frac{j_1(k)}{2(4D_\varphi^2+v^2)} (-vw_x^2 + 2D_\varphi w_xw_y) \\
	&- \frac{iv_0^*}{8(4D_\varphi^2+v^2)} \bigl[(-v\cos\alpha + 2D_\varphi\sin\alpha) (k\cdot w^*) \\
	&\qquad+ (v\sin\alpha + 2D_\varphi\cos\alpha) (k\cdot w_\perp^*)\bigr] \\
	&- \frac{iv_0^* [1+j_1(k)]}{4(4D_\varphi^2+v^2)} \bigl[(v\cos\alpha + 2D_\varphi\sin\alpha) (k\cdot w^*) \\
	&\qquad+ (-v\sin\alpha + 2D_\varphi\cos\alpha) (k\cdot w_\perp^*)\bigr],
\end{aligned}
\end{equation}
\begin{equation}
\begin{aligned}
	M&_{33} = -D_\varphi - \frac{(v_0^*)^2}{4(4D_\varphi^2+v^2)} \vert k\vert^2 D_\varphi + \frac{1}{2}j_1(k)\cos\alpha \\
	&+ iv\pi(k_x-k_y) - \frac{\cos\alpha - 2D_\varphi}{2} \\
	&- \frac{j_1(k)}{2(4D_\varphi^2+v^2)} (-vw_xw_y + 2D_\varphi w_y^2) \\
	&+ \frac{iv_0^*}{8(4D_\varphi^2+v^2)} \bigl[(-v\cos\alpha + 2D_\varphi\sin\alpha) (k\cdot w_\perp^*) \\
	&\qquad- (v\sin\alpha + 2D_\varphi\cos\alpha) (k\cdot w^*)] \\
	&- \frac{iv_0^* [1+j_1(k)]}{4(4D_\varphi^2+v^2)} \bigl[(v\cos\alpha + 2D_\varphi\sin\alpha) (k\cdot w_\perp^*) \\
	&\qquad- (-v\sin\alpha + 2D_\varphi\cos\alpha) (k\cdot w^*)\bigr].
\end{aligned}
\end{equation}

For the subsequent analysis, we consider two simplified cases. Namely, we investigate the longitudinal and transversal perturbations with respect to the direction of collective motion.

\subsubsection{Longitudinal Perturbations}

We consider the longitudinal perturbations of the form $k=(k_x,0)^T$, $\delta \hat{w} = (\delta\hat{w}_x,0)^T$ for the flow with the momentum field $w^*=(w_x^*,0)$. The eigenvalues of the resulting eigenvalue problem are
\begin{equation}
	\lambda_{\pm} = \frac{M_{11} + M_{22} \pm \sqrt{D}}{2},
\end{equation}
where the discriminant is $D = (M_{11} + M_{22})^2 - 4(M_{11}M_{22} - M_{12}M_{21})$ and the required coefficients of the stability matrix read
\begin{equation}
	M_{11} = iv\pi k_x,\quad M_{12} = iv_0^* k_x,
\end{equation}
\begin{equation}
\begin{aligned}
	M_{21} &= \frac{iv_0^*}{2} k_x + \frac{1}{2}(1-j_1) \Vert w^*\Vert \cos\alpha \\
	&+ \frac{\cos\alpha-2D_\varphi}{4D_\varphi} j_1(k) \Bigl[ 4D_\varphi \Vert w^*\Vert \\
	&- iv_0^* k_x (2D_\varphi\cos\alpha - v\sin\alpha) \Bigr],
\end{aligned}
\end{equation}
\begin{equation}
\begin{aligned}
	M_{22} &= -D_\varphi - \frac{(v_0^*)^2 k_x^2 D_\varphi}{4(4D_\varphi^2+v^2)} + \frac{1}{2}j_1(k)\cos\alpha \\
	&- \left[ \frac{1}{2} + j_1(k) \right](\cos\alpha - 2D_\varphi) + iv\pi k_x \\
	&- \frac{iv_0^* k_x\Vert w^*\Vert}{4(4D_\varphi^2+v^2)} \biggl[ \frac{1}{2} (2D_\varphi\cos\alpha + v\sin\alpha) \\
	&- [1+j_1(k)] (2D_\varphi \cos\alpha - v\sin\alpha) \biggr].
\end{aligned}
\end{equation}

One can show that the eigenvalue $\lambda_+$ is a hydrodynamic mode since it becomes zero in the limit of small wave numbers, while the other eigenvalue $\lambda_-=-(\cos\alpha-2D_\varphi)$ is always negative since the condition $D_\varphi<\frac{1}{2}\cos\alpha$ is the existence condition for the given solution. The presence of a hydrodynamic mode might lead to the long wave number instability of the traveling wave solution. It is what we investigate in the following.

\begin{figure}[b]
	\includegraphics[width=0.5\textwidth]{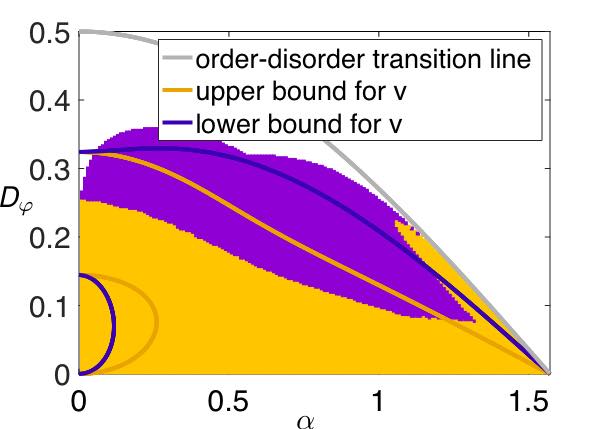}
	\caption
	{
		\label{fig:hydrodynamic_stability_analysis} Instability regions given by the analysis of the hydrodynamic equations \EqRef{sm:hydrodynamic_equations_traveling_wave}. The gray line denotes the order-disorder transition line $D_\varphi=\frac{1}{2}\cos\alpha$. The colored lines are found analytically by restricting the analysis to longitudinal perturbations solely. They enclose a line, below which long wavelength perturbations acting on the traveling wave solution appear. The blue (yellow) line is defined by \EqRef{sm:condition_for_long_wave_length_perturbations} with $v=-\sin\alpha$ ($v=-\frac{1}{2}\sin\alpha$). 		
		The violet and yellow regions are obtained by numerically solving an eigenvalue problem for \EqRef{sm:linearized_equations_for_perturbations_of_hydrodynamaic_variables}. The violet region includes parameters for which at most one Fourier mode becomes unstable for each wave vector. The yellow region includes parameters for which there are at most two Fourier modes that become unstable for each wave vector.
		Other parameters are $\tilde{\varrho}=1$, $\varrho=0.01$, $v_0=0.01$.
	}
\end{figure}

Expanding the eigenvalue of the hydrodynamic mode to the second order in $k_x$ around $k_x=0$, we find
\begin{equation}
\begin{aligned}
	\lambda&_+(k_x) = i(v\pi + v_0^* \Vert w^*\Vert) k_x \\
	&+ \frac{(v_0^*)^2}{2} \Biggl[ \frac{1}{4D_\varphi}\left( 32D_\varphi^2 + D_\varphi\cos\alpha + 8v^2 + \frac{1}{2}v\sin\alpha \right) \\
	&- \frac{1}{\cos\alpha-2D_\varphi} \Biggr] k_x^2 + \mathcal{O}(k_x^3).
\end{aligned}
\end{equation}
If the long wave length perturbations act on the solution, it is signified by $\text{Re}\lambda_+(k_x)>0$. Since the group velocity $v$ enters the expression, we cannot draw conclusions about instabilities in the system as such because this parameter is not independent but implicitly depends upon the system parameters. However, we know from the analysis of self-consistent equations \EqRef{sm:order_parameter_self_consistent_real},\EqRef{sm:order_parameter_self_consistent_imaginary} that next to the order-disorder transition line the critical group velocity attained along that line is $v = -\frac{1}{2}\sin\alpha$.
Moreover, we know from the analysis of the particle model (see the main text) that the lower bound for the group velocity may be assumed $v=-\sin\alpha$, which is the rate of change of each particles' direction of motion in the case of complete phase synchronization when $D_\varphi\rightarrow0$. Knowing those two bounds, we could guess an approximate boundary of the parameter region, where spatial nonhomogeneities should occur (cf. Fig.~\ref{fig:hydrodynamic_stability_analysis}, blue and yellow lines).

The condition for the emergence of the long wave length perturbations in the longitudinal direction (with respect to the direction of collective motion) is given by
\begin{equation}
\label{sm:condition_for_long_wave_length_perturbations}
\begin{aligned}
	(\cos\alpha - 2D_\varphi)\Biggl( 32&D_\varphi^2 + D_\varphi\cos\alpha \\
	&+ 8v^2 + \frac{v}{2}\sin\alpha \Biggr) - 4D_\varphi > 0,
\end{aligned}
\end{equation}
where the group velocity is bounded by $-\sin\alpha < v < -\frac{1}{2}\sin\alpha$ (cf. Fig.~\ref{fig:hydrodynamic_stability_analysis}). Recall that the hydrodynamic equations, we are working with, are valid for sufficiently high diffusion levels, i.e., close to the order-disorder transition line. Thus, we see from Fig.~\ref{fig:hydrodynamic_stability_analysis} that the long wavelength perturbations are expected to be observed only for high enough values of $\alpha$. Moreover, we conclude that the long wavelength perturbations do not arise at the order-disorder transition line.

\begin{figure*}[]
	\includegraphics[width=1.0\textwidth]{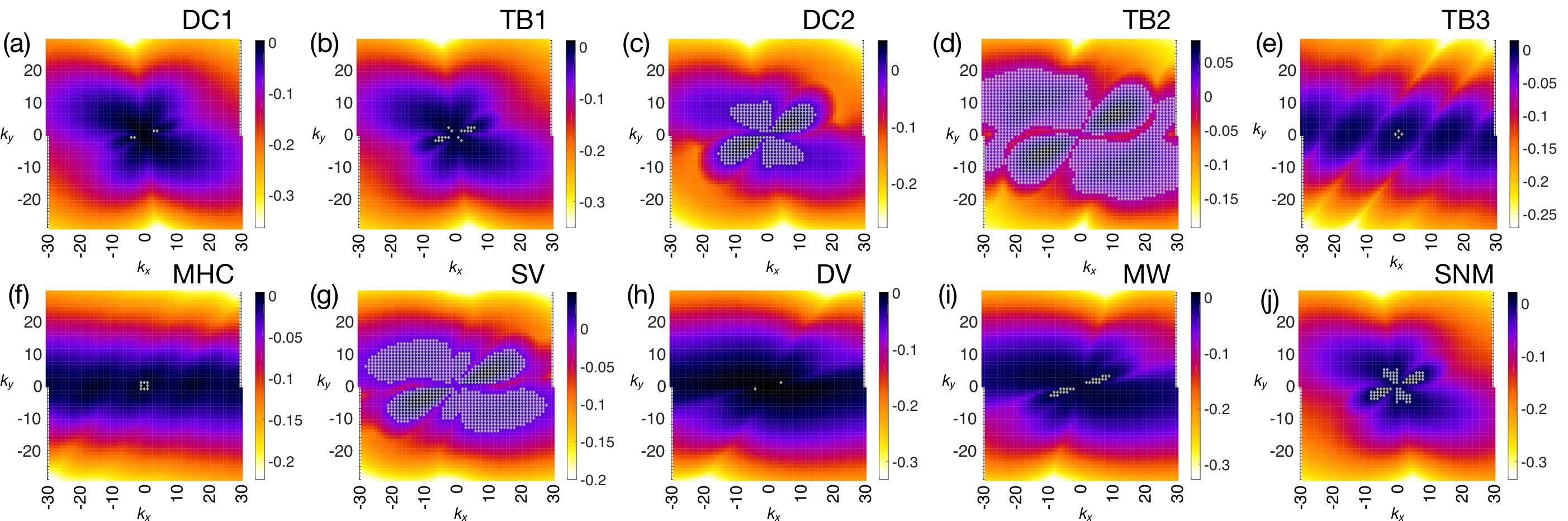}
	\caption
	{
		\label{fig:dispersion_relations} Dispersion relations $\lambda(k_x,k_y) = \max\limits_{n\in\mathbb{Z}} \lambda_n(k_x,k_y)$ obtained by solving an eigenvalue problem for \EqRef{sm:infinite_stability_matrix_for_traveling_wave}. Grey markers indicate wave vectors, at which there is a Fourier mode with a positive real part. Parameters: $N=5\cdot10^4$, $\tilde{\varrho}=1$, $v_0=0.01$, (a) $\varrho=0.01,\alpha=0.78,D_\varphi=0.2075$, (b) $\varrho=0.01,\alpha=0.9,D_\varphi=0.18$, (c) $\varrho=0.01,\alpha=1.3,D_\varphi=0.06$, (d) $\varrho=0.01,\alpha=1.45,D_\varphi=0.01$, (e) $\varrho=0.4,\alpha=1.45,D_\varphi=0.005$, (f) $\varrho=0.2,\alpha=1.36,D_\varphi=0.005$, (g) $\varrho=0.01,\alpha=1.3,D_\varphi=0.02$, (h) $\varrho=0.01,\alpha=1.0,D_\varphi=0.0375$, (i) $\varrho=0.01,\alpha=1.0,D_\varphi=0.0575$, and (j) $\varrho=0.01,\alpha=1.07,D_\varphi=0.145$.
	}
\end{figure*}

The last result tells us that there might be the following scenarios for the system behavior. First, at the order-disorder transition line, the traveling wave solution might be stable. Second, at that line, short wavelength perturbations might appear. Third, at that line, long wavelength perturbations transversal to the direction of collective motion might appear. We discard the first case because we know from the kinetic theory that the traveling wave solution is unstable for high enough $\alpha$ at the order-disorder transition line. The analytic confirmation of the existence of short wavelength perturbations seems to be unfeasible since at $D_\varphi=\frac{1}{2}\cos\alpha$, the magnitude of the momentum field, given by \EqRef{sm:solution_hydrodynamic_polarization_traveling_wave}, is proportional $\Vert w^*\Vert\propto D_\varphi^{1/2}$ and we cannot perform the respective expansion. Thus, we next look whether we could gain some insight about perturbations transversal to the direction of collective motion.

\begin{figure*}[]
	\includegraphics[width=1.0\textwidth]{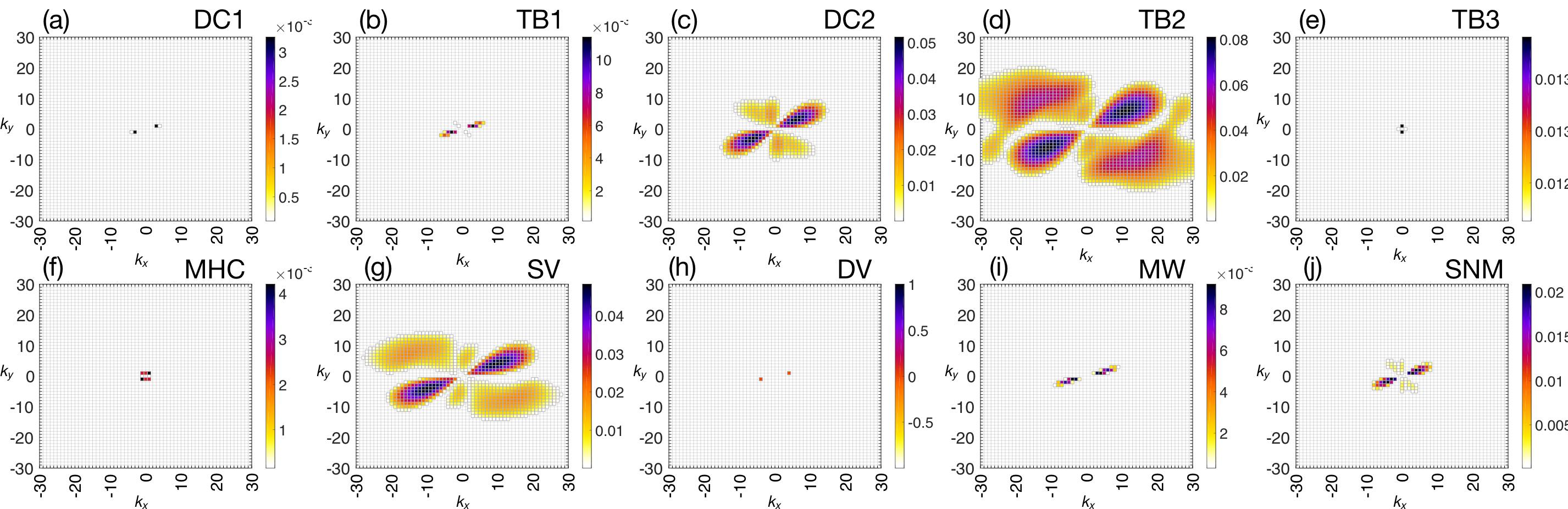}
	\caption
	{
		\label{fig:dispersion_relations_unstable} Positive real parts of each dispersion relation from Fig.~\ref{fig:dispersion_relations}. Parameters are the same as in Fig.~\ref{fig:dispersion_relations}.
	}
\end{figure*}

\subsubsection{Transversal Perturbations}

We consider the transversal perturbations of the form $k=(0,k_y)^T$, $\delta \hat{w} = (0,\delta\hat{w}_y)$, i.e., orthogonal to the direction of collective motion. The coefficients of the matrix read

\begin{equation}
	M_{11} = -iv\pi k_y,
\quad M_{13} = iv_0^* k_y,
\end{equation}

\begin{equation}
\begin{aligned}
	M_{31} &= \frac{i}{2}v_0^* k_y - \frac{1}{2}[1-j_1(k)]\Vert w^*\Vert\sin\alpha \\
	&+ \frac{\cos\alpha-2D_\varphi}{4D_\varphi} j_1(k) \Bigl[ -2v\Vert w^*\Vert \\
	&+ iv_0^* k_y (-v\sin\alpha + 2D_\varphi\cos\alpha) \Bigr],
\end{aligned}
\end{equation}

\begin{equation}
\begin{aligned}
	M_{33} &= -D_\varphi - \frac{(v_0^*)^2 k_y^2 D_\varphi}{4(4D_\varphi^2+v^2)} + \frac{1}{2}j_1(k)\cos\alpha \\
	&- \frac{1}{2}(\cos\alpha - 2D_\varphi) - iv\pi k_y \\
	&+ \frac{iv_0^* k_y\Vert w^*\Vert}{4(4D_\varphi^2+v^2)} \biggl[ \frac{1}{2} (-v\cos\alpha + 2D_\varphi\sin\alpha) \\
	&- [1+j_1(k)] (v\cos\alpha + 2D_\varphi\sin\alpha) \biggr].
\end{aligned}
\end{equation}

The restriction to consider only the perturbations transversal to the direction of collective motion reveals that the dispersion relations are then proportional to the square root of the wave number $\propto\sqrt{k_y}$ in the limit $k_y\rightarrow0$. This fact does not allow us to consider the expansion of the dispersion relations near small wave numbers.

\section{Solutions of Eigenvalue Problems for Kinetic Equations\label{sec:solutions_of_eigenvalue_problems}}

As found previously, close to the order-disorder transition line, we do not observe any instabilities if we restrict ourselves to consider only longitudinal perturbations. However, further away from that line, we have proved that longitudinal perturbations do arise. Because the insight on instability mechanisms of a traveling wave solution is quite limited analytically using the hydrodynamic equations \EqRef{sm:hydrodynamic_equations_traveling_wave}, we need to solve an eigenvalue problem for the complete system \EqRef{sm:linearized_equations_for_perturbations_of_hydrodynamaic_variables} numerically. The results of previous analytical studies as well as such a numerical integration can be found in Fig.~\ref{fig:hydrodynamic_stability_analysis}. According to the approximation \EqRef{sm:independence_of_wavevectors_approximation}, we must restrict ourselves to consider small particle velocities. We thus fix $v_0=0.01$ and assume $\tilde{\varrho}=1$ for simplicity. By solving the eigenvalue problem for \EqRef{sm:linearized_equations_for_perturbations_of_hydrodynamaic_variables} numerically, we obtain dispersion relations $\lambda_n(k_x,k_y)\in\mathbb{C}, n,k_x,k_y\in\mathbb{Z}$. We have considered wave vectors as $k_x,k_y\in[-30,30]$ without restricting their direction. As a result, in Fig.~\ref{fig:hydrodynamic_stability_analysis}, we see that there are indeed spatially nonhomogeneous perturbations acting on \EqRef{sm:solution_hydrodynamic_polarization_traveling_wave} close to the order-disorder transition line for $\alpha$ sufficiently high. We also see that close to that line, there are two unstable modes. Still, this is quite a limited insight on the structure of the phase diagram. Therefore, at this point, we proceed to the solution of an eigenvalue problem from the kinetic theory, to have a clearer picture of the phase diagram.

We solve an eigenvalue problem for \EqRef{sm:infinite_stability_matrix_for_traveling_wave} and the resulting phase diagrams are presented in the main text. As one might expect, the unstable wave vectors are not bound to be either in the longitudinal or in transversal directions with respect to the direction of collective motion. In fact, most of unstable wave vectors lie away from those axes as can be seen in Figs. \ref{fig:dispersion_relations} and \ref{fig:dispersion_relations_unstable} where we have gathered the most exemplary dispersion relations out of an instability region. The corresponding spatially nonhomogeneous particle dynamics are presented in the main text and in \ApRef{sec:spatially_nonhomogeneous_particle_dynamics}.

\bibliography{continuum_limit_dynamics_pre}

\begin{thebibliography}{61}%
\makeatletter
\providecommand \@ifxundefined [1]{%
 \@ifx{#1\undefined}
}%
\providecommand \@ifnum [1]{%
 \ifnum #1\expandafter \@firstoftwo
 \else \expandafter \@secondoftwo
 \fi
}%
\providecommand \@ifx [1]{%
 \ifx #1\expandafter \@firstoftwo
 \else \expandafter \@secondoftwo
 \fi
}%
\providecommand \natexlab [1]{#1}%
\providecommand \enquote  [1]{``#1''}%
\providecommand \bibnamefont  [1]{#1}%
\providecommand \bibfnamefont [1]{#1}%
\providecommand \citenamefont [1]{#1}%
\providecommand \href@noop [0]{\@secondoftwo}%
\providecommand \href [0]{\begingroup \@sanitize@url \@href}%
\providecommand \@href[1]{\@@startlink{#1}\@@href}%
\providecommand \@@href[1]{\endgroup#1\@@endlink}%
\providecommand \@sanitize@url [0]{\catcode `\\12\catcode `\$12\catcode
  `\&12\catcode `\#12\catcode `\^12\catcode `\_12\catcode `\%12\relax}%
\providecommand \@@startlink[1]{}%
\providecommand \@@endlink[0]{}%
\providecommand \url  [0]{\begingroup\@sanitize@url \@url }%
\providecommand \@url [1]{\endgroup\@href {#1}{\urlprefix }}%
\providecommand \urlprefix  [0]{URL }%
\providecommand \Eprint [0]{\href }%
\providecommand \doibase [0]{https://doi.org/}%
\providecommand \selectlanguage [0]{\@gobble}%
\providecommand \bibinfo  [0]{\@secondoftwo}%
\providecommand \bibfield  [0]{\@secondoftwo}%
\providecommand \translation [1]{[#1]}%
\providecommand \BibitemOpen [0]{}%
\providecommand \bibitemStop [0]{}%
\providecommand \bibitemNoStop [0]{.\EOS\space}%
\providecommand \EOS [0]{\spacefactor3000\relax}%
\providecommand \BibitemShut  [1]{\csname bibitem#1\endcsname}%
\let\auto@bib@innerbib\@empty
\bibitem [{\citenamefont {Vicsek}\ and\ \citenamefont
  {Zafeiris}(2012)}]{vicsek:phys_rep}%
  \BibitemOpen
  \bibfield  {author} {\bibinfo {author} {\bibfnamefont {T.}~\bibnamefont
  {Vicsek}}\ and\ \bibinfo {author} {\bibfnamefont {A.}~\bibnamefont
  {Zafeiris}},\ }\bibfield  {title} {\bibinfo {title} {Collective motion},\
  }\href {https://doi.org/http://dx.doi.org/10.1016/j.physrep.2012.03.004}
  {\bibfield  {journal} {\bibinfo  {journal} {Physics Reports}\ }\textbf
  {\bibinfo {volume} {517}},\ \bibinfo {pages} {71 } (\bibinfo {year}
  {2012})}\BibitemShut {NoStop}%
\bibitem [{\citenamefont {Vicsek}\ \emph {et~al.}(1995)\citenamefont {Vicsek},
  \citenamefont {Czir\'ok}, \citenamefont {Ben-Jacob}, \citenamefont {Cohen},\
  and\ \citenamefont {Shochet}}]{vicsek:prl}%
  \BibitemOpen
  \bibfield  {author} {\bibinfo {author} {\bibfnamefont {T.}~\bibnamefont
  {Vicsek}}, \bibinfo {author} {\bibfnamefont {A.}~\bibnamefont {Czir\'ok}},
  \bibinfo {author} {\bibfnamefont {E.}~\bibnamefont {Ben-Jacob}}, \bibinfo
  {author} {\bibfnamefont {I.}~\bibnamefont {Cohen}},\ and\ \bibinfo {author}
  {\bibfnamefont {O.}~\bibnamefont {Shochet}},\ }\bibfield  {title} {\bibinfo
  {title} {Novel type of phase transition in a system of self-driven
  particles},\ }\href {https://doi.org/10.1103/PhysRevLett.75.1226} {\bibfield
  {journal} {\bibinfo  {journal} {Phys. Rev. Lett.}\ }\textbf {\bibinfo
  {volume} {75}},\ \bibinfo {pages} {1226} (\bibinfo {year}
  {1995})}\BibitemShut {NoStop}%
\bibitem [{\citenamefont {Romanczuk}\ \emph {et~al.}(2012)\citenamefont
  {Romanczuk}, \citenamefont {B{\"a}r}, \citenamefont {Ebeling}, \citenamefont
  {Lindner},\ and\ \citenamefont {Schimansky-Geier}}]{romanczuk:epjst}%
  \BibitemOpen
  \bibfield  {author} {\bibinfo {author} {\bibfnamefont {P.}~\bibnamefont
  {Romanczuk}}, \bibinfo {author} {\bibfnamefont {M.}~\bibnamefont {B{\"a}r}},
  \bibinfo {author} {\bibfnamefont {W.}~\bibnamefont {Ebeling}}, \bibinfo
  {author} {\bibfnamefont {B.}~\bibnamefont {Lindner}},\ and\ \bibinfo {author}
  {\bibfnamefont {L.}~\bibnamefont {Schimansky-Geier}},\ }\bibfield  {title}
  {\bibinfo {title} {Active brownian particles},\ }\href
  {https://doi.org/10.1140/epjst/e2012-01529-y} {\bibfield  {journal} {\bibinfo
   {journal} {The European Physical Journal Special Topics}\ }\textbf {\bibinfo
  {volume} {202}},\ \bibinfo {pages} {1} (\bibinfo {year} {2012})}\BibitemShut
  {NoStop}%
\bibitem [{\citenamefont {Chat{\'e}}\ \emph {et~al.}(2008)\citenamefont
  {Chat{\'e}}, \citenamefont {Ginelli}, \citenamefont {Gr{\'e}goire},
  \citenamefont {Peruani},\ and\ \citenamefont {Raynaud}}]{chate:epjb}%
  \BibitemOpen
  \bibfield  {author} {\bibinfo {author} {\bibfnamefont {H.}~\bibnamefont
  {Chat{\'e}}}, \bibinfo {author} {\bibfnamefont {F.}~\bibnamefont {Ginelli}},
  \bibinfo {author} {\bibfnamefont {G.}~\bibnamefont {Gr{\'e}goire}}, \bibinfo
  {author} {\bibfnamefont {F.}~\bibnamefont {Peruani}},\ and\ \bibinfo {author}
  {\bibfnamefont {F.}~\bibnamefont {Raynaud}},\ }\bibfield  {title} {\bibinfo
  {title} {Modeling collective motion: variations on the {V}icsek model},\
  }\href {https://doi.org/10.1140/epjb/e2008-00275-9} {\bibfield  {journal}
  {\bibinfo  {journal} {The European Physical Journal B}\ }\textbf {\bibinfo
  {volume} {64}},\ \bibinfo {pages} {451} (\bibinfo {year} {2008})}\BibitemShut
  {NoStop}%
\bibitem [{\citenamefont {Mishra}\ \emph {et~al.}(2010)\citenamefont {Mishra},
  \citenamefont {Baskaran},\ and\ \citenamefont {Marchetti}}]{mishra:pre}%
  \BibitemOpen
  \bibfield  {author} {\bibinfo {author} {\bibfnamefont {S.}~\bibnamefont
  {Mishra}}, \bibinfo {author} {\bibfnamefont {A.}~\bibnamefont {Baskaran}},\
  and\ \bibinfo {author} {\bibfnamefont {M.~C.}\ \bibnamefont {Marchetti}},\
  }\bibfield  {title} {\bibinfo {title} {Fluctuations and pattern formation in
  self-propelled particles},\ }\href
  {https://doi.org/10.1103/PhysRevE.81.061916} {\bibfield  {journal} {\bibinfo
  {journal} {Phys. Rev. E}\ }\textbf {\bibinfo {volume} {81}},\ \bibinfo
  {pages} {061916} (\bibinfo {year} {2010})}\BibitemShut {NoStop}%
\bibitem [{\citenamefont {Farrell}\ \emph {et~al.}(2012)\citenamefont
  {Farrell}, \citenamefont {Marchetti}, \citenamefont {Marenduzzo},\ and\
  \citenamefont {Tailleur}}]{farrell:prl}%
  \BibitemOpen
  \bibfield  {author} {\bibinfo {author} {\bibfnamefont {F.~D.~C.}\
  \bibnamefont {Farrell}}, \bibinfo {author} {\bibfnamefont {M.~C.}\
  \bibnamefont {Marchetti}}, \bibinfo {author} {\bibfnamefont {D.}~\bibnamefont
  {Marenduzzo}},\ and\ \bibinfo {author} {\bibfnamefont {J.}~\bibnamefont
  {Tailleur}},\ }\bibfield  {title} {\bibinfo {title} {Pattern formation in
  self-propelled particles with density-dependent motility},\ }\href
  {https://doi.org/10.1103/PhysRevLett.108.248101} {\bibfield  {journal}
  {\bibinfo  {journal} {Phys. Rev. Lett.}\ }\textbf {\bibinfo {volume} {108}},\
  \bibinfo {pages} {248101} (\bibinfo {year} {2012})}\BibitemShut {NoStop}%
\bibitem [{\citenamefont {Nagai}\ \emph {et~al.}(2015)\citenamefont {Nagai},
  \citenamefont {Sumino}, \citenamefont {Montagne}, \citenamefont {Aranson},\
  and\ \citenamefont {Chat\'e}}]{nagai2015}%
  \BibitemOpen
  \bibfield  {author} {\bibinfo {author} {\bibfnamefont {K.~H.}\ \bibnamefont
  {Nagai}}, \bibinfo {author} {\bibfnamefont {Y.}~\bibnamefont {Sumino}},
  \bibinfo {author} {\bibfnamefont {R.}~\bibnamefont {Montagne}}, \bibinfo
  {author} {\bibfnamefont {I.~S.}\ \bibnamefont {Aranson}},\ and\ \bibinfo
  {author} {\bibfnamefont {H.}~\bibnamefont {Chat\'e}},\ }\bibfield  {title}
  {\bibinfo {title} {Collective motion of self-propelled particles with
  memory},\ }\href {https://doi.org/10.1103/PhysRevLett.114.168001} {\bibfield
  {journal} {\bibinfo  {journal} {Phys. Rev. Lett.}\ }\textbf {\bibinfo
  {volume} {114}},\ \bibinfo {pages} {168001} (\bibinfo {year}
  {2015})}\BibitemShut {NoStop}%
\bibitem [{\citenamefont {Solon}\ \emph {et~al.}(2015)\citenamefont {Solon},
  \citenamefont {Chat\'e},\ and\ \citenamefont {Tailleur}}]{solon:prl}%
  \BibitemOpen
  \bibfield  {author} {\bibinfo {author} {\bibfnamefont {A.~P.}\ \bibnamefont
  {Solon}}, \bibinfo {author} {\bibfnamefont {H.}~\bibnamefont {Chat\'e}},\
  and\ \bibinfo {author} {\bibfnamefont {J.}~\bibnamefont {Tailleur}},\
  }\bibfield  {title} {\bibinfo {title} {From phase to microphase separation in
  flocking models: The essential role of nonequilibrium fluctuations},\ }\href
  {https://doi.org/10.1103/PhysRevLett.114.068101} {\bibfield  {journal}
  {\bibinfo  {journal} {Phys. Rev. Lett.}\ }\textbf {\bibinfo {volume} {114}},\
  \bibinfo {pages} {068101} (\bibinfo {year} {2015})}\BibitemShut {NoStop}%
\bibitem [{\citenamefont {O'Keeffe}\ \emph {et~al.}(2017)\citenamefont
  {O'Keeffe}, \citenamefont {Hong},\ and\ \citenamefont
  {Strogatz}}]{keeffe:nature_comm}%
  \BibitemOpen
  \bibfield  {author} {\bibinfo {author} {\bibfnamefont {K.~P.}\ \bibnamefont
  {O'Keeffe}}, \bibinfo {author} {\bibfnamefont {H.}~\bibnamefont {Hong}},\
  and\ \bibinfo {author} {\bibfnamefont {S.~H.}\ \bibnamefont {Strogatz}},\
  }\bibfield  {title} {\bibinfo {title} {Oscillators that sync and swarm},\
  }\href {https://doi.org/10.1038/s41467-017-01190-3} {\bibfield  {journal}
  {\bibinfo  {journal} {Nature Communications}\ }\textbf {\bibinfo {volume}
  {8}},\ \bibinfo {pages} {1504} (\bibinfo {year} {2017})}\BibitemShut
  {NoStop}%
\bibitem [{\citenamefont {Degond}\ \emph
  {et~al.}(2014{\natexlab{a}})\citenamefont {Degond}, \citenamefont {Dimarco},\
  and\ \citenamefont {Mac}}]{degond:math_models_and_methods}%
  \BibitemOpen
  \bibfield  {author} {\bibinfo {author} {\bibfnamefont {P.}~\bibnamefont
  {Degond}}, \bibinfo {author} {\bibfnamefont {G.}~\bibnamefont {Dimarco}},\
  and\ \bibinfo {author} {\bibfnamefont {T.~B.~N.}\ \bibnamefont {Mac}},\
  }\bibfield  {title} {\bibinfo {title} {Hydrodynamics of the
  {K}uramoto-{V}icsek model of rotating self-propelled particles},\ }\href
  {https://doi.org/10.1142/S0218202513400095} {\bibfield  {journal} {\bibinfo
  {journal} {Mathematical Models and Methods in Applied Sciences}\ }\textbf
  {\bibinfo {volume} {24}},\ \bibinfo {pages} {277} (\bibinfo {year}
  {2014}{\natexlab{a}})}\BibitemShut {NoStop}%
\bibitem [{\citenamefont {Denk}\ \emph {et~al.}(2016)\citenamefont {Denk},
  \citenamefont {Huber}, \citenamefont {Reithmann},\ and\ \citenamefont
  {Frey}}]{denk:prl}%
  \BibitemOpen
  \bibfield  {author} {\bibinfo {author} {\bibfnamefont {J.}~\bibnamefont
  {Denk}}, \bibinfo {author} {\bibfnamefont {L.}~\bibnamefont {Huber}},
  \bibinfo {author} {\bibfnamefont {E.}~\bibnamefont {Reithmann}},\ and\
  \bibinfo {author} {\bibfnamefont {E.}~\bibnamefont {Frey}},\ }\bibfield
  {title} {\bibinfo {title} {Active curved polymers form vortex patterns on
  membranes},\ }\href {https://doi.org/10.1103/PhysRevLett.116.178301}
  {\bibfield  {journal} {\bibinfo  {journal} {Phys. Rev. Lett.}\ }\textbf
  {\bibinfo {volume} {116}},\ \bibinfo {pages} {178301} (\bibinfo {year}
  {2016})}\BibitemShut {NoStop}%
\bibitem [{\citenamefont {Chen}\ \emph {et~al.}(2017)\citenamefont {Chen},
  \citenamefont {Liu}, \citenamefont {Shi}, \citenamefont {Chat{\'e}},\ and\
  \citenamefont {Wu}}]{chen:nature}%
  \BibitemOpen
  \bibfield  {author} {\bibinfo {author} {\bibfnamefont {C.}~\bibnamefont
  {Chen}}, \bibinfo {author} {\bibfnamefont {S.}~\bibnamefont {Liu}}, \bibinfo
  {author} {\bibfnamefont {X.-q.}\ \bibnamefont {Shi}}, \bibinfo {author}
  {\bibfnamefont {H.}~\bibnamefont {Chat{\'e}}},\ and\ \bibinfo {author}
  {\bibfnamefont {Y.}~\bibnamefont {Wu}},\ }\bibfield  {title} {\bibinfo
  {title} {Weak synchronization and large-scale collective oscillation in dense
  bacterial suspensions},\ }\href {https://doi.org/10.1038/nature20817}
  {\bibfield  {journal} {\bibinfo  {journal} {Nature}\ }\textbf {\bibinfo
  {volume} {542}},\ \bibinfo {pages} {210 EP } (\bibinfo {year}
  {2017})}\BibitemShut {NoStop}%
\bibitem [{\citenamefont {Liebchen}\ and\ \citenamefont
  {Levis}(2017)}]{liebchen:prl}%
  \BibitemOpen
  \bibfield  {author} {\bibinfo {author} {\bibfnamefont {B.}~\bibnamefont
  {Liebchen}}\ and\ \bibinfo {author} {\bibfnamefont {D.}~\bibnamefont
  {Levis}},\ }\bibfield  {title} {\bibinfo {title} {Collective behavior of
  chiral active matter: Pattern formation and enhanced flocking},\ }\href
  {https://doi.org/10.1103/PhysRevLett.119.058002} {\bibfield  {journal}
  {\bibinfo  {journal} {Phys. Rev. Lett.}\ }\textbf {\bibinfo {volume} {119}},\
  \bibinfo {pages} {058002} (\bibinfo {year} {2017})}\BibitemShut {NoStop}%
\bibitem [{\citenamefont {Levis}\ \emph {et~al.}(2019)\citenamefont {Levis},
  \citenamefont {Pagonabarraga},\ and\ \citenamefont {Liebchen}}]{levis:prr}%
  \BibitemOpen
  \bibfield  {author} {\bibinfo {author} {\bibfnamefont {D.}~\bibnamefont
  {Levis}}, \bibinfo {author} {\bibfnamefont {I.}~\bibnamefont
  {Pagonabarraga}},\ and\ \bibinfo {author} {\bibfnamefont {B.}~\bibnamefont
  {Liebchen}},\ }\bibfield  {title} {\bibinfo {title} {Activity induced
  synchronization: Mutual flocking and chiral self-sorting},\ }\href
  {https://doi.org/10.1103/PhysRevResearch.1.023026} {\bibfield  {journal}
  {\bibinfo  {journal} {Phys. Rev. Research}\ }\textbf {\bibinfo {volume}
  {1}},\ \bibinfo {pages} {023026} (\bibinfo {year} {2019})}\BibitemShut
  {NoStop}%
\bibitem [{\citenamefont {Lei}\ \emph {et~al.}(2019)\citenamefont {Lei},
  \citenamefont {Ciamarra},\ and\ \citenamefont {Ni}}]{lei:sci_adv}%
  \BibitemOpen
  \bibfield  {author} {\bibinfo {author} {\bibfnamefont {Q.-L.}\ \bibnamefont
  {Lei}}, \bibinfo {author} {\bibfnamefont {M.~P.}\ \bibnamefont {Ciamarra}},\
  and\ \bibinfo {author} {\bibfnamefont {R.}~\bibnamefont {Ni}},\ }\bibfield
  {title} {\bibinfo {title} {Nonequilibrium strongly hyperuniform fluids of
  circle active particles with large local density fluctuations},\ }\bibfield
  {journal} {\bibinfo  {journal} {Science Advances}\ }\textbf {\bibinfo
  {volume} {5}},\ \href {https://doi.org/10.1126/sciadv.aau7423}
  {10.1126/sciadv.aau7423} (\bibinfo {year} {2019})\BibitemShut {NoStop}%
\bibitem [{\citenamefont {Souslov}\ \emph {et~al.}(2017)\citenamefont
  {Souslov}, \citenamefont {van Zuiden}, \citenamefont {Bartolo},\ and\
  \citenamefont {Vitelli}}]{souslov:nature_physics}%
  \BibitemOpen
  \bibfield  {author} {\bibinfo {author} {\bibfnamefont {A.}~\bibnamefont
  {Souslov}}, \bibinfo {author} {\bibfnamefont {B.~C.}\ \bibnamefont {van
  Zuiden}}, \bibinfo {author} {\bibfnamefont {D.}~\bibnamefont {Bartolo}},\
  and\ \bibinfo {author} {\bibfnamefont {V.}~\bibnamefont {Vitelli}},\
  }\bibfield  {title} {\bibinfo {title} {Topological sound in active-liquid
  metamaterials},\ }\href {https://doi.org/10.1038/nphys4193} {\bibfield
  {journal} {\bibinfo  {journal} {Nature Physics}\ }\textbf {\bibinfo {volume}
  {13}},\ \bibinfo {pages} {1091} (\bibinfo {year} {2017})}\BibitemShut
  {NoStop}%
\bibitem [{\citenamefont {Han}\ \emph {et~al.}(2017)\citenamefont {Han},
  \citenamefont {Yan}, \citenamefont {Granick},\ and\ \citenamefont
  {Luijten}}]{han:pnas}%
  \BibitemOpen
  \bibfield  {author} {\bibinfo {author} {\bibfnamefont {M.}~\bibnamefont
  {Han}}, \bibinfo {author} {\bibfnamefont {J.}~\bibnamefont {Yan}}, \bibinfo
  {author} {\bibfnamefont {S.}~\bibnamefont {Granick}},\ and\ \bibinfo {author}
  {\bibfnamefont {E.}~\bibnamefont {Luijten}},\ }\bibfield  {title} {\bibinfo
  {title} {Effective temperature concept evaluated in an active colloid
  mixture},\ }\href {https://doi.org/10.1073/pnas.1706702114} {\bibfield
  {journal} {\bibinfo  {journal} {Proceedings of the National Academy of
  Sciences}\ }\textbf {\bibinfo {volume} {114}},\ \bibinfo {pages} {7513}
  (\bibinfo {year} {2017})}\BibitemShut {NoStop}%
\bibitem [{\citenamefont {Tociu}\ \emph {et~al.}(2019)\citenamefont {Tociu},
  \citenamefont {Fodor}, \citenamefont {Nemoto},\ and\ \citenamefont
  {Vaikuntanathan}}]{tociu:prx}%
  \BibitemOpen
  \bibfield  {author} {\bibinfo {author} {\bibfnamefont {L.}~\bibnamefont
  {Tociu}}, \bibinfo {author} {\bibfnamefont {E.}~\bibnamefont {Fodor}},
  \bibinfo {author} {\bibfnamefont {T.}~\bibnamefont {Nemoto}},\ and\ \bibinfo
  {author} {\bibfnamefont {S.}~\bibnamefont {Vaikuntanathan}},\ }\bibfield
  {title} {\bibinfo {title} {How dissipation constrains fluctuations in
  nonequilibrium liquids: Diffusion, structure, and biased interactions},\
  }\href {https://doi.org/10.1103/PhysRevX.9.041026} {\bibfield  {journal}
  {\bibinfo  {journal} {Phys. Rev. X}\ }\textbf {\bibinfo {volume} {9}},\
  \bibinfo {pages} {041026} (\bibinfo {year} {2019})}\BibitemShut {NoStop}%
\bibitem [{\citenamefont {Nourhani}\ \emph {et~al.}(2015)\citenamefont
  {Nourhani}, \citenamefont {Crespi},\ and\ \citenamefont
  {Lammert}}]{nourhani:prl}%
  \BibitemOpen
  \bibfield  {author} {\bibinfo {author} {\bibfnamefont {A.}~\bibnamefont
  {Nourhani}}, \bibinfo {author} {\bibfnamefont {V.~H.}\ \bibnamefont
  {Crespi}},\ and\ \bibinfo {author} {\bibfnamefont {P.~E.}\ \bibnamefont
  {Lammert}},\ }\bibfield  {title} {\bibinfo {title} {Guiding chiral
  self-propellers in a periodic potential},\ }\href
  {https://doi.org/10.1103/PhysRevLett.115.118101} {\bibfield  {journal}
  {\bibinfo  {journal} {Phys. Rev. Lett.}\ }\textbf {\bibinfo {volume} {115}},\
  \bibinfo {pages} {118101} (\bibinfo {year} {2015})}\BibitemShut {NoStop}%
\bibitem [{\citenamefont {Narinder}\ \emph {et~al.}(2018)\citenamefont
  {Narinder}, \citenamefont {Bechinger},\ and\ \citenamefont
  {Gomez-Solano}}]{narinder:prl}%
  \BibitemOpen
  \bibfield  {author} {\bibinfo {author} {\bibfnamefont {N.}~\bibnamefont
  {Narinder}}, \bibinfo {author} {\bibfnamefont {C.}~\bibnamefont
  {Bechinger}},\ and\ \bibinfo {author} {\bibfnamefont {J.~R.}\ \bibnamefont
  {Gomez-Solano}},\ }\bibfield  {title} {\bibinfo {title} {Memory-induced
  transition from a persistent random walk to circular motion for achiral
  microswimmers},\ }\href {https://doi.org/10.1103/PhysRevLett.121.078003}
  {\bibfield  {journal} {\bibinfo  {journal} {Phys. Rev. Lett.}\ }\textbf
  {\bibinfo {volume} {121}},\ \bibinfo {pages} {078003} (\bibinfo {year}
  {2018})}\BibitemShut {NoStop}%
\bibitem [{\citenamefont {Lauga}\ \emph {et~al.}(2006)\citenamefont {Lauga},
  \citenamefont {DiLuzio}, \citenamefont {Whitesides},\ and\ \citenamefont
  {Stone}}]{lauga:biophysical}%
  \BibitemOpen
  \bibfield  {author} {\bibinfo {author} {\bibfnamefont {E.}~\bibnamefont
  {Lauga}}, \bibinfo {author} {\bibfnamefont {W.~R.}\ \bibnamefont {DiLuzio}},
  \bibinfo {author} {\bibfnamefont {G.~M.}\ \bibnamefont {Whitesides}},\ and\
  \bibinfo {author} {\bibfnamefont {H.~A.}\ \bibnamefont {Stone}},\ }\bibfield
  {title} {\bibinfo {title} {Swimming in circles: Motion of bacteria near solid
  boundaries},\ }\bibfield  {booktitle} {\emph {\bibinfo {booktitle}
  {Biophysical Journal}},\ }\href {https://doi.org/10.1529/biophysj.105.069401}
  {\bibfield  {journal} {\bibinfo  {journal} {Biophysical Journal}\ }\textbf
  {\bibinfo {volume} {90}},\ \bibinfo {pages} {400} (\bibinfo {year}
  {2006})}\BibitemShut {NoStop}%
\bibitem [{\citenamefont {Lemelle}\ \emph {et~al.}(2010)\citenamefont
  {Lemelle}, \citenamefont {Palierne}, \citenamefont {Chatre},\ and\
  \citenamefont {Place}}]{lemelle:bacteriology}%
  \BibitemOpen
  \bibfield  {author} {\bibinfo {author} {\bibfnamefont {L.}~\bibnamefont
  {Lemelle}}, \bibinfo {author} {\bibfnamefont {J.-F.}\ \bibnamefont
  {Palierne}}, \bibinfo {author} {\bibfnamefont {E.}~\bibnamefont {Chatre}},\
  and\ \bibinfo {author} {\bibfnamefont {C.}~\bibnamefont {Place}},\ }\bibfield
   {title} {\bibinfo {title} {Counterclockwise circular motion of bacteria
  swimming at the air-liquid interface},\ }\href
  {https://doi.org/10.1128/JB.00397-10} {\bibfield  {journal} {\bibinfo
  {journal} {Journal of Bacteriology}\ }\textbf {\bibinfo {volume} {192}},\
  \bibinfo {pages} {6307} (\bibinfo {year} {2010})}\BibitemShut {NoStop}%
\bibitem [{\citenamefont {Sumino}\ \emph {et~al.}(2012)\citenamefont {Sumino},
  \citenamefont {Nagai}, \citenamefont {Shitaka}, \citenamefont {Tanaka},
  \citenamefont {Yoshikawa}, \citenamefont {Chat{\'e}},\ and\ \citenamefont
  {Oiwa}}]{sumino:nature}%
  \BibitemOpen
  \bibfield  {author} {\bibinfo {author} {\bibfnamefont {Y.}~\bibnamefont
  {Sumino}}, \bibinfo {author} {\bibfnamefont {K.~H.}\ \bibnamefont {Nagai}},
  \bibinfo {author} {\bibfnamefont {Y.}~\bibnamefont {Shitaka}}, \bibinfo
  {author} {\bibfnamefont {D.}~\bibnamefont {Tanaka}}, \bibinfo {author}
  {\bibfnamefont {K.}~\bibnamefont {Yoshikawa}}, \bibinfo {author}
  {\bibfnamefont {H.}~\bibnamefont {Chat{\'e}}},\ and\ \bibinfo {author}
  {\bibfnamefont {K.}~\bibnamefont {Oiwa}},\ }\bibfield  {title} {\bibinfo
  {title} {Large-scale vortex lattice emerging from collectively moving
  microtubules},\ }\href {https://doi.org/10.1038/nature10874} {\bibfield
  {journal} {\bibinfo  {journal} {Nature}\ }\textbf {\bibinfo {volume} {483}},\
  \bibinfo {pages} {448} (\bibinfo {year} {2012})}\BibitemShut {NoStop}%
\bibitem [{\citenamefont {Ērglis}\ \emph {et~al.}(2007)\citenamefont
  {Ērglis}, \citenamefont {Wen}, \citenamefont {Ose}, \citenamefont {Zeltins},
  \citenamefont {Sharipo}, \citenamefont {Janmey},\ and\ \citenamefont
  {Cēbers}}]{erglis:biophysical_journal}%
  \BibitemOpen
  \bibfield  {author} {\bibinfo {author} {\bibfnamefont {K.}~\bibnamefont
  {Ērglis}}, \bibinfo {author} {\bibfnamefont {Q.}~\bibnamefont {Wen}},
  \bibinfo {author} {\bibfnamefont {V.}~\bibnamefont {Ose}}, \bibinfo {author}
  {\bibfnamefont {A.}~\bibnamefont {Zeltins}}, \bibinfo {author} {\bibfnamefont
  {A.}~\bibnamefont {Sharipo}}, \bibinfo {author} {\bibfnamefont {P.~A.}\
  \bibnamefont {Janmey}},\ and\ \bibinfo {author} {\bibfnamefont
  {A.}~\bibnamefont {Cēbers}},\ }\bibfield  {title} {\bibinfo {title}
  {Dynamics of magnetotactic bacteria in a rotating magnetic field},\ }\href
  {https://doi.org/https://doi.org/10.1529/biophysj.107.107474} {\bibfield
  {journal} {\bibinfo  {journal} {Biophysical Journal}\ }\textbf {\bibinfo
  {volume} {93}},\ \bibinfo {pages} {1402 } (\bibinfo {year}
  {2007})}\BibitemShut {NoStop}%
\bibitem [{\citenamefont {Cēbers}(2011)}]{cebers:jmmm}%
  \BibitemOpen
  \bibfield  {author} {\bibinfo {author} {\bibfnamefont {A.}~\bibnamefont
  {Cēbers}},\ }\bibfield  {title} {\bibinfo {title} {Diffusion of
  magnetotactic bacterium in rotating magnetic field},\ }\href
  {https://doi.org/https://doi.org/10.1016/j.jmmm.2010.09.017} {\bibfield
  {journal} {\bibinfo  {journal} {Journal of Magnetism and Magnetic Materials}\
  }\textbf {\bibinfo {volume} {323}},\ \bibinfo {pages} {279 } (\bibinfo {year}
  {2011})}\BibitemShut {NoStop}%
\bibitem [{\citenamefont {Riedel}\ \emph {et~al.}(2005)\citenamefont {Riedel},
  \citenamefont {Kruse},\ and\ \citenamefont {Howard}}]{riedel:science}%
  \BibitemOpen
  \bibfield  {author} {\bibinfo {author} {\bibfnamefont {I.~H.}\ \bibnamefont
  {Riedel}}, \bibinfo {author} {\bibfnamefont {K.}~\bibnamefont {Kruse}},\ and\
  \bibinfo {author} {\bibfnamefont {J.}~\bibnamefont {Howard}},\ }\bibfield
  {title} {\bibinfo {title} {A self-organized vortex array of hydrodynamically
  entrained sperm cells},\ }\href {https://doi.org/10.1126/science.1110329}
  {\bibfield  {journal} {\bibinfo  {journal} {Science}\ }\textbf {\bibinfo
  {volume} {309}},\ \bibinfo {pages} {300} (\bibinfo {year}
  {2005})}\BibitemShut {NoStop}%
\bibitem [{\citenamefont {Friedrich}\ and\ \citenamefont
  {J{\"u}licher}(2007)}]{friedrich:pnas}%
  \BibitemOpen
  \bibfield  {author} {\bibinfo {author} {\bibfnamefont {B.~M.}\ \bibnamefont
  {Friedrich}}\ and\ \bibinfo {author} {\bibfnamefont {F.}~\bibnamefont
  {J{\"u}licher}},\ }\bibfield  {title} {\bibinfo {title} {Chemotaxis of sperm
  cells},\ }\href {https://doi.org/10.1073/pnas.0703530104} {\bibfield
  {journal} {\bibinfo  {journal} {Proceedings of the National Academy of
  Sciences}\ }\textbf {\bibinfo {volume} {104}},\ \bibinfo {pages} {13256}
  (\bibinfo {year} {2007})}\BibitemShut {NoStop}%
\bibitem [{\citenamefont {Kastberger}\ \emph {et~al.}(2008)\citenamefont
  {Kastberger}, \citenamefont {Schmelzer},\ and\ \citenamefont
  {Kranner}}]{kastberger:plos_one}%
  \BibitemOpen
  \bibfield  {author} {\bibinfo {author} {\bibfnamefont {G.}~\bibnamefont
  {Kastberger}}, \bibinfo {author} {\bibfnamefont {E.}~\bibnamefont
  {Schmelzer}},\ and\ \bibinfo {author} {\bibfnamefont {I.}~\bibnamefont
  {Kranner}},\ }\bibfield  {title} {\bibinfo {title} {Social waves in giant
  honeybees repel hornets},\ }\href
  {https://doi.org/10.1371/journal.pone.0003141} {\bibfield  {journal}
  {\bibinfo  {journal} {PLoS ONE}\ }\textbf {\bibinfo {volume} {3}},\ \bibinfo
  {pages} {1} (\bibinfo {year} {2008})}\BibitemShut {NoStop}%
\bibitem [{\citenamefont {Kuramoto}\ and\ \citenamefont
  {Battogtokh}(2002)}]{kuramoto2002}%
  \BibitemOpen
  \bibfield  {author} {\bibinfo {author} {\bibfnamefont {Y.}~\bibnamefont
  {Kuramoto}}\ and\ \bibinfo {author} {\bibfnamefont {D.}~\bibnamefont
  {Battogtokh}},\ }\bibfield  {title} {\bibinfo {title} {Coexistence of
  coherence and incoherence in nonlocally coupled phase oscillators},\
  }\href@noop {} {\bibfield  {journal} {\bibinfo  {journal} {Nonlinear
  Phenomena in Complex Systems}\ }\textbf {\bibinfo {volume} {5}},\ \bibinfo
  {pages} {380} (\bibinfo {year} {2002})}\BibitemShut {NoStop}%
\bibitem [{\citenamefont {Abrams}\ and\ \citenamefont
  {Strogatz}(2004)}]{abrams2004}%
  \BibitemOpen
  \bibfield  {author} {\bibinfo {author} {\bibfnamefont {D.~M.}\ \bibnamefont
  {Abrams}}\ and\ \bibinfo {author} {\bibfnamefont {S.~H.}\ \bibnamefont
  {Strogatz}},\ }\bibfield  {title} {\bibinfo {title} {Chimera states for
  coupled oscillators},\ }\href {https://doi.org/10.1103/PhysRevLett.93.174102}
  {\bibfield  {journal} {\bibinfo  {journal} {Phys. Rev. Lett.}\ }\textbf
  {\bibinfo {volume} {93}},\ \bibinfo {pages} {174102} (\bibinfo {year}
  {2004})}\BibitemShut {NoStop}%
\bibitem [{\citenamefont {Omel'chenko}(2018)}]{omelchenko:nonlinearity}%
  \BibitemOpen
  \bibfield  {author} {\bibinfo {author} {\bibfnamefont {O.~E.}\ \bibnamefont
  {Omel'chenko}},\ }\bibfield  {title} {\bibinfo {title} {The mathematics
  behind chimera states},\ }\href {https://doi.org/10.1088/1361-6544/aaaa07}
  {\bibfield  {journal} {\bibinfo  {journal} {Nonlinearity}\ }\textbf {\bibinfo
  {volume} {31}},\ \bibinfo {pages} {R121} (\bibinfo {year}
  {2018})}\BibitemShut {NoStop}%
\bibitem [{\citenamefont {Omel'chenko}\ \emph {et~al.}(2012)\citenamefont
  {Omel'chenko}, \citenamefont {Wolfrum}, \citenamefont {Yanchuk},
  \citenamefont {Maistrenko},\ and\ \citenamefont {Sudakov}}]{omelchenko:pre}%
  \BibitemOpen
  \bibfield  {author} {\bibinfo {author} {\bibfnamefont {O.~E.}\ \bibnamefont
  {Omel'chenko}}, \bibinfo {author} {\bibfnamefont {M.}~\bibnamefont
  {Wolfrum}}, \bibinfo {author} {\bibfnamefont {S.}~\bibnamefont {Yanchuk}},
  \bibinfo {author} {\bibfnamefont {Y.~L.}\ \bibnamefont {Maistrenko}},\ and\
  \bibinfo {author} {\bibfnamefont {O.}~\bibnamefont {Sudakov}},\ }\bibfield
  {title} {\bibinfo {title} {Stationary patterns of coherence and incoherence
  in two-dimensional arrays of non-locally-coupled phase oscillators},\ }\href
  {https://doi.org/10.1103/PhysRevE.85.036210} {\bibfield  {journal} {\bibinfo
  {journal} {Phys. Rev. E}\ }\textbf {\bibinfo {volume} {85}},\ \bibinfo
  {pages} {036210} (\bibinfo {year} {2012})}\BibitemShut {NoStop}%
\bibitem [{\citenamefont {Kruk}\ \emph {et~al.}(2018)\citenamefont {Kruk},
  \citenamefont {Maistrenko},\ and\ \citenamefont {Koeppl}}]{kruk:aps2018}%
  \BibitemOpen
  \bibfield  {author} {\bibinfo {author} {\bibfnamefont {N.}~\bibnamefont
  {Kruk}}, \bibinfo {author} {\bibfnamefont {Y.}~\bibnamefont {Maistrenko}},\
  and\ \bibinfo {author} {\bibfnamefont {H.}~\bibnamefont {Koeppl}},\
  }\bibfield  {title} {\bibinfo {title} {Self-propelled chimeras},\ }\href
  {https://doi.org/10.1103/PhysRevE.98.032219} {\bibfield  {journal} {\bibinfo
  {journal} {Phys. Rev. E}\ }\textbf {\bibinfo {volume} {98}},\ \bibinfo
  {pages} {032219} (\bibinfo {year} {2018})}\BibitemShut {NoStop}%
\bibitem [{\citenamefont {Dobrushin}(1979)}]{dobrushin}%
  \BibitemOpen
  \bibfield  {author} {\bibinfo {author} {\bibfnamefont {R.~L.}\ \bibnamefont
  {Dobrushin}},\ }\bibfield  {title} {\bibinfo {title} {Vlasov equations},\
  }\href {https://doi.org/10.1007/BF01077243} {\bibfield  {journal} {\bibinfo
  {journal} {Functional Analysis and Its Applications}\ }\textbf {\bibinfo
  {volume} {13}},\ \bibinfo {pages} {115} (\bibinfo {year} {1979})}\BibitemShut
  {NoStop}%
\bibitem [{\citenamefont {Lancellotti}(2005)}]{lancellotti:ttsp}%
  \BibitemOpen
  \bibfield  {author} {\bibinfo {author} {\bibfnamefont {C.}~\bibnamefont
  {Lancellotti}},\ }\bibfield  {title} {\bibinfo {title} {On the {V}lasov limit
  for systems of nonlinearly coupled oscillators without noise},\ }\href
  {https://doi.org/10.1080/00411450508951152} {\bibfield  {journal} {\bibinfo
  {journal} {Transport Theory and Statistical Physics}\ }\textbf {\bibinfo
  {volume} {34}},\ \bibinfo {pages} {523} (\bibinfo {year} {2005})}\BibitemShut
  {NoStop}%
\bibitem [{\citenamefont {Kipnis}\ and\ \citenamefont
  {Landim}(1998)}]{kipnis1998scaling}%
  \BibitemOpen
  \bibfield  {author} {\bibinfo {author} {\bibfnamefont {C.}~\bibnamefont
  {Kipnis}}\ and\ \bibinfo {author} {\bibfnamefont {C.}~\bibnamefont
  {Landim}},\ }\href {https://books.google.de/books?id=eLUNVIoYcOoC} {\emph
  {\bibinfo {title} {Scaling Limits of Interacting Particle Systems}}},\
  Grundlehren der mathematischen Wissenschaften\ (\bibinfo  {publisher}
  {Springer Berlin Heidelberg},\ \bibinfo {year} {1998})\BibitemShut {NoStop}%
\bibitem [{\citenamefont {Laney}(1998)}]{laney:comp_gas_dyn}%
  \BibitemOpen
  \bibfield  {author} {\bibinfo {author} {\bibfnamefont {C.}~\bibnamefont
  {Laney}},\ }\href {https://books.google.de/books?id=r-bYw-JjKGAC} {\emph
  {\bibinfo {title} {Computational Gasdynamics}}},\ Computational Gasdynamics\
  (\bibinfo  {publisher} {Cambridge University Press},\ \bibinfo {year}
  {1998})\BibitemShut {NoStop}%
\bibitem [{\citenamefont {Risken}\ and\ \citenamefont {Frank}(1996)}]{risken}%
  \BibitemOpen
  \bibfield  {author} {\bibinfo {author} {\bibfnamefont {H.}~\bibnamefont
  {Risken}}\ and\ \bibinfo {author} {\bibfnamefont {T.}~\bibnamefont {Frank}},\
  }\href {https://books.google.de/books?id=MG2V9vTgSgEC} {\emph {\bibinfo
  {title} {The Fokker-Planck Equation: Methods of Solution and
  Applications}}},\ Springer Series in Synergetics\ (\bibinfo  {publisher}
  {Springer Berlin Heidelberg},\ \bibinfo {year} {1996})\BibitemShut {NoStop}%
\bibitem [{\citenamefont {Archer}\ and\ \citenamefont
  {Rauscher}(2004)}]{archer:jpa}%
  \BibitemOpen
  \bibfield  {author} {\bibinfo {author} {\bibfnamefont {A.~J.}\ \bibnamefont
  {Archer}}\ and\ \bibinfo {author} {\bibfnamefont {M.}~\bibnamefont
  {Rauscher}},\ }\bibfield  {title} {\bibinfo {title} {Dynamical density
  functional theory for interacting {B}rownian particles: stochastic or
  deterministic?},\ }\href {https://doi.org/10.1088/0305-4470/37/40/001}
  {\bibfield  {journal} {\bibinfo  {journal} {Journal of Physics A:
  Mathematical and General}\ }\textbf {\bibinfo {volume} {37}},\ \bibinfo
  {pages} {9325} (\bibinfo {year} {2004})}\BibitemShut {NoStop}%
\bibitem [{\citenamefont {Spohn}(1991)}]{spohn:springer}%
  \BibitemOpen
  \bibfield  {author} {\bibinfo {author} {\bibfnamefont {H.}~\bibnamefont
  {Spohn}},\ }\href {https://doi.org/10.1007/978-3-642-84371-6} {\emph
  {\bibinfo {title} {Large Scale Dynamics of Interacting Particles}}},\
  \bibinfo {edition} {1st}\ ed.\ (\bibinfo  {publisher} {Springer},\ \bibinfo
  {year} {1991})\BibitemShut {NoStop}%
\bibitem [{\citenamefont {Bertini}\ \emph {et~al.}(2010)\citenamefont
  {Bertini}, \citenamefont {Giacomin},\ and\ \citenamefont
  {Pakdaman}}]{bertini:journal_of_stat_phys}%
  \BibitemOpen
  \bibfield  {author} {\bibinfo {author} {\bibfnamefont {L.}~\bibnamefont
  {Bertini}}, \bibinfo {author} {\bibfnamefont {G.}~\bibnamefont {Giacomin}},\
  and\ \bibinfo {author} {\bibfnamefont {K.}~\bibnamefont {Pakdaman}},\
  }\bibfield  {title} {\bibinfo {title} {Dynamical aspects of mean field plane
  rotators and the {K}uramoto model},\ }\href
  {https://doi.org/10.1007/s10955-009-9908-9} {\bibfield  {journal} {\bibinfo
  {journal} {Journal of Statistical Physics}\ }\textbf {\bibinfo {volume}
  {138}},\ \bibinfo {pages} {270} (\bibinfo {year} {2010})}\BibitemShut
  {NoStop}%
\bibitem [{\citenamefont {Giacomin}\ \emph {et~al.}(2012)\citenamefont
  {Giacomin}, \citenamefont {Pakdaman},\ and\ \citenamefont
  {Pellegrin}}]{giacomin:nonlinearity}%
  \BibitemOpen
  \bibfield  {author} {\bibinfo {author} {\bibfnamefont {G.}~\bibnamefont
  {Giacomin}}, \bibinfo {author} {\bibfnamefont {K.}~\bibnamefont {Pakdaman}},\
  and\ \bibinfo {author} {\bibfnamefont {X.}~\bibnamefont {Pellegrin}},\
  }\bibfield  {title} {\bibinfo {title} {Global attractor and asymptotic
  dynamics in the {K}uramoto model for coupled noisy phase oscillators},\
  }\href {http://stacks.iop.org/0951-7715/25/i=5/a=1247} {\bibfield  {journal}
  {\bibinfo  {journal} {Nonlinearity}\ }\textbf {\bibinfo {volume} {25}},\
  \bibinfo {pages} {1247} (\bibinfo {year} {2012})}\BibitemShut {NoStop}%
\bibitem [{\citenamefont {Gupta}\ \emph {et~al.}(2018)\citenamefont {Gupta},
  \citenamefont {Campa},\ and\ \citenamefont {Ruffo}}]{gupta:first_order}%
  \BibitemOpen
  \bibfield  {author} {\bibinfo {author} {\bibfnamefont {S.}~\bibnamefont
  {Gupta}}, \bibinfo {author} {\bibfnamefont {A.}~\bibnamefont {Campa}},\ and\
  \bibinfo {author} {\bibfnamefont {S.}~\bibnamefont {Ruffo}},\ }\bibinfo
  {title} {Oscillators with first-order dynamics},\ in\ \href
  {https://doi.org/10.1007/978-3-319-96664-9_2} {\emph {\bibinfo {booktitle}
  {Statistical Physics of Synchronization}}}\ (\bibinfo  {publisher} {Springer
  International Publishing},\ \bibinfo {address} {Cham},\ \bibinfo {year}
  {2018})\ pp.\ \bibinfo {pages} {39--80}\BibitemShut {NoStop}%
\bibitem [{\citenamefont {Mardia}\ and\ \citenamefont
  {Jupp}(2009)}]{mardia2009directional}%
  \BibitemOpen
  \bibfield  {author} {\bibinfo {author} {\bibfnamefont {K.}~\bibnamefont
  {Mardia}}\ and\ \bibinfo {author} {\bibfnamefont {P.}~\bibnamefont {Jupp}},\
  }\href {https://books.google.de/books?id=PTNiCm4Q-M0C} {\emph {\bibinfo
  {title} {Directional Statistics}}},\ Wiley Series in Probability and
  Statistics\ (\bibinfo  {publisher} {Wiley},\ \bibinfo {year}
  {2009})\BibitemShut {NoStop}%
\bibitem [{\citenamefont {Toner}\ and\ \citenamefont {Tu}(1998)}]{toner:pre}%
  \BibitemOpen
  \bibfield  {author} {\bibinfo {author} {\bibfnamefont {J.}~\bibnamefont
  {Toner}}\ and\ \bibinfo {author} {\bibfnamefont {Y.}~\bibnamefont {Tu}},\
  }\bibfield  {title} {\bibinfo {title} {Flocks, herds, and schools: A
  quantitative theory of flocking},\ }\href
  {https://doi.org/10.1103/PhysRevE.58.4828} {\bibfield  {journal} {\bibinfo
  {journal} {Phys. Rev. E}\ }\textbf {\bibinfo {volume} {58}},\ \bibinfo
  {pages} {4828} (\bibinfo {year} {1998})}\BibitemShut {NoStop}%
\bibitem [{\citenamefont {Bertin}\ \emph {et~al.}(2009)\citenamefont {Bertin},
  \citenamefont {Droz},\ and\ \citenamefont {Gr{\'{e}}goire}}]{bertin:jpa}%
  \BibitemOpen
  \bibfield  {author} {\bibinfo {author} {\bibfnamefont {E.}~\bibnamefont
  {Bertin}}, \bibinfo {author} {\bibfnamefont {M.}~\bibnamefont {Droz}},\ and\
  \bibinfo {author} {\bibfnamefont {G.}~\bibnamefont {Gr{\'{e}}goire}},\
  }\bibfield  {title} {\bibinfo {title} {Hydrodynamic equations for
  self-propelled particles: microscopic derivation and stability analysis},\
  }\href {https://doi.org/10.1088/1751-8113/42/44/445001} {\bibfield  {journal}
  {\bibinfo  {journal} {Journal of Physics A: Mathematical and Theoretical}\
  }\textbf {\bibinfo {volume} {42}},\ \bibinfo {pages} {445001} (\bibinfo
  {year} {2009})}\BibitemShut {NoStop}%
\bibitem [{\citenamefont {Gro{\ss}mann}\ \emph {et~al.}(2013)\citenamefont
  {Gro{\ss}mann}, \citenamefont {Schimansky-Geier},\ and\ \citenamefont
  {Romanczuk}}]{grossmann:iop}%
  \BibitemOpen
  \bibfield  {author} {\bibinfo {author} {\bibfnamefont {R.}~\bibnamefont
  {Gro{\ss}mann}}, \bibinfo {author} {\bibfnamefont {L.}~\bibnamefont
  {Schimansky-Geier}},\ and\ \bibinfo {author} {\bibfnamefont {P.}~\bibnamefont
  {Romanczuk}},\ }\bibfield  {title} {\bibinfo {title} {Self-propelled
  particles with selective attraction{\textendash}repulsion interaction: from
  microscopic dynamics to coarse-grained theories},\ }\href
  {https://doi.org/10.1088/1367-2630/15/8/085014} {\bibfield  {journal}
  {\bibinfo  {journal} {New Journal of Physics}\ }\textbf {\bibinfo {volume}
  {15}},\ \bibinfo {pages} {085014} (\bibinfo {year} {2013})}\BibitemShut
  {NoStop}%
\bibitem [{\citenamefont {Degond}\ \emph
  {et~al.}(2014{\natexlab{b}})\citenamefont {Degond}, \citenamefont
  {Frouvelle},\ and\ \citenamefont {Liu}}]{degond:arma}%
  \BibitemOpen
  \bibfield  {author} {\bibinfo {author} {\bibfnamefont {P.}~\bibnamefont
  {Degond}}, \bibinfo {author} {\bibfnamefont {A.}~\bibnamefont {Frouvelle}},\
  and\ \bibinfo {author} {\bibfnamefont {J.-G.}\ \bibnamefont {Liu}},\
  }\bibfield  {title} {\bibinfo {title} {Phase transitions, hysteresis, and
  hyperbolicity for self-organized alignment dynamics},\ }\href
  {https://doi.org/10.1007/s00205-014-0800-7} {\bibfield  {journal} {\bibinfo
  {journal} {Archive for Rational Mechanics and Analysis}\ }\textbf {\bibinfo
  {volume} {216}},\ \bibinfo {pages} {63} (\bibinfo {year}
  {2014}{\natexlab{b}})}\BibitemShut {NoStop}%
\bibitem [{\citenamefont {Peshkov}\ \emph {et~al.}(2012)\citenamefont
  {Peshkov}, \citenamefont {Aranson}, \citenamefont {Bertin}, \citenamefont
  {Chat\'e},\ and\ \citenamefont {Ginelli}}]{peshkov:prl}%
  \BibitemOpen
  \bibfield  {author} {\bibinfo {author} {\bibfnamefont {A.}~\bibnamefont
  {Peshkov}}, \bibinfo {author} {\bibfnamefont {I.~S.}\ \bibnamefont
  {Aranson}}, \bibinfo {author} {\bibfnamefont {E.}~\bibnamefont {Bertin}},
  \bibinfo {author} {\bibfnamefont {H.}~\bibnamefont {Chat\'e}},\ and\ \bibinfo
  {author} {\bibfnamefont {F.}~\bibnamefont {Ginelli}},\ }\bibfield  {title}
  {\bibinfo {title} {Nonlinear field equations for aligning self-propelled
  rods},\ }\href {https://doi.org/10.1103/PhysRevLett.109.268701} {\bibfield
  {journal} {\bibinfo  {journal} {Phys. Rev. Lett.}\ }\textbf {\bibinfo
  {volume} {109}},\ \bibinfo {pages} {268701} (\bibinfo {year}
  {2012})}\BibitemShut {NoStop}%
\bibitem [{sup()}]{supplemental_material}%
  \BibitemOpen
  \href@noop {} {\bibinfo {title} {See {S}upplemental {M}aterial at [url will
  be inserted by publisher] for movies of particle dynamics}}\BibitemShut
  {NoStop}%
\bibitem [{bcs()}]{bcs_youtube_channel}%
  \BibitemOpen
  \href@noop {} {\bibinfo {title}
  {\href{https://www.youtube.com/playlist?list=PLjL7stT6PH4xdc4X5Ee7xAr2vm49uIHvW}{\nolinkurl{https://www.youtube.com/playlist?list=pljl7stt6ph4xdc4x5ee7xar2vm49uihvw}}}}\BibitemShut
  {NoStop}%
\bibitem [{fig()}]{figshare}%
  \BibitemOpen
  \href@noop {} {\bibinfo {title}
  {\href{https://figshare.com/projects/Traveling_Bands_Clouds_and_Vortices_of_Chiral_Active_Matter/82163}{\nolinkurl{https://figshare.com/projects/traveling_bands_clouds_and_vortices_of_chiral_active_matter/82163}}}}\BibitemShut
  {NoStop}%
\bibitem [{\citenamefont {Platen}\ and\ \citenamefont
  {Bruti-Liberati}(2010)}]{platen}%
  \BibitemOpen
  \bibfield  {author} {\bibinfo {author} {\bibfnamefont {E.}~\bibnamefont
  {Platen}}\ and\ \bibinfo {author} {\bibfnamefont {N.}~\bibnamefont
  {Bruti-Liberati}},\ }\href@noop {} {\emph {\bibinfo {title} {Numerical
  Solution of Stochastic Differential Equations with Jumps in Finance}}}\
  (\bibinfo  {publisher} {Springer-Verlag Berlin Heidelberg},\ \bibinfo {year}
  {2010})\BibitemShut {NoStop}%
\bibitem [{\citenamefont {Gupta}\ \emph
  {et~al.}(2014{\natexlab{a}})\citenamefont {Gupta}, \citenamefont {Campa},\
  and\ \citenamefont {Ruffo}}]{gupta:pre}%
  \BibitemOpen
  \bibfield  {author} {\bibinfo {author} {\bibfnamefont {S.}~\bibnamefont
  {Gupta}}, \bibinfo {author} {\bibfnamefont {A.}~\bibnamefont {Campa}},\ and\
  \bibinfo {author} {\bibfnamefont {S.}~\bibnamefont {Ruffo}},\ }\bibfield
  {title} {\bibinfo {title} {Nonequilibrium first-order phase transition in
  coupled oscillator systems with inertia and noise},\ }\href
  {https://doi.org/10.1103/PhysRevE.89.022123} {\bibfield  {journal} {\bibinfo
  {journal} {Phys. Rev. E}\ }\textbf {\bibinfo {volume} {89}},\ \bibinfo
  {pages} {022123} (\bibinfo {year} {2014}{\natexlab{a}})}\BibitemShut
  {NoStop}%
\bibitem [{\citenamefont {Spohn}\ and\ \citenamefont
  {Neunzert}(1981)}]{spohn:mmas}%
  \BibitemOpen
  \bibfield  {author} {\bibinfo {author} {\bibfnamefont {H.}~\bibnamefont
  {Spohn}}\ and\ \bibinfo {author} {\bibfnamefont {H.}~\bibnamefont
  {Neunzert}},\ }\bibfield  {title} {\bibinfo {title} {On the {V}lasov
  hierarchy},\ }\href {https://doi.org/10.1002/mma.1670030131} {\bibfield
  {journal} {\bibinfo  {journal} {Mathematical Methods in the Applied
  Sciences}\ }\textbf {\bibinfo {volume} {3}},\ \bibinfo {pages} {445}
  (\bibinfo {year} {1981})}\BibitemShut {NoStop}%
\bibitem [{\citenamefont {Braun}\ and\ \citenamefont {Hepp}(1977)}]{braun:cmp}%
  \BibitemOpen
  \bibfield  {author} {\bibinfo {author} {\bibfnamefont {W.}~\bibnamefont
  {Braun}}\ and\ \bibinfo {author} {\bibfnamefont {K.}~\bibnamefont {Hepp}},\
  }\bibfield  {title} {\bibinfo {title} {The {V}lasov dynamics and its
  fluctuations in the $1/{N}$ limit of interacting classical particles},\
  }\href {https://projecteuclid.org:443/euclid.cmp/1103901139} {\bibfield
  {journal} {\bibinfo  {journal} {Comm. Math. Phys.}\ }\textbf {\bibinfo
  {volume} {56}},\ \bibinfo {pages} {101} (\bibinfo {year} {1977})}\BibitemShut
  {NoStop}%
\bibitem [{\citenamefont {Marconi}\ and\ \citenamefont
  {Tarazona}(1999)}]{marconi:jcp}%
  \BibitemOpen
  \bibfield  {author} {\bibinfo {author} {\bibfnamefont {U.~M.~B.}\
  \bibnamefont {Marconi}}\ and\ \bibinfo {author} {\bibfnamefont
  {P.}~\bibnamefont {Tarazona}},\ }\bibfield  {title} {\bibinfo {title}
  {Dynamic density functional theory of fluids},\ }\href
  {https://doi.org/10.1063/1.478705} {\bibfield  {journal} {\bibinfo  {journal}
  {The Journal of Chemical Physics}\ }\textbf {\bibinfo {volume} {110}},\
  \bibinfo {pages} {8032} (\bibinfo {year} {1999})}\BibitemShut {NoStop}%
\bibitem [{\citenamefont {Olver}\ \emph {et~al.}(2010)\citenamefont {Olver},
  \citenamefont {Lozier}, \citenamefont {Boisvert},\ and\ \citenamefont
  {Clark}}]{olver:nist_handbook}%
  \BibitemOpen
  \bibfield  {author} {\bibinfo {author} {\bibfnamefont {F.~W.}\ \bibnamefont
  {Olver}}, \bibinfo {author} {\bibfnamefont {D.~W.}\ \bibnamefont {Lozier}},
  \bibinfo {author} {\bibfnamefont {R.~F.}\ \bibnamefont {Boisvert}},\ and\
  \bibinfo {author} {\bibfnamefont {C.~W.}\ \bibnamefont {Clark}},\ }\href@noop
  {} {\emph {\bibinfo {title} {NIST Handbook of Mathematical Functions}}},\
  \bibinfo {edition} {1st}\ ed.\ (\bibinfo  {publisher} {Cambridge University
  Press},\ \bibinfo {address} {New York, NY, USA},\ \bibinfo {year}
  {2010})\BibitemShut {NoStop}%
\bibitem [{\citenamefont {Pearce}(1981)}]{pearce:jsp}%
  \BibitemOpen
  \bibfield  {author} {\bibinfo {author} {\bibfnamefont {P.~A.}\ \bibnamefont
  {Pearce}},\ }\bibfield  {title} {\bibinfo {title} {Mean-field bounds on the
  magnetization for ferromagnetic spin models},\ }\href
  {https://doi.org/10.1007/BF01022189} {\bibfield  {journal} {\bibinfo
  {journal} {Journal of Statistical Physics}\ }\textbf {\bibinfo {volume}
  {25}},\ \bibinfo {pages} {309} (\bibinfo {year} {1981})}\BibitemShut
  {NoStop}%
\bibitem [{\citenamefont {Joshi}\ and\ \citenamefont
  {Bissu}(1991)}]{joshi:australian_math_soc}%
  \BibitemOpen
  \bibfield  {author} {\bibinfo {author} {\bibfnamefont {C.~M.}\ \bibnamefont
  {Joshi}}\ and\ \bibinfo {author} {\bibfnamefont {S.~K.}\ \bibnamefont
  {Bissu}},\ }\bibfield  {title} {\bibinfo {title} {Some inequalities of
  {B}essel and modified {B}essel functions},\ }\href
  {https://doi.org/10.1017/S1446788700032791} {\bibfield  {journal} {\bibinfo
  {journal} {Journal of the Australian Mathematical Society. Series A. Pure
  Mathematics and Statistics}\ }\textbf {\bibinfo {volume} {50}},\ \bibinfo
  {pages} {333–342} (\bibinfo {year} {1991})}\BibitemShut {NoStop}%
\bibitem [{\citenamefont {Gupta}\ \emph
  {et~al.}(2014{\natexlab{b}})\citenamefont {Gupta}, \citenamefont {Campa},\
  and\ \citenamefont {Ruffo}}]{gupta:jsm}%
  \BibitemOpen
  \bibfield  {author} {\bibinfo {author} {\bibfnamefont {S.}~\bibnamefont
  {Gupta}}, \bibinfo {author} {\bibfnamefont {A.}~\bibnamefont {Campa}},\ and\
  \bibinfo {author} {\bibfnamefont {S.}~\bibnamefont {Ruffo}},\ }\bibfield
  {title} {\bibinfo {title} {Kuramoto model of synchronization: equilibrium and
  nonequilibrium aspects},\ }\href
  {http://stacks.iop.org/1742-5468/2014/i=8/a=R08001} {\bibfield  {journal}
  {\bibinfo  {journal} {Journal of Statistical Mechanics: Theory and
  Experiment}\ }\textbf {\bibinfo {volume} {2014}},\ \bibinfo {pages} {R08001}
  (\bibinfo {year} {2014}{\natexlab{b}})}\BibitemShut {NoStop}%
\end{thebibliography}%

%
\end{document}